%% file: root.tex
\documentclass{ieeeconf}    % Enable this line and disable the
\usepackage{graphicx}          % Include this line if your
\usepackage{amsmath,amssymb,amsfonts}
\usepackage[utf8]{inputenc}
\usepackage[english]{babel}
\usepackage{hyperref}

\let\labelindent\relax

\usepackage{enumitem}
\usepackage{subcaption}
\usepackage{mathrsfs}
\usepackage{makecell}
\usepackage{tikz}
\usepackage{pgfplots}

\pgfplotsset{compat=newest}
\pgfplotsset{plot coordinates/math parser=false}
\pgfplotsset{every axis plot/.append style={line width=1.25pt}, every axis/.append style={line width=0.7pt},  every axis plot post/.append style={every mark/.append style={line width=0.75pt}}}
\usetikzlibrary{arrows}

\newcommand{\R}{\ensuremath{\mathbb{R}}}
\newcommand{\N}{\ensuremath{\mathbb{N}}}
\newcommand{\X}{\ensuremath{\mathbb{X}}}
\newcommand{\V}{\ensuremath{\mathbb{V}}}

\newcommand{\Linf}{\ensuremath{\mathcal{L}_\infty}}
\newcommand{\W}{\ensuremath{\mathcal{W}}}
\newcommand{\A}{\ensuremath{\mathcal{A}}}
\newcommand{\PC}{\ensuremath{\mathcal{PC}}}

\newcommand{\yest}{\ensuremath{\widehat{\kern-1pt\widetilde{y}}\kern1pt}}
\newcommand{\xest}{\ensuremath{\widehat{\kern-1pt\widetilde{x}}\kern1pt}}

\newcommand{\what}{\ensuremath{\widehat{\kern-0.5pt w}}}
\newcommand{\xhat}{\ensuremath{\widehat{\kern-0.5pt x}}}
\def\qedsymbol{$\blacksquare$}

\DeclareMathOperator*{\dom}{dom}
\DeclareMathOperator*{\ess}{ess}
\DeclareMathOperator{\proj}{\Pi}
\DeclareMathOperator{\diag}{diag}
\DeclareMathOperator{\interior}{int}

\newtheorem{theorem}{Theorem}
\newtheorem{lemma}{Lemma}
\newtheorem{definition}{Definition}
\newtheorem{remark}{Remark}
\newtheorem{assumption}{Assumption}

\newtheorem{corollary}{Corollary}
\newtheorem{proposition}{Proposition}
\let\leq\leqslant
\let\geq\geqslant

\begin{document}

\title{Robustifying Event-Triggered Control to Measurement Noise}%: A Space-Regularization Approach}

\author{K.J.A. Scheres, R. Postoyan, W.P.M.H. Heemels
\thanks{This work is supported by the ANR via grant HANDY 18-CE40-0010.}
\thanks{Koen Scheres and Maurice Heemels are with the Control Systems Technology section, Department Mechanical Engineering, Eindhoven University of Technology, The Netherlands.
    {\tt\small \{k.j.a.scheres,m.heemels\}@tue.nl}}%
\thanks{Romain Postoyan is with the Universit\'e de Lorraine, CNRS, CRAN, F-54000 Nancy, France.
        {\tt\small romain.postoyan@univ-lorraine.fr}}%
}
\maketitle

\begin{abstract}                          % Abstract of not more than 200 words.
    While many event-triggered control strategies are available in the literature, most of them are designed ignoring the presence of measurement noise. As measurement noise is omnipresent in practice and can have detrimental effects, for instance, by inducing Zeno behavior in the closed-loop system and with that the lack of a positive lower bound on the inter-event times, rendering the event-triggered control design practically useless, it is of great importance to address this gap in the literature. To do so, we present a general framework for set stabilization of (distributed) event-triggered control systems affected by additive measurement noise. It is shown that, under general conditions, Zeno-free static as well as dynamic triggering rules can be designed such that the closed-loop system satisfies an input-to-state practical set stability property. We ensure Zeno-freeness by proving the existence of a uniform strictly positive lower-bound on the minimum inter-event time. The general framework is applied to point stabilization and consensus problems as particular cases, where we show that, under similar assumptions as the original work, existing schemes can be redesigned to robustify them to measurement noise. Consequently, using this framework, noise-robust triggering conditions can be designed both from the ground up \emph{and} by simple redesign of several important existing schemes. Simulation results are provided that illustrate the strengths of this novel approach.
\end{abstract}

\section{Introduction}\label{ch:1}
In recent years, event-triggered control (ETC), see, e.g., \cite{Heemels2012,Nowzari2019} and the references therein, has been studied extensively as a resource-aware sampling paradigm, as an alternative to periodic time-triggered control, reducing the computational burden and/or the communication bandwidth of the control strategies, while still ensuring important closed-loop stability and performance properties. In ETC, the sampling or transmission instants are decided on the basis of well-designed state- or output-based triggering conditions, rendering these instants not necessarily periodic. The general idea in ETC is to allow more flexibility in the sampling and communication processes and adapt the communication to the system needs according to the desired objectives.

Most literature on ETC for continuous-time systems assumes that perfect state or output information is available for control, see, e.g., \cite{Tabuada2007,Girard2015}. In most physical systems, this assumption is typically not satisfied as essentially all sensors are susceptible to measurement noises. The presence of measurement noises may cause the absence of a positive lower bound on the inter-event times, and, even Zeno behavior (an infinite number of transmission times in a finite time interval) if not carefully handled, as demonstrated in, e.g., \cite{Borgers2014}. If an ETC scheme exhibits Zeno behavior, it is practically useless as it cannot be implemented and certainly is not saving communication resources compared to time-triggered periodic control. Therefore, establishing strong Zeno-freeness is important, not only for implementability and saving resources, as an analytic lower bound on the inter-event times determines the worst-case scenario in terms of resource utilization, but also for the stability analysis and proofs to be complete and meaningful.

Few solutions have been proposed in the literature to address this problem, see, e.g., \cite{Mousavi2019,Abdelrahim2017}. However, these rely on restrictive assumptions on the noise signal, in particular, the noise has to be differentiable and its derivative has to be bounded in an $\mathcal{L}_\infty$ sense. Moreover, the ensured input-to-state stability (ISS) or $\mathcal{L}_p$-stability of the closed-loop system holds with respect to the noise \emph{and} its time-derivative. When dealing with real sensors, the differentiability condition and global boundedness of the derivative of the noise may not be natural assumptions. The observer-based approach in \cite{Lehmann2011} overcomes this issue, but these results only apply to linear systems and involve multiple additional internal models, thereby requiring extra processing power and energy to run. An alternative result is studied in \cite{Selivanov2016}, where a periodic event-triggered controller (PETC), in the sense that the triggering rule is only evaluated at some periodically spaced discrete instants, is run simultaneously with a continuous event-triggered controller (CETC), and transmission occurs when the triggering conditions of both controllers hold. The downside to this particular method is that, if the state is close to the origin, transmissions occur periodically, hence, the communication benefit of ETC might not be preserved. ETC design under measurement noise becomes even harder when designing distributed event-triggered controllers for consensus \cite{Nowzari2019}. We are aware of only one work dealing with measurement noise in this context, \cite{Hardouin2019}, where the control input is integrated to estimate an upper-bound for the error. Due to the use of an upper-bound on the error and the use of an absolute fixed threshold condition, the amount of controller updates (network bandwidth) required may become relatively large compared to other ETC consensus algorithms, see, e.g., \cite{Dolk2019}. With this in mind, there is a strong need for event-triggered controllers applicable to systems where the available output information is corrupted by (additive) measurement noise, where more natural conditions are imposed on the type of noise signals.

In this context, we are interested in a general framework to design event-triggered controllers robust to measurement noise. The framework that we present is based on space-regularized (fixed threshold) ETC, see, e.g., \cite{Miskowicz2006}, in line with classical event generators, such as \cite{Tabuada2007,Girard2015,Donkers2012}. To analyze the resulting ETC closed-loop system, we present a new hybrid model, in which a jump models a transmission. The model does not involve the derivative of the noise as opposed to \cite{Mousavi2019,Abdelrahim2017}. This new model is instrumental, and, based on it, we provide general prescriptive conditions, under which both static and dynamic triggering rules can be designed, to ensure an input-to-state practical stability property, while ruling out Zeno phenomena. In particular, we will show that applying space-regularization, i.e., enlarging the ``flow set'' of the hybrid model, needs to be done with care to ensure the existence of a strictly positive (semi-global) minimum inter-event time, which only requires that an upper-bound of the noise level is known. Our results apply to the general scenario where $N$ plants, possibly interconnected, are each controlled by an event-triggered controller. We thereby cover both classical point stabilization problems ($N=1$) as in, e.g., \cite{Tabuada2007,Girard2015,Donkers2012,Dolk2017}, where we also extend the results to output-feedback control, and consensus problems ($N>1$) as in, e.g., \cite{Dolk2017b}, in a unified way. Moreover, we explain how our framework leads to modifications of the triggering rules presented in \cite{Tabuada2007,Girard2015} to ensure Zeno-freeness in presence of measurement noise. We also apply the framework to consensus seeking problems, where we show that we can maintain ``long'' inter-event times even in the presence of measurement noise. We show this, for instance, for ``robustified'' versions of \cite{Dolk2019,Garcia2013}. Lastly, we present numerical case studies to show the effectiveness of our technique and to demonstrate the implications of applying space-regularization.

This work generalizes the results of our preliminary version \cite{Scheres2020}. Compared to \cite{Scheres2020}, where only static state-feedback controllers were considered, we include several extensions, such as output-feedback controllers, more general holding functions and dynamical disturbances. Moreover, the full proofs are provided here, which are not available in \cite{Scheres2020}. %Additionally, this work includes new results on the existence and completeness of solutions for hybrid general systems with inputs, which is of independent interest.

The remainder of this paper is structured as follows. In Section \ref{ch:2}, we present the preliminaries and notational conventions. Section \ref{ch:3} contains the problem statement. We present the hybrid model and the framework in Section \ref{ch:4}. The main results are given in Section \ref{ch:5}. We apply the framework to case studies in Section \ref{ch:6}. Finally, we illustrate the obtained results numerically in Section \ref{ch:7}, and provide conclusions in Section \ref{ch:8}.

\section{Preliminaries}\label{ch:2}
\subsection{Notation}
The sets of all non-negative and positive integers are denoted $\N$ and $\N_{>0}$, respectively. The fields of all reals and all non-negative reals are indicated by $\R$ and $\R_{\geq0}$, respectively. The identity matrix of size $N\times N$ is denoted by $I_N$, and the vectors in $\R^N$ whose elements are all ones or zeros are denoted by $\mathbf{1}_N$ and $\mathbf{0}_N$, respectively; the index of $\mathbf{0}_N$ or $\mathbf{1_N}$ is dropped when clear from the context. The $N\times M$ zero matrix is denoted $\mathbf{0}_{N,M}$. For $N$ vectors $x_i\in\R^{n_i}$, we use the notation $(x_1,x_2,\ldots,x_N)$ to denote $\begin{bmatrix}x_1^\top&x_2^\top&\ldots&x_N^\top\end{bmatrix}^\top$. Given a pair of square matrices $A_1,\ldots,A_n$, we denote by $\diag(A_1,\ldots,A_n)$ the block-diagonal matrix where the main diagonal blocks consist of the matrices $A_1$ to $A_n$ and all other blocks are zero matrices. By $\langle\cdot,\cdot\rangle$ and $|\cdot|$ we denote the usual inner product of real vectors and the Euclidean norm, respectively. %For a symmetric matrix $A$, by $A\succ0$ ($A\succeq0$) and $A\prec0$ ($A\preceq0$) we denote that $A$ is positive definite (positive semidefinite) and negative definite (negative semidefinite), respectively.
%For $\mathcal{W}\subseteq\R^{n}$ and $n\in\N_{>0}$, we define the class of inputs ${\mathcal{L}}_{\mathcal{W}}$ as all functions (defined for all $t\in\R_{\geq0}$) that are Lebesgue measurable and locally essentially bounded with $w(t)\in\mathcal{W}$ for all $t\in\R_{\geq0}$.
For a measurable signal $w:\R_{\geq0}\to\R^{n_w}$, we denote by $\|w\|_\infty:=\ess\sup_{t\in\R_{\geq0}}|w(t)|$ its $\mathcal{L}_\infty$-norm provided it is finite, in which case we write $w\in\mathcal{L}_\infty$. A function $w : \R_{\geq0} \rightarrow \R^{n_w}$ is said to be c\`adl\`ag ``continue \`a droite, limite \`a gauche'', denoted by $w \in\mathcal{PC}$, when there exists a sequence $\{t_i\}_{i\in\N}$ with $t_{i+1}>t_i >t_0 =0$ for all $i\in\N$ and $t_i\rightarrow\infty$ when $i\rightarrow\infty$ such that $w$ is continuous on $(t_i,t_{i+1})$ where $\lim_{t\uparrow t_i} w(t)$ exists for all $i\in\N_{>0}$ and $\lim_{t\downarrow t_i} w(t)$ exists for all $i\in\N$ with $\lim_{t\downarrow t_i} w(t) = w(t_i)$, i.e., $w$ is piecewise continuous, right continuous and left limits exist for each $t_i$, $i\in\N_{>0}$. Given a set $\mathcal{W}\subseteq\R^{n_w}$, we denote by $\PC_\W$ the set of functions $\{w\in\PC\mid w(t)\in\W\text{ for all }t\in\R_{\geq0}\}$. Note that continuous functions are contained in $\mathcal{PC}$, $\{t_i\}_{i\in\mathcal{N}}$ can be chosen arbitrarily then.
For any $x\in\R^N$, the distance to a closed non-empty set $\mathcal{A}$ is denoted by $|x|_\mathcal{A}:=\min_{y\in\mathcal{A}}|x-y|$. The interior of a set $\mathcal{A}$ is denoted $\interior\mathcal{A}$. %The closure of a set $\mathcal{A}$ is denoted by $\overline{\mathcal{A}}$.
Given a vector $x\in\R^{n_x}$ and a set $\mathcal{A}\subseteq\R^{n_x}\times\R^{n_y}$, $\proj_x(\A)$ denotes the projection of $\A$ onto the $x$-plane $\R^{n_x}$, i.e., $\proj_x(\A)=\{z\in\R^{n_x}\mid \exists y\in\R^{n_y}\text{ s.t. }(z,y)\in\A\}$.
%We use the usual definitions for comparison functions.
%A function $\alpha:\R_{\geq0}\to\R_{\geq0}$ is a class-$\mathcal{K}$ function, if it is continuous, strictly increasing and $\alpha(0) = 0$ and it is a class-$\mathcal{K}_\infty$ function, if, in addition, $\alpha(r)\to\infty$ as $r\to\infty$. A function $\beta:\R_{\geq0}\times\R_{\geq0}\to\R_{\geq0}$ is a class-$\mathcal{KL}$ function if it is continuous and, for each fixed $s\geq0$, the mapping $\beta(\cdot,s)$ is a class-$\mathcal{K}$ function and, for each fixed $r$, the mapping $\beta(r,s)$ is decreasing with respect to $s$ and $\beta(r,s)\to0$ as $s\to\infty$.
We consider $\mathcal{K}$, $\mathcal{K}_\infty$ and $\mathcal{KL}$ functions as defined in \cite[Chapter 3]{hybridsystems}.
By $\land$ and $\lor$ we denote the logical \emph{and} and \emph{or} operators respectively. %We use $U^{\circ}(x;v)$ to denote the generalized directional derivative of Clarke of a locally Lipschitz function $U$ at $x$ in the direction $v$, \emph{i.e.}, $U^{\circ}(x;v):=\lim\sup_{h\rightarrow 0^+,\, y\rightarrow x}(U(y+hv)-U(y))/h$, which reduces to the standard directional derivative $\left\langle \nabla U(x),v\right\rangle$ when $U$ is continuously differentiable.

\subsection{Graph theory}
A weighted graph $\mathscr{G}:=(\mathscr{V},\mathscr{E},A)$ consists of a vertex set $\mathscr{V}:=\{1,2,...,N\}$, a set of edges $\mathscr{E}\subset\mathscr{V}\times\mathscr{V}$ and an adjacency matrix $A\in\R^{N\times N}$. %The number of vertices or cardinality of $\mathcal{V}$ is denoted by $N\in\mathbb{N}_{>0}$.
An ordered pair $(i,j)\in\mathscr{E}$, with $i,j\in\mathscr{V}$, is an edge from $i$ to $j$. For an edge $(i,j)\in\mathscr{E}$, $i$ is called the \emph{in}-neighbor of $j$, and $j$ is called the \emph{out}-neighbor of $i$. All $(i,j)\in\mathscr{E}$ have an associated weight, denoted $\alpha_{ij}\in\R_{>0}$. The adjacency matrix $A:=(a_{i,j})$, $i,j\in\mathscr{V}$ of a graph is defined as
\begin{equation}
a_{i,j}:=\begin{cases}
\alpha_{ij}&\mathrm{if}\;(i,j)\in\mathscr{E},\\
0&\mathrm{otherwise}.
\end{cases}
\end{equation}
The set $\mathscr{V}_i^\mathrm{in}$ of the in-neighbors of $i$ is defined as $\mathscr{V}_i^\mathrm{in}:=\{j\in\mathscr{V}\;|\;(j,i)\in\mathscr{E}\}$ and the set of out-neighbors as $\mathscr{V}_i^\mathrm{out}:=\{j\in\mathscr{V}\;|\;(i,j)\in\mathscr{E}\}$. An undirected graph is a graph where, for any edge $(i,j)\in\mathscr{E}$, $(j,i)$ is also in $\mathscr{E}$. A sequence of edges $(i,j)\in\mathscr{E}$ connecting two vertices is called a directed path. For a connected graph $\mathscr{G}$, there exists a path between any two vertices in $\mathscr{V}$. The in-degree is defined as $d_i^\mathrm{in}:=\sum_{j\in\mathscr{V}^\mathrm{in}_i}\alpha_{ji}$ and the out-degree as $d_i^\mathrm{out}:=\sum_{j\in\mathscr{V}^\mathrm{out}_i}\alpha_{ij}$. The in-degree matrix $D^\mathrm{in}$ and out-degree matrix $D^\mathrm{out}$ are diagonal matrices with $d_i^\mathrm{in}$ respectively $d_i^\mathrm{out}$ as the $i$th diagonal element. A weight-balanced digraph (directed graph) is a digraph where $d_i^\mathrm{out}=d_i^\mathrm{in}$ for all $i$. The Laplacian $L$ of a graph $\mathscr{G}$ is defined as $L:=D^\mathrm{out}-A$. For an undirected graph, $D^\mathrm{in}:=D^\mathrm{out}$.

\subsection{Hybrid systems}\label{sec:prelim_hs}
Based on the formalism of \cite{hybridsystems}, we model hybrid systems $\mathcal{H}(F,\mathcal{C},G,\mathcal{D},\X,\V)$ as
\begin{equation}
    \begin{cases}
        \begin{aligned}
            \dot{\xi}&\in F(\xi,\nu)\\
            \xi^+&\in G(\xi,\nu)
        \end{aligned}
        &\quad
        \begin{aligned}
            (\xi,\nu)&\in\mathcal{C},\\
            (\xi,\nu)&\in\mathcal{D},
        \end{aligned}
    \end{cases}\label{eq:hybridsys}
\end{equation}
where $\xi\in\X\subseteq\R^{n_\xi}$ denotes the state, $\nu$ an external input taking values in $\V\subseteq\R^{n_v}$, $\mathcal{C}\subseteq\X\times\V$ the flow set, $\mathcal{D}\subseteq\X\times\V$ the jump set, $F:\X\times\V\rightrightarrows\R^{n_\xi}$ the (set-valued) flow map and $G:\X\times\V\rightrightarrows\R^{n_\xi}$ the (set-valued) jump map. Sets $\mathcal{C}$ and $\mathcal{D}$ are assumed to be closed. We refer to \cite{hybridsystems} for notions related to \eqref{eq:hybridsys} such as hybrid time domains or hybrid arcs. For a hybrid time domain $E$, $\sup_t E:=\sup\left\{t\in\R_{\geq0}:\exists j\in\N\;\text{such that}\;(t,j)\in E\right\}$, $\sup_j E:=\sup\left\{j\in\N:\exists t\in\R_{\geq0}\;\text{such that}\;(t,j)\in E\right\}$ and $\sup E:=(\sup_t E,\sup_j E)$. We consider the notion of solutions proposed in \cite{Heemels2021}.

\begin{definition}[\cite{Heemels2021}] Given $\nu\in\PC_\V$, a hybrid arc $\phi$ is a \emph{solution}\footnote{This corresponds to the notion of \emph{e-solution}, see \cite[Definition 4.3]{Heemels2021}.} to $\mathcal{H}$, if
    \begin{itemize}
        \item[(S1)] for all $j\in\N$ such that $I^j:=\{t \mid (t,j) \in \dom \phi\}$ has nonempty interior, it holds that $\dot\phi(t,j)\in F(\phi(t,j),\nu(t))$ for almost all $t\in\interior I^j$ and $(\phi(t,j),\nu(t))\in\mathcal{C}$ for all $t\in\interior I^j$;
        \item[(S2)] for all $(t,j)\in\dom\phi$ such that $(t,j+1)\in\dom\phi$, $(\phi(t,j),\nu(t))\in\mathcal{D}$ and $\phi(t,j+1)\in G(\phi(t,j),\nu(t))$.
    \end{itemize}
\end{definition}
We will also use the following definitions.
\begin{definition}
    Given an input $\nu\in\PC_\V$, a solution $\phi$ is called {\em non-trivial}, if $\dom\phi$ contains at least two points. We say that $\phi$ is {\em maximal}, if there does not exist another solution $\psi$  to $\mathcal{H}$ for the same input $\nu$ such that $\dom\phi$ is a proper subset of $\dom\psi$ (i.e., $\dom\phi\subset\dom\psi$, but $\dom\phi\neq\dom\psi$) and $\phi(t,j)=\psi(t,j)$ for all $(t,j)\in\dom\phi$. We denote the set of all maximal solutions to $\mathcal{H}$ for input $\nu$ by $\mathcal{S}_\mathcal{H}(\nu)$. We say that the solution $\phi$ is {\em complete}, if $\dom\phi$ is unbounded, and we say that it is {\em $t$-complete}, if $\sup_t\dom \phi = \infty$. We say that $\mathcal{H}$ is \emph{persistently flowing} if all maximal solutions for all $\nu\in\mathcal{PC}_\V$ are t-complete.
\end{definition}

\begin{remark}\label{rem:measurableinput}
    If $\mathcal{C}$ and $\mathcal{D}$ do not depend on the input $\nu$, the inputs can be taken as measurable functions instead of piecewise continuous; see \cite{Heemels2021} for further insights on this point.
\end{remark}

In this paper, we are interested in systems $\mathcal{H}$ that are persistently flowing and we focus on the stability notions below.
\begin{definition}
    \label{def:isps}For a persistently flowing hybrid system $\mathcal{H}$, a non-empty closed set $\mathcal{A}\subset\R^{n_\xi}$ is \emph{input-to-state practically stable} (ISpS), if there exist $\gamma\in\mathcal{K}$, $\beta\in\mathcal{KL}$ and $d\in\R_{\geq0}$ such that for any input $\nu\in\PC_\V$ and any associated solution $\xi$,
\begin{equation}
    |\xi(t,j)|_\mathcal{A}\leq\beta(|\xi(0,0)|_\mathcal{A},t)+\gamma(\|\nu\|_\infty)+d,\label{eq:ispsdef}
\end{equation}
for all $(t,j)\in\dom\xi$. If \eqref{eq:ispsdef} holds with $d=0$, then $\mathcal{A}$ is said to be \emph{input-to-state stable} (ISS) for $\mathcal{H}$.
\end{definition}
To prove that a given non-empty, closed set $\mathcal{A}$ is IS(p)S, we will use the following Lyapunov conditions; recall that, when $\mathcal{H}$ is persistently flowing, necessarily $G(D)\times\V\subset\mathcal{C}\cup\mathcal{D}$, otherwise, not all maximal solutions would be complete. %We use below the notation $\Pi_\xi(\mathcal{Q})$ for a given set $\mathcal{Q}\subseteq\X\times\V$, where $\Pi_\xi(\mathcal{Q}):=\{\xi\in\X\mid\exists\nu\in\V,(\xi,\nu)\in\mathcal{Q}\}$.
\begin{proposition} \label{prop:lyapunovisps} Suppose $\mathcal{H}$ is persistently flowing and let $\mathcal{A}\subset\R^{n_\xi}$ be a non-empty closed set. If there exist $V:\dom V\to\R_{\geq0}$, $\alpha,\underline{\alpha},\overline{\alpha}\in\mathcal{K}_\infty$, $\gamma\in\mathcal{K}$ and $c\in\R_{\geq0}$ such that
\begin{enumerate}[label={\roman*)}]%\itemsep0ex
    \item $\Pi_\xi(\mathcal{C}\cup\mathcal{D})\subset\dom V$ and $V$ is continuously differentiable on an open set containing $\Pi_\xi(\mathcal{C})$,
    \item for any $\xi\in\X$,
    \[\underline{\alpha}(|\xi|_\mathcal{A})\leq V(\xi)\leq\overline{\alpha}(|\xi|_\mathcal{A}),\]
    \item for all $(\xi,\nu)\in\mathcal{C}$ and all $f\in F(\xi,\nu)$,
    \[\left\langle\nabla V(\xi),f\right\rangle\leq-\alpha(|\xi|_\mathcal{A})+\gamma(|\nu|)+c,\]
    \item for all $(\xi,\nu)\in\mathcal{D}$ and all $g\in G(\xi,\nu)$,
    \[V(g)-V(\xi)\leq0,\]
\end{enumerate}
then $\mathcal{A}$ is ISpS. If, moreover, $c=0$ in item iii), then $\A$ is ISS.
\end{proposition}
\textbf{Sketch of proof}
Let $\nu\in\PC_\V$ and $\xi$ be an associated solution to $\mathcal{H}$, $(t,j)\in\dom\xi$ and $0=t_{0}\leq t_{1}\leq\ldots\leq t_{j+1}=t$ the jump times of $\xi$ which satisfy $ \dom \xi \cap ([0,t]\times\{0,1,\ldots,j\}) =\allowbreak {\underset{i\in\{0,1,\ldots,j\}}\bigcup}[t_{i},t_{i+1}]\times\{i\}$.
For each $i\in\{1,2,\ldots,j\}$ and for almost all $s\in[t_{i},t_{i+1}]$, item 3) of Proposition \ref{prop:lyapunovisps} implies that $\left\langle\nabla V(\xi(s,i)),\dot\xi(s,i)\right\rangle\leq-\alpha(|\xi(s,i)|_\mathcal{A})+\gamma(|\nu(s)|)+c$. Similar arguments as in \cite[Lemma 2.14]{Sontag1995} complete the proof as 1) $V$ does not increase at jumps due to item iv) of Proposition \ref{prop:lyapunovisps}, 2) item ii) holds, and 3) $\mathcal{H}$ is persistently flowing. \hfill\qedsymbol

\section{Problem formulation}\label{ch:3}
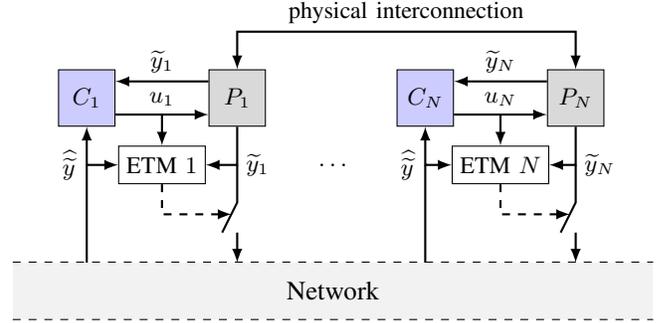
\begin{figure}[t]
    \tikzstyle{controller} = [draw, fill=blue!20, rectangle, minimum height=0.75cm, minimum width=0.75cm]
    \tikzstyle{dots} = [draw=none, rectangle, minimum height=0.75cm, minimum width=0.75cm]
    \tikzstyle{plant} = [draw, fill=gray!30, rectangle, minimum height=0.75cm, minimum width=0.75cm]
    \tikzstyle{network} = [draw=none, fill=gray!10, rectangle, minimum height=0.75cm, minimum width=8.5cm]
    \tikzstyle{generator} = [draw, rectangle, minimum height=0.5cm, minimum width=1cm]
    \small
    \centering
    \begin{tikzpicture}[node distance=1.5cm,>=latex]
        \node [controller] (c1) {$C_1$};
        \node [plant,right of=c1, node distance=2cm] (p1) {$P_1$};

        \draw [->, thick] (c1.330)--(p1.210) node[midway, above] (u1) {$u_1$};
        \draw [<-, thick] (c1.30)--(p1.150) node[midway, above] (y1b) {$\widetilde{y}_1$};
        \node [generator,below of=u1, node distance=0.875cm] (ETM1) {ETM $1$};

        \draw [->, thick] (u1)--(ETM1) {};
        \draw [->, thick] (p1)|-(ETM1) node[midway, right] (y1) {$\widetilde{y}_1$};

        \draw [-, thick] (y1.west) -- ++(0,-0.5) -- ++(-0.2,-0.4) {};
        \draw [->, thick] (y1.west) + (0,-0.9) -- ++(0,-1.3) {};
        \draw [->, dashed, thick] (ETM1.south) |- ++(0.925,-0.4) {};

        \draw [<->, thick] (c1.south)|-(ETM1.west) node[midway,left] (yhat1) {$\yest$};
        \draw [-, thick] (yhat1.east) -- ++(0,-1.5) {};

        \node[draw=none, right of=y1, node distance=1cm] (dots) {$\ldots$};

        \node [controller, right of=p1, node distance=2.5cm] (cN) {$C_N$};
        \node [plant,right of=cN, node distance=2cm] (pN) {$P_N$};

        \draw [->, thick] (cN.330)--(pN.210) node[midway, above] (uN) {$u_N$};
        \draw [<-, thick] (cN.30)--(pN.150) node[midway, above] (yNb) {$\widetilde{y}_N$};
        \node [generator,below of=uN, node distance=0.875cm] (ETMN) {ETM $N$};

        \draw [->, thick] (uN)--(ETMN) {};
        \draw [->, thick] (pN)|-(ETMN) node[midway, right] (yN) {$\widetilde{y}_N$};

        \draw [-, thick] (yN.west) -- ++(0,-0.5) -- ++(-0.2,-0.4) {};
        \draw [->, thick] (yN.west) + (0,-0.9) -- ++(0,-1.3) {};
        \draw [->, dashed, thick] (ETMN.south) |- ++(0.925,-0.4) {};

        \draw [<->, thick] (cN.south)|-(ETMN.west) node[midway,left] (yhatN) {$\yest$};
        \draw [-, thick] (yhatN.east) -- ++(0,-1.5) {};

        \node [network, below of=dots, node distance=1.675cm] (net) {\normalsize Network};
        \draw[dashed](net.north west)--(net.north east) (net.south west)--(net.south east);

        %\draw[<->, dashdotted] (p1.north east) to [out=30,in=150] node[above]{physical interconnection} (pN.north west);
        \draw[<->, thick] (p1.north) -- ++(0,0.5) -| (pN.north);
        \path (p1.north) -- (pN.north) node[midway,above,yshift=0.5cm] (label) {physical interconnection};

    \end{tikzpicture}
    \caption{Networked control setup with Event-Triggering Mechanism (ETM). ETM $i$ determines when the current noisy output $\widetilde{y}_i$ is transmitted over the network.}
    \label{fig:nc}
\end{figure}
We consider a collection of $N\in\N_{>0}$ interconnected plants ${P}_1,{P}_2,\ldots,{P}_N$. Each plant ${P}_i$, $i\in\mathcal{N}:=\{1,2,\ldots,N\}$, is equipped with a sensor that communicates its measured output, which is affected by measurement noise, to the controllers ${C}_1,{C}_2,\ldots,{C}_N$ via a digital packet-based network, see Fig. \ref{fig:nc}. Plant ${P}_i$, $i\in\mathcal{N}$, has state $x_{p,i}\in\R^{n_{x_{p,i}}}$, \emph{ideal} output $y_{i}\in\R^{n_{y,i}}$ and \emph{measured} output $\widetilde{y}_i\in\R^{n_{y,i}}$, affected by noise, with dynamics
\begin{equation}
    P_i:\begin{cases}
        \begin{aligned}
            \dot{x}_{p,i}&=f_{p,i}(x_p,u_i,v_i),\\
            y_i&=g_{p,i}(x_{p,i}),\\
            \widetilde{y}_i&=g_{p,i}(x_{p,i})+w_i=y_i+w_i,
        \end{aligned}
        \label{eq:sysdyn}
    \end{cases}
\end{equation}
where $u_i\in\R^{n_{u,i}}$ is the control input of ${P}_i$, $x_p:=(x_{p,1},x_{p,2},\ldots,x_{p,N})$ is the concatenated plant state, $v_i\in\R^{n_{v,i}}$ is a process disturbance, $w_i\in\R^{n_{y,i}}$ is the $i$-th measurement noise, $f_{p,i}:\R^{n_p}\times\R^{n_{u,i}}\times\R^{n_{v,i}}\to\R^{n_{x_{p,i}}}$ is continuous and $g_{p,i}:\R^{n_{x_{p,i}}}\to\R^{n_{y,i}}$ is continuously differentiable, where $n_p:=\sum_{i\in\mathcal{N}}n_{x_{p,i}}$. Note that $f_{p,i}$ may depend on the states of other plants, and, as such, physical couplings are allowed, as illustrated in Fig. \ref{fig:nc}. We assume that the process disturbances $v_i$ and the measurement noises $w_i$ satisfy the following assumption.
\begin{assumption}
    For each $i\in\mathcal{N}$, $v_i\in\Linf\cap\PC$ and $w_i\in\mathcal{PC}_{\mathcal{W}_i}$, where $\mathcal{W}_i:=\left\{w_i\in\R^{n_{y,i}}\;\big|\;|w_i|\leq\overline w_i\right\}$ for some known $\overline w_i\in\R_{\geq0}$.\label{ass:noise}
\end{assumption}
Assumption \ref{ass:noise} imposes natural boundedness conditions on the process disturbance and the noise and it does not impose restrictions on the existence or boundedness of their derivatives, as was required in, e.g., \cite{Mousavi2019,Abdelrahim2017}.

The controller ${C}_i$ with state $x_{c,i}\in\R^{n_{x_{c,i}}}$ and $n_{x_{c,i}}\in\N$, $i\in\mathcal{N}$, is given by
\begin{equation}
    C_i:\begin{cases}
        \begin{aligned}
            \dot{x}_{c,i}&=f_{c,i}(x_{c,i},\widetilde{y}_i,\yest\,),\\
            u_i&=g_{c,i}(x_{c,i},\widetilde{y}_i,\yest\,),
        \end{aligned}
    \end{cases}\label{eq:ctrldyn}
\end{equation}
with $f_{c,i}:\R^{n_{x_{c,i}}}\times\R^{n_{y,i}}\times\R^{n_y}\to\R^{n_{x_{c,i}}}$ and $g_{c,i}:\R^{n_{x_{c,i}}}\times\R^{n_{y,i}}\times\R^{n_y}\to\R^{n_{u,i}}$ continuous maps, $n_y:=\sum_{i\in\mathcal{N}}n_{y,i}$ and where $\yest$ denotes the sampled ``networked'' version of the outputs, which will be made more precise next. Note that static controllers are included in \eqref{eq:ctrldyn} by taking $n_{x_{c,i}}=0$.%ontrollers that satisfy \eqref{eq:ctrldyn} are, e.g., static controllers (in which case $n_{x_{c,i}}=0$), dynamic controllers such as proportional integral controllers, and observer-based controllers.

The $i$-th sensor, $i\in\mathcal{N}$, broadcasts its output $\widetilde{y}_i$ to the controllers $C_1, C_2, \ldots, C_N$ over the digital network. The corresponding transmissions occur at time instants $t_k^i$, $k\in\N$, which are generated by a local Event-Triggering Mechanism (ETM), which is to be designed. Because of the packet-based communication over the network, the $i$-th controller, which depends on the outputs of $P_j$, $j\in\mathcal{N}$, does not have continuous access to $\widetilde{y}:=(\widetilde{y}_1,\widetilde{y}_2,\ldots,\widetilde{y}_N)$, but only to its estimate $\yest:=(\,\yest_1,\yest_2,\ldots,\yest_N)$ and to its local output $\widetilde{y}_i$. When ETM $i\in\mathcal{N}$, transmits the measured output of plant $i$ over the network, $\yest_i$ is updated according to
\begin{equation}
    \yest_i((t_k^i)^+)=\widetilde{y}_i(t_k^i).
\end{equation}
In between transmissions the estimate evolves according to a holding function $f_{h,i}:\R^{n_{y,i}}\to\R^{n_{y,i}}$, $f_{h,i}$ continuous, i.e.,
\begin{equation}
    \dot{\kern1pt\yest}_i=f_{h,i}(\,\yest_i).\label{eq:estmodel}
\end{equation}
Consequently, each local controller that uses $y_i$ should implement the holding function $f_{h,i}$ locally. Of course, when a controller $C_j$ does not depend on $\tilde{y}_i$, $C_j$ does not need to generate $\yest_i$, but, to derive the model, we proceed as if it would for the sake of notational convenience. The case of a Zero-Order-Hold (ZOH), for instance, corresponds to the choice $f_{h,i}=0$. For modeling purposes, we define $\widehat{y}_i$ and $\what_i$, where
\begin{equation}
    \begin{aligned}
        \widehat{y}_i((t^i_k)^+)&=y(t^i_k),&\dot{\kern-1.5pt\widehat{y}}_i&=f_{h,i}(\widehat{y}_i+\what_i),\\
        \what_i((t^i_k)^+)&=w(t^i_k),&\dot{\what}_i&=0.
    \end{aligned}\label{eq:modelsplit}
\end{equation}
Hence, $\what_i$ is the value of $w_i$ at the last transmission instant of ETM $i$. Due to the aforementioned definitions, we obtain that $\yest_i=\widehat{y}_i+\what_i$.

We define the \emph{ideal} network-induced error $e_i$ as the difference between the sampled output $\widehat{y}_i$ \emph{without} measurement noise and the current output $y_i$ \emph{without} measurement noise:
\begin{equation}\label{eq:e}
    e_i:=\widehat{y}_i-y_i.
\end{equation}
Note that $e_i$ is \emph{not} known by the ETM, and therefore, cannot be used by the corresponding local triggering condition for determining $t_k^i$, $k\in\N$. Hence, we also define the \emph{measured} network-induced error $\widetilde{e}_i$ as the difference between the estimated output $\yest_i$ and the current measured output $\widetilde{y}_i$, which are both affected by noise, i.e.,
\begin{equation}
    \widetilde{e}_i:=\yest_i-\widetilde{y}_i=e_i+\what_i-w_i.
    \label{eq:etil}
\end{equation}
The local ETM at plant $i$ does have access to $\widetilde{e}_i$. We denote the concatenated variables corresponding to \eqref{eq:e} and \eqref{eq:etil} as $e:=(e_1,e_2,\ldots,e_N)$ and $\widetilde{e}:=(\widetilde{e}_1,\widetilde{e}_2,\ldots,\widetilde{e}_N)$, respectively.

We proceed by emulation and assume that the controllers $C_1,C_2,\ldots,C_N$ are designed such that, in closed loop with the plants $P_1,P_2,\ldots,P_N$, the closed-loop system satisfies an input-to-state stability property in the absence of a communication network (i.e., under perfect communication in the sense that $\yest=\widetilde{y}$). We will formalize these properties in Section \ref{ch:5} below. Based on these controllers, which can be designed with any (nonlinear) design tool as long as the assumption stated in Section \ref{ch:5} holds, our objective is to determine the transmission times $t_k^i$, $k\in\N$, for any $i\in\mathcal{N}$, based on suitable local ETMs, to ensure that:
\begin{enumerate}[label={(\alph*)}]
    \item the combined closed-loop system \eqref{eq:sysdyn}, \eqref{eq:ctrldyn} satisfies an input-to-state practical stability property in the presence of measurement noise and process disturbances;
    \item there exists a strictly positive lower-bound on the time between any two transmissions generated by ETM $i$, i.e., for any initial condition there exists a $T_i>0$ such that $t_{k+1}^i-t_k^i\geq T_i$ for all $k\in\N$, $i\in\mathcal{N}$.
\end{enumerate}
This problem formulation is further formalized in the next section in terms of hybrid systems concepts.

\section{Hybrid model}\label{ch:4}
We model the overall system as a hybrid system $\mathcal{H}$ as in Section \ref{sec:prelim_hs}, for which a jump corresponds to the broadcasting of one of the noisy outputs $\widetilde{y}_i$, $i\in\mathcal{N}$, over the network. We allow the local triggering (transmission) conditions to depend on a local auxiliary variable denoted $\eta_i\in\R_{\geq0}$, $i\in\mathcal{N}$, as is the case in dynamic triggering \cite{Girard2015,Dolk2017}. The dynamics of $\eta_i$ is designed in the following. We will also consider static triggering conditions as a special case, in which case $\eta_i$ is not needed.

We define $\eta:=(\eta_1,\eta_2,\ldots,\eta_N)\in\R^N_{\geq0}$, and stack the variables $x:=(x_1,x_2,\ldots,x_N)$ with $x_i:=(x_{p,i},x_{c,i})$, $e:=(e_1,e_2,\ldots,e_N)$ and  $\what:=(\what_1,\what_2,\ldots,\what_N)$ in
\begin{equation}\label{eq:chi}
    \chi:=(x,e,\what).
\end{equation}
The full state for $\mathcal{H}$ becomes $\xi:=(\chi,\eta)=(x,e,\what,\eta)\in\X$, where $\X:=\R^{n_x}\times\R^{n_y}\times\W\times\R^{N}_{\geq0}$, $\W:=\W_1\times\W_2\times\ldots\times\W_N$ with $\W_i$ as defined in Assumption \ref{ass:noise} and $n_x:=\sum_{i\in\mathcal{N}}\left(n_{x_{p,i}}+n_{x_{c,i}}\right)$. We define the concatenated exogenous inputs $\nu:=(v,w)\in\V$, where $\V:=\mathcal{V}\times\W$, $\mathcal{V}:=\R^{n_{v,1}}\times\R^{n_{v,2}}\times\ldots\times\R^{n_{v,N}}$ and with $v:=(v_1,v_2,\ldots,v_N)$ and $w:=(w_1,w_2,\ldots,w_N)$. The flow map $F:\X\times\V\to\X$ can then be written as
\begin{multline}\label{eq:flowmap}
        F(\xi,\nu):=\\(f(x,e,\what,v,w),g(x,e,\what,v,w),\mathbf{0}_{n_y},\bar\Psi(\widetilde{y},\yest,\widetilde{e},u,\eta)).
\end{multline}
Based on \eqref{eq:sysdyn}, \eqref{eq:ctrldyn} and \eqref{eq:etil}, we obtain $f(x,e,\what,v,w):=(f_1(x,e,\what,v_1,w_1),f_2(x,e,\what,v_2,w_2),\allowbreak\ldots,\allowbreak f_N(x,e,\what,v_N,w_N))$, where $f_i:\R^{n_x}\times\R^{n_y}\times\W\times\R^{n_{v,i}}\times\allowbreak\R^{n_{y,i}}\to\R^{n_{x,i}}$ is given by
\begin{align*}
    f_i&(x,e,\what,v_i,w_i):=\\&\begin{bmatrix}
    f_{p,i}\left(x_p,g_{c,i}(x_{c,i},g_{p,i}(x_{p,i})+w_i,g_p(x_p)+e+\what),v_i\right)\\
    f_{c,i}\left(x_{c,i},g_{p,i}(x_{p,i})+w_i,g_p(x_p)+e+\what\right)
    \end{bmatrix}
\end{align*}
with $g_p(x_p):=(g_{p,1}(x_{p,1}),g_{p,2}(x_{p,2}),\ldots,g_{p,N}(x_{p,N}))$. Based on \eqref{eq:sysdyn}, \eqref{eq:estmodel} and \eqref{eq:etil}, we obtain $g(x,e,\what,v,w):=(g_1(x,e,\what,v_1,w_1),g_2(x,e,\what,v_2,w_2),\ldots,\allowbreak g_N(x,e,\what,v_N,w_N))$, where $g_i(x,e,\what,v_i,w_i):=f_{h,i}(g_{p,i}(x_{p,i})+e_i+\what_i)-f_{y,i}(x,e,\what,v_i,w_i)$
with
\begin{align*}
    &f_{y,i}(x,e,\what,v_i,w_i):=\\
    &\frac{\partial g_{p,i}}{\partial x_{p,i}}f_{p,i}\left(x_p,g_{c,i}(x_{c,i},g_{p,i}(x_{p,i})+w_i,g_p(x_p)+e+\what),v_i\right).\nonumber
\end{align*}
The function $\bar\Psi$ defines the dynamics of the local triggering variables $\eta$, and it is defined as  $\bar\Psi(\widetilde{y},\yest,\widetilde{e},u,\eta):=(\bar\Psi_1(\eta_1,o_1),\bar\Psi_2(\eta_2,o_2),\ldots,\linebreak\bar\Psi_N(\eta_N,o_N))$, where $o_i:=(\widetilde{y}_i,\yest,\widetilde{e}_i,u_i)\in\R^{n_{o,i}}$ with $n_{o,i}:=3n_{y,i}+n_{u,i}$ is the locally available information at ETM $i$, and
\begin{equation}\label{eq:barpsi}
    \bar\Psi_i(\eta_i,o_i)=\Psi_i(o_i)-\varphi_i(\eta_i),
\end{equation}
with $\varphi_i$, $i\in\mathcal{N}$, any class-$\mathcal{K}_\infty$ function, and $\Psi_i$ in \eqref{eq:barpsi} will be constructed in the following.

The flow set $\mathcal{C}\subseteq\X\times\V$ is given by
\begin{equation}
    \mathcal{C}:=\bigcap_{i\in\mathcal{N}}\mathcal{C}_i
\end{equation}
with
\begin{equation}\label{eq:flowset}
    \mathcal{C}_i:=\left\{(\xi,\nu)\in\X\times\V\mid\eta_i+\theta_i\Psi_i(o_i)\geq0\right\},
\end{equation}
where $\theta_i\in\R_{\geq0}$ is a nonnegative tuning parameter. By selecting a larger $\theta_i$, the first triggering occurs earlier than when $\theta_i$ is small, given the same initial condition, see \cite[Proposition 3.2]{Girard2015}, where this is shown for the particular case of state-feedback and specific functions $\Psi_i$, $i\in\mathcal{N}$, related to \cite{Tabuada2007}. Generally, enlarging $\theta_i$ results in faster convergence but smaller inter-event times compared selecting $\theta_i$ smaller, which allows to tune bandwidth usage versus performance, see \cite{Girard2015} for more details.

The jump set is given by
\begin{equation}
    \mathcal{D}:=\bigcup_{i\in\mathcal{N}}\mathcal{D}_i
\end{equation}
with
\begin{equation}\label{eq:jumpset}
    \begin{aligned}
        \mathcal{D}_i:=\big\{(\xi,\nu)\in\X\times\V\mid\;\eta_i+\theta_i\Psi_i(o_i)\leq0\qquad&\\
        \land\;\Psi_i(o_i)\leq0&\big\}.
    \end{aligned}
    % \mathcal{D}_i:=\big\{(\xi,\nu)\in\X\times\V\mid\;\eta_i+\theta_i\Psi_i(o_i)\leq0\big\}
\end{equation}
The jump map $G:\X\times\V\rightrightarrows\X$ is, for any $(\xi,\nu)\in\X\times\V$, given by
\begin{equation}
    G(\xi,\nu):=\bigcup_{i\in\mathcal{N}}G_i(\xi,\nu)
\end{equation}
with
\begin{equation}
    G_i(\xi,\nu):=\begin{cases}\left\{(x,\overline{\Gamma}_ie,\overline{\Gamma}_i\what+\Gamma_iw,\eta)\right\},&\text{when\;}(\xi,\nu)\in\mathcal{D}_i,\\\emptyset,&\text{when\;}(\xi,\nu)\not\in\mathcal{D}_i,\end{cases}\label{eq:jumpmap}
\end{equation}
where $\Gamma_i:=\diag\left(\Delta_{i,1},\Delta_{i,2},\ldots,\Delta_{i,N}\right)$ with
\begin{equation}
     \Delta_{i,j}:=\begin{cases}\mathbf{0}_{n_{y,j},n_{y,j}},&\text{if}\;i\neq j,\\I_{n_{y,j}},&\text{if}\;i=j,\end{cases}
\end{equation}
and where $\overline{\Gamma}_i:=I_{n_y}-\Gamma_i$.
For future use, we use the compact notation $F_\chi$ to denote the flow map of variable $\chi$ as
\begin{equation}\label{eq:fchi}
    F_\chi(\chi,\nu):=\left(f(x,e,\what,v,w),g(x,e,\what,v,w),\mathbf{0}_{n_y}\right),
\end{equation}
and the jump map $G_\chi$ defined as
\begin{equation}\label{eq:gchi}
    G_\chi(\chi,\nu):=\bigcup_{i\in\mathcal{N}}\left\{(x,\overline{\Gamma}_ie,\overline{\Gamma}_i\what+\Gamma_iw)\right\}.
\end{equation}
The maps $F_\chi$ and $G_\chi$ will be used to formulate suitable conditions on the physical system (in absence of a network) for the design of suitable event generators.

Because of the chosen modelling setup, in particular \eqref{eq:modelsplit} and \eqref{eq:etil}, $\mathcal{H}$ does \emph{not} depend on the time-derivative of $w$ as in \cite{Mousavi2019,Abdelrahim2017}. This modeling choice is instrumental to work under more general and more natural assumptions on the measurement noise, see Assumption \ref{ass:noise}.

% \begin{remark}
%     Since the definitions of $\mathcal{C}$ and $\mathcal{D}$ do not impose constraints on $v_i$, $i\in\mathcal{N}$, we can relax Assumption \ref{ass:noise} to allow measurable (but essentially bounded) process disturbances, i.e., $v_i\in\Linf$ for all $i\in\mathcal{N}$, see Remark \ref{rem:measurableinput}. Note that this does \emph{not} hold for the measurement noises $w_i$, as the flow and jump sets $\mathcal{C}$ and $\mathcal{D}$ depend on $w_i$ directly via $\Psi_i$.
% \end{remark}

To formalize objective (ii) stated at the end of Section \ref{ch:3}, we introduce, for any solution $\xi$ to $\mathcal{H}$ for $\nu\in\PC_\V$ and $i\in\mathcal{N}$, the set
\begin{align}
    &\mathcal{T}_i(\xi,\nu):=\Big\{(t,j)\in\dom\xi\mid(t,j+1)\in\dom\xi\;\land\\&(\xi(t,j),\nu(t))\in\mathcal{D}_i\land(\xi(t,j+1),\nu(t))\in G_i(\xi(t,j),\nu(t))\Big\}.\nonumber\label{eq:triggertimes}
\end{align}
In words, $\mathcal{T}_i(\xi,\nu)$ contains all hybrid times belonging to the hybrid time domain of a solution $\xi$ to $\mathcal{H}$ given input $\nu$ at which a jump occurs due to triggering condition $i$ ($\mathcal{D}_i$ and $G_i$). The next (new) definition will be considered in place of item (ii) at the end of Section \ref{ch:3} in the rest of this paper.
\begin{definition}\label{defn:sgimiet}
    % Given a non-empty closed set $\mathcal{A}_\mathcal{H}\subseteq\X$, system $\mathcal{H}$ has a \emph{semi-global individual minimum inter-event time (SGiMIET)} with respect to $\mathcal{A}_\mathcal{H}$, if, for all $\Delta\geq0$ and all $i\in\mathcal{N}$, there exists a $\tau_\mathrm{MIET}^i>0$ such that for any input $v\in\PC_\V$ and associated solution $\xi$ with $|\xi(0,0)|_{\A_\mathcal{H}}\leq\Delta$, for all $(t,j),(t',j')\in \mathcal{T}_i(\xi,\nu)$,
    System $\mathcal{H}$ has a \emph{semi-global individual minimum inter-event time (SGiMIET)} if, for all $\Delta\geq0$ and all $i\in\mathcal{N}$, there exists a $\tau_\mathrm{MIET}^i>0$ such that for any input $v\in\PC_\V$ and associated solution $\xi$ with $|\xi(0,0)|\leq\Delta$, for all $(t,j),(t',j')\in \mathcal{T}_i(\xi,\nu)$,
    \begin{equation}\label{eq:miet}
        %j<j'\Rightarrow t-t'\geq\tau_\mathrm{MIET}^i.
        t+j<t'+j'\quad\implies\quad t'-t\geq\tau_\text{MIET}^i.
    \end{equation}
    %If $\tau^i_\mathrm{MIET}$ can be chosen independently of $\Delta_v$, then we say that $\mathcal{H}$ has a \emph{uniform semi-global minimum inter-event time (USGiMIET)}. If additionally, $\tau^i_\mathrm{MIET}$ can be chosen independently of $\Delta$ for all $i\in\mathcal{N}$, then we say that $\mathcal{H}$ has a \emph{uniform global individual minimum inter-event time (UGiMIET)}.
    If $\tau^i_\mathrm{MIET}$ can be chosen independently of $\Delta$ for all $i\in\mathcal{N}$, then we say that $\mathcal{H}$ has a \emph{global individual minimum inter-event time (GiMIET)}.% Similarly, if $\tau^i_\text{MIET}$ can be chosen independently of $\Delta_v$, we say that $\mathcal{H}$ has a \emph{uniform (semi-)global individual minimum inter-event time (U(S)GiMIET)}.
\end{definition}
Definition \ref{defn:sgimiet} means that the (continuous) time between two successive transmission instants due to a trigger of condition $i$ are spaced by at least $\tau_\text{MIET}^i$ units of time, and that $\tau_\text{MIET}^i$ depends on the set of initial conditions in general. %Note that, due to the distributed nature of the problem, several different local triggering conditions can generate transmissions at the same time.

Using the hybrid model $\mathcal{H}$ and the terminology in Definition \ref{defn:sgimiet}, we can now formally state the problem formulation at the end of Section \ref{ch:3} as follows: For a given non-empty closed set $\A\subset\R^{n_x}\times\R^{n_y}\times\W$ (describing a target set for the state $\chi\in\R^{n_x}\times\R^{n_y}\times\W$), synthesize the functions $\Psi_i$, $i\in\mathcal{N}$, such that $\A_\mathcal{H}:=\{\xi\in\X\mid\chi\in\A\land\eta=\mathbf{0}\}$ is ISpS w.r.t. $\mathcal{A}_\mathcal{H}$ and $\mathcal{H}$ has the SGiMIET property for all $\theta_i\in\R_{\geq0}$, $i\in\mathcal{N}$.

\section{Main result}\label{ch:5}
\subsection{Main Assumption}
As indicated in Section \ref{ch:3}, we assume that the controllers ${C}_i$, $i\in\mathcal{N}$, are designed such that the closed-loop system satisfies suitable stability properties, as formalized below in terms of the data of $\mathcal{H}$. We show in Section \ref{ch:6} how these properties can be naturally obtained for several important case studies.
\begin{assumption}\label{ass:lyapcond}
    There exist $\alpha,\underline\alpha,\overline\alpha\in\mathcal{K}_\infty$, $\gamma\in\mathcal{K}$, $\beta_i\in\mathcal{K}$ and continuous functions $\delta_i:\R^{n_{o,i}}\rightarrow\R_{\geq0}$, where $n_{o,i}=3n_{y,i}+n_{u,i}$, for all $i\in\mathcal{N}$, a non-empty closed set $\mathcal{A}\subset\R^{n_x}\times\R^{n_y}\times\W$ and a continuously differentiable function $V:\R^{n_x}\times\R^{n_y}\times\W\to\R_{\geq0}$ such that
    \begin{enumerate}[label={\roman*)}]
        \item for any $\chi\in\R^{n_x}\times\R^{n_y}\times\W$,
        \begin{equation}
            \underline\alpha(|\chi|_{\A})\leq V(\chi)\leq\overline\alpha(|\chi|_{\A}),
        \end{equation}
        \item for all $\chi\in\R^{n_x}\times\R^{n_y}\times\W$ and $\nu\in\V$,
        \begin{equation}\label{eq:vbound}
            \begin{aligned}
                \left\langle\nabla V(\chi),F_\chi(\chi,\nu)\right\rangle\leq &-\alpha(|\chi|_{\A}) + \gamma(|\nu|) \\
                &+ \sum_{i\in\mathcal{N}}\left(\beta_i(|\widetilde{e}_i|) - \delta_i(o_i)\right),
            \end{aligned}
        \end{equation}
        \item for all $\chi\in\R^{n_x}\times\R^{n_y}\times\W$ and $\nu\in\V$ such that $(\xi,\nu)\in\mathcal{D}$ and $g\in G_\chi(\chi,\nu)$,
        \begin{equation}\label{eq:vbound2}
            V(g)-V(\chi)\leq0,
        \end{equation}
        \item for every $p>0$ and $\nu\in\PC_\V$, there exists $q\geq 0$ such that for all $\phi\in\mathcal{S}_\mathcal{H}(\nu,\mathcal{B})$\footnote{$\mathcal{S}_\mathcal{H}(\nu,\mathcal{B})$ denotes the set of maximal solutions $\phi$ for the hybrid system $\mathcal{H}$ with $\phi(0,0)\in\mathcal{B}$ and input $\nu$.} with $\mathcal{B}:=\{\xi\in\X\mid|\xi|\leq p\}$, $|\phi(t,j)|\leq q$ for all $(t,j)\in\dom\phi$.
        % \item the system $(\dot{x},\dot{e}) = \hat{F}(x,e,\what,\nu)$ is forward complete for all $\xi\in\X$ and $\nu\in\PC_\V$,
        % \item for any bounded set $\mathcal{B}\subseteq\R^{n_y}$, the system $\dot x = f(x,e,\what,v,w)$ is forward complete for all $x(0)=x_0\in\R^{n_x}$ and all $e\in\mathcal{PC}_\mathcal{B}$, $\what\in\W$ and $\nu\in\PC_\V$,

        % \item for any $p,q>0$, there exist $r\in\R_{>0}$ and $s\in\mathcal{K}_\infty$ such that for all $(\xi,\nu)\in\mathcal{C}$ with $|\chi|_\A\leq p$ and $|\xi(0,0)|\leq q$,
        % \begin{equation}\label{eq:deltabound}
            % \delta_i(o_i)\leq r+s(|\chi|_\A),
        % \end{equation}

        %\item for any $\kappa,\mu\in\R_{\geq0}$, there exist $m\in\R_{\geq0}$ and $\ell\in\mathcal{K}$ such that for all $(\xi,\nu)\in\mathcal{C}$ with $|\chi|_\A\leq\kappa$ and $|e|\leq\mu$,% and all $e\in\R^{n_y}$,
        %\begin{equation}\label{eq:gdyn}
            % |g(x,e,\what,v,w)|\leq m+\ell(|\nu|).
        % \end{equation}
    \end{enumerate}
    % with $F_\chi$ and $G_\chi$ as \eqref{eq:fchi} and \eqref{eq:gchi}, respectively.
\end{assumption}

%\note{Item v) does not hold for the consensus case, since we use $u_i$ there as $\delta_i(o_i)$, which does not satisfy this property in the case where $\proj_e(\A)$ is not compact (i.e., it does hold for Victor's case but not for the other two). We may need to think about this a little more, but this is the least conservative condition that I can think of.}
Items i)-iii) of Assumption \ref{ass:lyapcond} impose Lyapunov conditions on the $\chi$-system of $\mathcal{H}$. In particular, items i) and ii) imply that the $\chi$-system satisfies an input-to-state dissipativity property during flow w.r.t. the set $\mathcal{A}$, which directly connects to the controller robustly stabilizing $\mathcal{A}$. This property may be established by ignoring the network and treating $e_i$ as exogeneous inputs. % is typically satisfied due to our emulation-based approach. %The challenge is thus to (approximately) preserve this property (together with ensuring the existence of a SGiMIET) for system \eqref{eq:dHS}.
Item iii) implies that the Lyapunov function does not increase at jumps. This condition directly holds when the expression of $V$ only involves $x$, for instance, as $x$ evolves continuously over time and is thus not affected by jumps. As we will see in Section \ref{sec:Dolk}, in some existing ETC setups, the Lyapunov function involves $e$ and the auxiliary variables $\eta_i$ may be updated after a transmission (and not kept constant as we do here). In such cases, item iii) is required to ensure that transmissions do not destabilize the system. Items i)-iii) of Assumption \ref{ass:lyapcond} imply that, in absence of a digital network (and thus, $e_i=\widetilde{e}_i=0$ and $\what_i=w_i$), the set $\mathcal{A}$ is input-to-state stable with respect to inputs $\nu$.

Item iv) is a boundedness property of the solutions to $\mathcal{H}$; this condition directly follows for many relevant cases from items i)-iii) of Assumption \ref{ass:lyapcond}, such as when $\A$ is compact, or when $\proj_x(\A)$ is compact and a ZOH is employed. More details are given in Section \ref{sec:boundednessxi}, particularly Lemma \ref{lem:xibound}. In other cases, e.g., when $\proj_x(\A)$ is unbounded or when $\A$ is not compact and a non-zero holding-function is used, the dynamics of the system have to be exploited to establish item iv) of Assumption \ref{ass:lyapcond}, as we will show, e.g., in the consensus case study in Section \ref{sec:consensus}.

%The condition in item v) is used to ensure boundedness of the network-induced error $e$. Lastly, item vi) imposes a boundedness condition on $g$, which is used to ensure the (S)GiMIET property. Note that items v) and vi) are \emph{sufficient} conditions, as for specific problem settings different methods may be used to prove boundedness of the network-induced error or the existence of a (S)GiMIET. More details on this will follow in Remark \ref{rem:gbound}.
Again, a broad range of examples of systems verifying Assumption \ref{ass:lyapcond} are provided in Section \ref{ch:6}. Items i), ii) and iv) may be verified when designing the controllers $C_1,C_2,\ldots,C_N$ and holding functions $f_{h,i}$, $i\in\mathcal{N}$, at the first step of emulation. The challenge is to design the local triggering conditions to handle the potentially destabilizing effect due to the measurement noise and the true network-induced error $e_i$ in $\widetilde{e}_i$ in item ii) of Assumption \ref{ass:lyapcond}, which is addressed in the next subsection.

\subsection{Dynamic triggering rules}
The next theorem explains how to design the dynamics of the dynamic triggering variable $\eta$, in particular $\Psi_i$, $i\in\mathcal{N}$, in \eqref{eq:barpsi} to ensure the desired objectives based on Assumptions \ref{ass:noise}, \ref{ass:lyapcond}. Its proof is provided in the appendix.
\begin{theorem}
    Consider system $\mathcal{H}$ as given by \eqref{eq:flowmap}, \eqref{eq:flowset}, \eqref{eq:jumpset} and \eqref{eq:jumpmap} and suppose Assumptions \ref{ass:noise} and \ref{ass:lyapcond} hold. We define for all $i\in\mathcal{N}$, $\xi\in\mathbb X$ and all $\nu\in\V$
    \begin{equation}
        \Psi_i(o_i):=\delta_i(o_i)-\beta_i(|\widetilde{e}_i|)+c_i\label{eq:trigger_dynamic}
    \end{equation}
    with $c_i>\beta_i(2\overline{w}_i)$ being a tuning parameter and $\overline{w}_i$ and $\beta_i$ come from Assumptions \ref{ass:noise} and \ref{ass:lyapcond}, respectively. System $\mathcal{H}$ with \eqref{eq:trigger_dynamic} is persistently flowing, has a SGiMIET and the set $\mathcal{A}_\mathcal{H}:=\{\xi\in\X\mid \chi\in\mathcal{A}\land\eta=\mathbf{0}\}$ is ISpS for all $\theta_i\in\R_{\geq0}$, $i\in\mathcal{N}$.\label{thm:dynamictrigger}
\end{theorem}

Theorem \ref{thm:dynamictrigger} provides the expression of $\Psi_i$, $i\in\mathcal{N}$, which ensures that the ISS property of the set $\mathcal{A}$ guaranteed by Assumption \ref{ass:lyapcond} in the absence of the network, is approximately preserved in the presence of the digital network. Moreover, the existence of a strictly positive lower-bound on the inter-event times of each triggering mechanism is guaranteed. The interest of Theorem \ref{thm:dynamictrigger} lies in its basic nature, generality and in revealing the main concepts as a ``prescriptive framework,'' and we will show its broad applicability in several important applications in Section \ref{ch:6}.

The expression of $\Psi_i$ in \eqref{eq:trigger_dynamic} is based on so-called \emph{space-regularization}, as by introducing $c_i$, we enlarge the flow set to ensure the existence of a SGiMIET. While space-regularization is well known in the hybrid systems literature and has been used in different forms in event-triggered control, see, e.g., \cite{Borgers2014,Donkers2012,Aarzen1999,Miskowicz2006}, the selection of $c_i$ has to be done carefully in the presence of measurement noise, because the Zeno-freeness a priori only holds if $c_i$ satisfies the condition mentioned in Theorem \ref{thm:dynamictrigger}.

A few remarks are in order. First, note that the consequence of $c_i>\beta_i(2\overline{w}_i)$ in Theorem \ref{thm:dynamictrigger} is that we obtain a \emph{practical} stability notion, i.e., the constant $d$ in \eqref{eq:ispsdef} is non-zero, see Remark \ref{rem:d} below for more details. Second, interestingly, Theorem \ref{thm:dynamictrigger} does not require to make assumptions on the differentiability of $w_i$, and a fortiori on boundedness properties of $\dot w_i$, as was required in various works in ETC considering measurement noise, see, e.g., \cite{Mousavi2019,Abdelrahim2017}. \mbox{Additionally}, we may exploit the structure present in specific scenarios or ETC mechanisms to obtain less conservative bounds for the parameters $c_i$ and, in some cases, a global individual MIET (see Definition \ref{defn:sgimiet}), as opposed to a semiglobal one as in Theorem \ref{thm:dynamictrigger}, as will be illustrated in Section \ref{ch:6}.

\subsection{Static triggering rules}
We can derive similar results when the triggering conditions are static, i.e., when no variable $\eta_i$ is used to define the transmission instants. In this case, we obtain the hybrid system $\mathcal{H}_s$ defined as
\begin{equation}
    \begin{aligned}
        \dot\chi &= F_\chi(\chi,\nu), &\quad&(\chi,\nu)\in \mathcal{C}^s,\\
        \chi^+ &\in G_\chi(\chi,\nu),&&(\chi,\nu)\in \mathcal{D}^s,\label{eq:sHS}
    \end{aligned}
\end{equation}
where
\begin{equation}\label{eq:sflowjump}
    \begin{aligned}
        \mathcal{C}^s:=\bigcap_{i\in\mathcal{N}}\mathcal{C}_i^s,\quad\mathcal{D}^s:=\bigcup_{i\in\mathcal{N}}\mathcal{D}_i^s
    \end{aligned}
\end{equation}
with the sets $\mathcal{C}^s_i,\mathcal{D}^s_i$ as
\begin{equation}
    \begin{aligned}
        \mathcal{D}^s_i&:=\{(\chi,\nu)\in\R^{n_x}\times\R^{n_y}\times\W\times\V\;|\;\Psi_i(o_i)\leq0\},\\
        \mathcal{C}^s_i&:=\{(\chi,\nu)\in\R^{n_x}\times\R^{n_y}\times\W\times\V\;|\;\Psi_i(o_i)\geq0\},
    \end{aligned}
\end{equation}
where $\Psi_i(o_i)\leq0$ is a (local) static triggering condition, which is designed according to the following result.
\begin{corollary}\label{cor:statictrigger}
    Consider system \eqref{eq:sHS} and suppose Assumptions \ref{ass:noise} and \ref{ass:lyapcond} hold. We define for all $i\in\mathcal{N}$, $\chi\in\R^{n_x}\times\R^{n_y}\times\mathcal{W}$ and $\nu\in\V$%w\in\mathcal{W}$
    \begin{equation}
        \Psi_i(o_i):=\delta_i(o_i)+c_i-\beta_i(|\widetilde{e}_i|)\label{eq:trigger_static}
    \end{equation}
    with $c_i>\beta_i(2\overline{w}_i)$ a tuning parameter. The system $\mathcal{H}^s$ with \eqref{eq:trigger_static} is persistently flowing, has a SGiMIET, and the set $\A$ is ISpS.%\label{thm:dynamictrigger}
    %The set $\mathcal{A}$ is ISpS and system \eqref{eq:sHS} has a SGiMIET.
\end{corollary}
The proof of Corollary \ref{cor:statictrigger} follows similar steps as the proof of Theorem \ref{thm:dynamictrigger}, and is therefore omitted.
\begin{remark}
    Corollary \ref{cor:statictrigger} and Assumption \ref{ass:lyapcond} allow us to consider the special case where $\delta_i=0$ for some $i\in\mathcal{N}$. In this case, $\Psi_i$ is given by $\Psi_i(o_i):=c_i-\beta_i(|\widetilde{e}_i|)$, with $c_i>\beta_i(2\overline{w}_i)$ a tuning parameter. Note that triggering conditions of this form are often called absolute triggering conditions or fixed threshold policies in the event-triggered control literature, see, e.g., \cite{Borgers2014,Aarzen1999,Miskowicz2006}.
\end{remark}

\begin{remark}
    The parameters $c_i$ for $i\in\mathcal{N}$ in Theorem \ref{thm:dynamictrigger} and Corollary \ref{cor:statictrigger} are directly related to the constant $d$ in the ISpS definition \eqref{eq:ispsdef}. From the proof of Theorem \ref{thm:dynamictrigger}, it follows that \eqref{eq:ispsdef} holds with $d=\sum_{i\in\mathcal{N}}c_i$, where $c_i$ comes from \eqref{eq:trigger_dynamic}. Hence, for a tighter ultimate bound on $|\xi(t,j)|_{\mathcal{A}_\mathcal{H}}$, we require that the $c_i$'s are smaller. Recall, however, that due to Theorem \ref{thm:dynamictrigger}, $c_i$ is lower-bounded by $\beta_i(2\overline{w}_i)$, and thus the infimum value of $d$ is $d_\mathrm{min}=\sum_{i\in\mathcal{N}}\beta_i(2\overline{w}_i)$ to ensure proper SGiMIET and for handling measurement noise. On the other hand, selecting a small $c_i$ implies a small lower bound on the inter-event times, i.e., that the constants $\tau_\text{MIET}^i$ in \eqref{eq:miet} are small. Hence, this suggests a trade-off between large lower bounds on the inter-event times and ``asymptotic closeness'' to $\mathcal{A}_\mathcal{H}$ in terms of $d$, see \eqref{eq:ispsdef}, which is tunable via the selection of $c_i$, $i\in\mathcal{N}$.\label{rem:d}
\end{remark}

\begin{remark}
    We recover as a particular case the result of \cite[Remark V.3]{Borgers2014} when we specialize our results to the same setting, i.e., when a single linear plant model is considered and ZOH devices are implemented. Indeed if $\beta_i$ is the identity function as in \cite{Borgers2014}, we recover the lower-bound $c_i>2\overline{w}_i$ in Theorem \ref{thm:dynamictrigger}. The results here are more general as they apply to a broad range of nonlinear and distributed problem setups, as we will show in Section \ref{ch:5} below.
\end{remark}

\begin{remark}\label{rem:timeregularization}
    Due to the modeling similarities, this framework can also be applied to \cite{Dolk2017} with minimal adjustments. In \cite{Dolk2017}, time-regularization is used to design triggers for classes of nonlinear systems in absence of measurement noise. By including this framework, these triggering techniques can be made robust to measurement noise with minimal changes. Due to the additional (notational) burden of including time-regularization, and the fact that our technique works without time-regularization, we opted to omit time-regularization from this paper to ensure that the main message is not blurred by too many technicalities. However, we would like to point out that by using time-regularization, there is no need for a (strictly positive) lower bound on the constants $c_i$, and, hence, we can obtain ISS properties in this case by selecting $c_i=0$ for all $i\in\mathcal{N}$. As will be demonstrated in Section \ref{ch:7}, including space-regularization may still be beneficial to obtain more favorable inter-event times close to the desired stability set $\mathcal{A}$.
\end{remark}

\subsection{Boundedness of $\xi$}\label{sec:boundednessxi}
We provide the following lemma to ensure item iv) based on items i)-iii) of Assumption \ref{ass:lyapcond} for several common cases.
\begin{lemma}\label{lem:xibound}
    Consider system $\mathcal{H}$ as given by \eqref{eq:flowmap}, \eqref{eq:flowset}, \eqref{eq:jumpset} and \eqref{eq:jumpmap}, with the trigger dynamics as given by \eqref{eq:trigger_dynamic} in case of dynamic triggering and \eqref{eq:trigger_static} in case of static triggering. When Assumption \ref{ass:noise} and items i)-iii) of Assumption \ref{ass:lyapcond} hold, and one of the following conditions is met:
    \begin{itemize}
        \item $\mathcal{A}$ is compact;
        \item $\proj_x(\mathcal{A})$ is compact and $f_{h,i}=0$ for all $i\in\mathcal{N}$;
    \end{itemize}
    then item iv) of Assumption \ref{ass:lyapcond} also holds.
\end{lemma}
The proof of this lemma \ref{lem:xibound} is provided in the Appendix. Other cases for which Assumption \ref{ass:lyapcond} is satisfied (without having the conditions of Lemma \ref{lem:xibound}) are discussed in the next section (e.g., the consensus case).

\section{Case studies}\label{ch:6}
In this section, we revisit and extend several existing event-triggering techniques of the literature to handle measurement noise, exploiting the prescriptive framework laid down in the previous sections. We want to stress that a non-exhaustive sample of a few well-known techniques is considered, however, many more can be handled given the generality of our framework. To apply the framework, we prove that Assumption \ref{ass:lyapcond} is verified, which allows us to then directly apply Theorem \ref{thm:dynamictrigger} and Corollary \ref{cor:statictrigger}. At the end of this section, in Table \ref{tab:triggers}, the original triggering rules and their (robustified) counterparts are summarized.

\subsection{The nonlinear single-system case}\label{sec:girard}
In this section, we aim to stabilize the origin of a single plant using a dynamic output-feedback controller, thereby revisiting the techniques of \cite{Aarzen1999,Tabuada2007,Girard2015}, originally developed for static feedback laws ignoring measurement noise. As such, we consider a single plant $P$ and a single controller $C$, i.e., $N=1$ in Fig. \ref{fig:nc}, where the plant is given by
\begin{equation}\label{eq:sisoplant}
    P:\begin{cases}
        \begin{aligned}
            \dot x_p&=f_p(x,u,v)\\
            y&=g_p(x_p)\\
            % \widetilde{y}&=g_p(x_p)+w,
        \end{aligned}
    \end{cases}
\end{equation}
and the `ideal' non-networked feedback controller by
\begin{equation}\label{eq:sisocont}
    C:\begin{cases}
        \begin{aligned}
            \dot x_c&=f_c(x_c,y)\\
            u&=g_c(x_c,y).
        \end{aligned}
    \end{cases}
\end{equation}
The plant and controller states are concatenated as $x:=(x_p,x_c)$, whose dynamics are then given by
\begin{equation}
    \dot{x}=f(x,\varepsilon,v):=\begin{bmatrix}
    f_p(x_p,g_c(x_c,g_p(x_p)+\varepsilon),v)\\
    f_c(x_c,g_p(x_p)+\varepsilon)
    \end{bmatrix}.
\end{equation}
where $\varepsilon\in\R^{n_y}$ can be perceived as an additive measurement error. We employ in this case a ZOH, i.e., $f_h=0$ in \eqref{eq:estmodel}. We assume that the following properties hold.
\begin{assumption}
    Maps $f_p$, $f_c$ and $g_c$ are locally Lipschitz and $g_p$ is continuously differentiable. Additionally, there exist locally Lipschitz $\underline\alpha,\overline\alpha,\alpha,\varrho,\in\mathcal{K}_\infty$, $\vartheta\in\mathcal{K}$, a locally Lipschitz positive definite function $\zeta:\R_{\geq0}\to\R_{\geq0}$ and a continuously differentiable Lyapunov function $W:\R^n\rightarrow\R_{\geq0}$ satisfying, for all $x\in\R^{n_x}$, $w\in\R^{n_y}$ and $v\in\mathcal{V}$,% and any $d\in\R^{n_y}$,
    \begin{equation}
        \underline\alpha(|x|)\leq W(x)\leq\overline\alpha(|x|)
    \end{equation}
    and
    \begin{multline}\label{eq:Wder}
        \langle\nabla W(x),f(x,\varepsilon,v)\rangle\\
        \qquad\qquad\leq-\alpha(|x|)-\zeta(|y|)+\varrho(|\varepsilon|)+\vartheta(|v|).
    \end{multline}
    \label{ass:siso}
\end{assumption}
Assumption \ref{ass:siso} implies that the origin of $\dot x=f(x,\varepsilon,v)$ is ISS with respect to $(\varepsilon,v)$. We derive the following result from Assumption \ref{ass:siso}.
\begin{proposition}\label{prop:siso}
    Consider system \eqref{eq:sisoplant} with controller \eqref{eq:sisocont} and suppose Assumption \ref{ass:siso} holds. Then all conditions of Assumption \ref{ass:lyapcond} are met for $\mathcal{A}=\{\chi\in\R^{n_x}\times\R^{n_y}\times\W \mid x=\mathbf{0}\}$ with $\beta(s)=\varrho(2s)$ for $s\geq0$, $\delta(o)=\zeta(\frac{1}{2}|\widetilde{y}|)$ for $\widetilde{y}\in\R^{n_y}$ and $V(\chi)=W(x)$ as in Assumption \ref{ass:siso}.
\end{proposition}
\begin{proof}
    We take $V(\chi)=W(x)$ for all $\chi=(x,e,\what)\in\R^{n_x}\times\R^{n_y}\times\W$. By Assumption \ref{ass:siso}, items i) and iii) of Assumption \ref{ass:lyapcond} hold. Let $\chi=(x,e,\what)\in\R^{n_x}\times\R^{n_y}\times\mathcal{W}$. In view of the definition of $F_\chi$, \eqref{eq:etil} and \eqref{eq:Wder},
    %Assumption 3, by replacing $\varepsilon$ with $e+\what=\widetilde{e}+w$ (by ), we obtain
    \begin{equation}
        \begin{aligned}
            \langle\nabla V(\chi),&F_\chi(\chi,\nu)\rangle=\langle\nabla W(x),f(x,\widetilde{e}+w,v)\rangle\\
            &\leq - \alpha(|x|) -\zeta(|y|) + \varrho(|\widetilde{e}+w|)+\vartheta(|v|),
        \end{aligned}
    \end{equation}
    where we take $\varepsilon=\widetilde{e}+w$, such that the input to the feedback controller is $\yest=y+\widetilde{e}+w$. Next, we use the weak triangular inequality, see \cite{Jiang1994}, i.e., for any $\gamma\in\mathcal{K}$, $\gamma(a+b)\leq\gamma(2a)+\gamma(2b)$ for any $a,b\in\R_{\geq0}$, to obtain
    \begin{equation}
        \begin{aligned}
            \langle\nabla &V(\chi),F_\chi(\chi,\nu)\rangle\\&\leq - \alpha(|x|) -\zeta(|y|) + \varrho(2|\widetilde{e}|)+\varrho(2|w|)+\vartheta(|v|).
        \end{aligned}
    \end{equation}
    From the weak triangular inequality we also obtain $-\zeta(|y|)-\zeta(|w|)\leq-\zeta\left(\frac{1}{2}(|y|+|w|)\right)\leq-\zeta\left(\frac{1}{2}|y+w|\right)=-\zeta\left(\frac{1}{2}|\widetilde{y}|\right)$.
    Thus,
    \begin{equation}
        \begin{aligned}
            \langle\nabla V(\chi),F_\chi&(\chi,\nu)\rangle\\
            \leq& - \alpha(|x|) -\zeta(|y|)-\zeta(|w|) + \varrho(2|\widetilde{e}|)\\&+\varrho(2|w|)+\zeta(|w|)+\vartheta(|v|)\\
            \leq& - \alpha(|x|) -\zeta\left(\tfrac{1}{2}|\widetilde{y}|\right) + \varrho(2|\widetilde{e}|)+\gamma(|\nu|)
        \end{aligned}
    \end{equation}
    for some $\gamma\in\mathcal{K}$ and where we recall that $\nu=(v,w)$. Hence item ii) of Assumption \ref{ass:lyapcond} holds. Since $\proj_x(\A)=\{\mathbf{0}\}$ (which is compact) and $f_{h}=0$, item iv) of Assumption \ref{ass:lyapcond} holds as well due to Lemma \ref{lem:xibound}. %all maps being locally Lipschitz, the attractor being compact w.r.t $x$ and due to the ZOH holding device.
\end{proof}

Proposition \ref{prop:siso} implies that, for any bounded measurement noise as defined by Assumption \ref{ass:noise}, the trigger dynamics defined in Theorem \ref{thm:dynamictrigger} and the static trigger defined in Corollary \ref{cor:statictrigger} render the origin of the closed-loop system ISpS with the SGiMIET property.

As a special case of Proposition \ref{prop:siso}, when the output of the system is the full state, i.e., when $y=x_p$, and when the controller is static, i.e., when $u=k(x_p)$ as in \cite{Tabuada2007,Girard2015}, the conditions on Assumption \ref{ass:siso} can be relaxed as follows.
\begin{assumption}
    The maps $f_p$ and $k$ are Lipschitz continuous on compacts. Additionally, there exist $\underline\alpha,\overline\alpha,\alpha,\zeta,\in\mathcal{K}_\infty$, $\vartheta\in\mathcal{K}$ and a continuously differentiable Lyapunov function $W:\R^n\rightarrow\R$ satisfying, for any $x\in\R^{n_x}$,
    \begin{equation}
        \begin{gathered}
            \underline\alpha(|x|)\leq W(x)\leq\overline\alpha(|x|),\\
            \begin{aligned}
                \langle\nabla W(x),&f(x,\varepsilon,v)\rangle\leq-\alpha(|x|)+\varrho(|\varepsilon|)+\varsigma(|v|),
            \end{aligned}
            %\dot V\leq-\alpha(|x|)-\delta(|y|)+\gamma(|d|),
        \end{gathered}
    \end{equation}
    implying that the origin of $\dot x=f(x,\varepsilon,v)$ is ISS with respect to $\varepsilon$ and $v$.
    \label{ass:tabuada}
\end{assumption}
We derive the following result from Assumption \ref{ass:tabuada}.

%we recover the setting of \cite{Tabuada2007,Girard2015}. In that case, the term $\zeta(|y|)$
\begin{corollary}\label{cor:siso}
    Consider system $\dot{x}=f_p(x,u,v)$ with controller $u=k(x)$ and suppose Assumption \ref{ass:tabuada} holds. Then all conditions of Assumption \ref{ass:lyapcond} are met for $\mathcal{A}=\{\chi\in\R^{n_x}\times\R^{n_y}\times\W \mid x=\mathbf{0}\}$ with $\beta(s)=\varrho(2s)$ for $s\geq0$, $\delta(o)=\sigma\alpha(\frac{1}{2}|\widetilde{y}|)$ for $\widetilde{y}\in\R^{n_x}$, with $\sigma\in(0,1)$ a tuning parameter, and $V(\chi)=W(x)$ as in Assumption \ref{ass:tabuada}.
\end{corollary}
\begin{proof}
    Let $x\in\R^{n_x}$. By noting that $y=x$, we obtain, for any $\sigma\in(0,1)$,% and by using the triangular inequalities
    \begin{equation}
        \begin{aligned}
            -\alpha(|x|)=&-(1-\sigma)\alpha(|x|)-\sigma\alpha(|x|)\\
            =&-(1-\sigma)\alpha(|x|)-\sigma\alpha(|y|).%\\
            % \leq&-(1-\sigma)\alpha(|x|)-\sigma\alpha\left(\tfrac{1}{2}|\widetilde{y}|\right)+\sigma\alpha(|w|).
        \end{aligned}
    \end{equation}
    The result is then obtained by following similar steps as the proof of Proposition \ref{prop:siso}.
\end{proof}
\begin{remark}
    Corollary \ref{cor:siso} is a generalization of the setting considered in \cite{Aarzen1999,Tabuada2007,Girard2015}, towards the inclusions of measurement noise as well as process disturbances. Indeed, if the measurement noise and process disturbances are absent, we recover the exact cases as \cite{Aarzen1999,Tabuada2007,Girard2015}. Thus, triggers designed by these methods can be made robust to measurement noise (and process noise) by applying the presented framework.
\end{remark}
\begin{remark}
    For a single system (i.e. when $N=1$), the event-triggered system $\mathcal{H}$ has a global MIET when the system and controller dynamics are linear.
\end{remark}

\subsection{Consensus for multi-agent systems}\label{sec:consensus}
A specific field of interest for ETC is consensus of multi-agent systems. We study several event-triggering control schemes in this context next. We focus on single integrator systems, where each plant $P_i$, which we call agent in this section, has dynamics $\dot{x}_i=u_i$, with $x_i,u_i\in\R$, and the output $y_i=x_i$. However, the ideas in this paper apply to more general settings as well.

For a network topology described by a connected weight-balanced digraph $\mathscr{G}$ with Laplacian $L$, it is known that agents achieve consensus when the \emph{ideal} (static) control law
\begin{equation}
\bar{u}_i=\sum_{j\in\mathscr{V}^\text{in}_i}(x_i-x_j),
\label{eq:control2}
\end{equation}
with $\mathscr{V}_i^\text{in}$ the in-neighbors of agent $i$, is applied, see \cite{Dimarogonas2012}. In vector notation, this is written as $\bar{u}=-Lx$, where $\bar{u}:=(u_1,u_2,\ldots,u_N)$ and $L$ is the Laplacian matrix of the graph. We use the noisy sampled states for each agent instead of the actual states, resulting in the \emph{actual} control law
\begin{equation}\label{eq:consensusui}
    u_i=\sum_{j\in\mathscr{V}^\text{in}_i}(\yest_i-\yest_j)=\sum_{j\in\mathscr{V}^\text{in}_i}(x_i+e_i+\what_i-x_j-e_j-\what_j),
\end{equation}
written in vector notation as
\begin{equation}\label{eq:consensusu}
    u=-L(x+e+\what).
\end{equation}
Hence, the closed-loop system dynamics are
\begin{equation}
    %\dot x=f(x,k(x+e+\what))=-Lx-Le-L\what,\label{eq:consensusdynamics}
    \dot x=-Lx-Le-L\what.\label{eq:consensusdynamics}
\end{equation}
We employ a ZOH as the holding function, which results in the dynamics for the hybrid system as
\begin{equation}
    F_\chi(\chi)=(-Lx-Le-L\what,Lx+Le+L\what,\mathbf{0}_N).\label{eq:fdynconsensus}
\end{equation}
We are interested in stability properties of the consensus set
\begin{equation}\label{eq:consensusset}
    \mathcal{A}:=\left\{\chi\in\R^{N^2}\times\W\mid x_1=x_2=\ldots=x_N\right\}.
\end{equation}

We show that the results of Section \ref{ch:5} can be applied to render the ETC schemes of \cite{Garcia2013,Dolk2019,Berneburg} robust to measurement noise.

\subsubsection{Decentralized strategy for undirected graphs \cite{Garcia2013}}\label{sec:Garcia}
For this case we consider an undirected, connected graph. The event generator is of particular interest, since the original paper does \emph{not} show that solutions are Zeno-free, as noted in \cite{Nowzari2019}. By applying the proposed results, we can design robust distributed triggering rules such that the system $\mathcal{H}$ has the SGiMIET property and thus does not exhibit Zeno behavior.

We consider the Lyapunov function candidate $W(x)=\frac{1}{2}x^\top Lx$ for $x\in\R^{N}$. Note that, due to the undirected graph, $L^\top=L$. Using \eqref{eq:consensusdynamics}, for all $x\in\R^{N}$,
\begin{equation}\label{eq:Vflowu}
    \begin{aligned}
        \langle\nabla W(x)&,f(x,e,\what)\rangle%=-x^\top LL(x+e+\what)\\
        =-(x+e+\what)^\top L^\top Lx\\
        =&-(x+e+\what)^\top L^\top L(x+e+\what-e-\what)\\
        %&=-(x+e+\what)^\top L^\top L(x+e+\what)+(x+e+\what)^\top L^\top Le+(x+e+\what)^\top L^\top L\what)\\
        %&=-u^\top L^\top L(u-e-\what)\\
        =&-u^\top u-u^\top L(e+\what)\\
        =&-u^\top u-u^\top L(\widetilde{e}+w),
    \end{aligned}
\end{equation}
where we use \eqref{eq:etil} to substitute $e+\what$ by $\widetilde{e}+w$. Following \cite{Garcia2013}, using Young's inequality, we obtain for some $a\in(0,\frac{1}{2N_i})$, where $N_i$ denotes the number of neighbors for agent $i$, i.e., $N_i=\operatorname{card}\mathscr{V}_i^\text{in}$, that
\begin{equation}
    \left\langle\nabla W(x),f(x,e,\what)\right\rangle\leq\sum_{i\in\mathcal{N}}-(1-2aN_i)u_i^2+\tfrac{1}{a}N_i\left(\widetilde{e}_i^2+w_i^2\right).\label{eq:consensuslyapu}
\end{equation}
Similarly, by using the first expression in \eqref{eq:Vflowu}, we can also bound it as
\begin{equation}
    \left\langle\nabla W(x),f(x,e,\what)\right\rangle\leq\sum_{i\in\mathcal{N}}-(1-2aN_i)z_i^2+\tfrac{1}{a}N_i\left(\widetilde{e}_i^2+w_i^2\right)\label{eq:consensuslyapz}
\end{equation}
where $z_i:=L_ix$, and $L_i$ denotes the $i$-th row of the matrix $L$.
With these preliminaries in place, we are ready to state the next proposition, with which we show that Assumption \ref{ass:lyapcond} holds.
\begin{proposition}
    Assumption \ref{ass:lyapcond} holds for \eqref{eq:gchi} and \eqref{eq:fdynconsensus} with $\A$ as defined in \eqref{eq:consensusset} with $\beta_i(s)=\frac{1}{a}N_is^2$ and $\delta_i(o_i)=\sigma_i(1-2aN_i)u_i^2$, where $N_i$ denotes the number of neighbors of agent $i$ and $a\in(0,\frac{1}{2N_i})$, $\sigma_i\in(0,1)$ are tuning parameters.\label{prop:garcia}
\end{proposition}
Proposition \ref{prop:garcia} implies that, for any bounded measurement noise as defined by Assumption 1, the triggering conditions defined by Theorem \ref{thm:dynamictrigger} and Corollary \ref{cor:statictrigger} render the hybrid system $\mathcal{H}$ ISpS w.r.t. $\mathcal{A}_\mathcal{H}$ with the SGiMIET property.

\begin{proof}
    We use the Lyapunov function $V(\chi)=W(x)=x^\top Lx$ for any $\chi=(x,e,\what)\in\R^{N^2}\times\W$. According to \cite[Lemma 1]{Dolk2019}, for this Lyapunov function, there exist $0<\underline{\beta}<\overline{\beta}$ such that $\underline{\beta}|\chi|_\A\leq V(\chi)\leq\overline{\beta}|\chi|_\A$, hence item i) of Assumption \ref{ass:lyapcond} holds. Additionally, item iii) holds as $x$ is not affected by jumps. For item ii) of Assumption \ref{ass:lyapcond}, let $x\in\R^{N}$ and recall that \eqref{eq:consensuslyapu} and \eqref{eq:consensuslyapz} hold. Moreover, note that $u_i$ as in \eqref{eq:consensusui} is included in $o_i$ as it is locally available. Then, for any $\sigma_i\in(0,1)$, it holds that
    \begin{equation}\label{eq:lyapconsensus}
        \begin{aligned}
            &\left\langle\nabla V(\chi),F_\chi(\chi)\right\rangle=\left\langle\nabla W(x),f(x,e,\what)\right\rangle\\
            &\leq\textstyle\sum_{i\in\mathcal{N}}-(1-\sigma_i)(1-2aN_i)z_i^2+\tfrac{1}{a}N_iw_i^2\\
            &~\;\;\;-\sigma_i(1-2aN_i)u_i^2+\tfrac{1}{a}N_i\widetilde{e}_i^2\\
            &\leq-\alpha(|\chi|_\A)+\gamma(|w|)+\textstyle\sum_{i\in\mathcal{N}}-\sigma_i(1-2aN_i)u_i^2+\tfrac{1}{a}N_i\widetilde{e}_i^2
        \end{aligned}
    \end{equation}
    for some $\alpha\in\mathcal{K}_\infty$ and $\gamma\in\mathcal{K}$, where $\alpha$ can be obtained from \cite[(3)]{Nowzari2019} and the sandwich bounds. Hence, item ii) of Assumption \ref{ass:lyapcond} holds. To prove that item iv) of Assumption \ref{ass:lyapcond} holds, we cannot use Lemma \ref{lem:xibound}. However, observe that $\bar x=\tfrac{1}{N}\sum_{i\in\mathcal{N}}x_i$ is invariant under the dynamics \eqref{eq:consensusdynamics} as the graph is undirected, i.e., $\dot{\bar{x}}=0$, and hence, $\mathcal{S}:=\{x\in\R^N\mid \bar x=\tfrac{1}{N}\sum_{i\in\mathcal{N}}x_i \}$ is forward invariant for a fixed $\bar{x}\in\R^N$. Let $p>0$ and $\nu\in\PC_\V$. Items i)-iii) of Assumption \ref{ass:lyapcond} are sufficient to prove \eqref{eq:ispsdef} (with the disregard of $t$-completeness), see the proof of Lemma \ref{lem:xibound}. Since $\mathcal{S}\cap\proj_x(\mathcal{A})=\big\{x\in\R^N\mid x_1=\ldots=x_N=\bar x(0,0)=\allowbreak\frac{1}{N}\sum_{i\in\N}x_i(0,0)\big\}$ is compact when $|\xi(0,0)|$ is bounded, it is trivial to see that the $x$-part of the trajectories $x(t,j)$ lie in a compact set for all $|\xi(0,0)|\leq p$. Due to the use of the ZOH, the network-induced error is then necessarily upper-bounded by the maximum of the distance between two points in the compact set of trajectories of $x$ and the value of $e(0,0)$ (see the proof of Lemma \ref{lem:xibound}, case 2). Since this set is compact, the network-induced error cannot grow unbounded. Moreover, $\what\in\W$ which is compact. Lastly, $\proj_\eta(\A_\mathcal{H})=\{\mathbf{0}\}$, therefore, $\eta$ remains bounded over the trajectories of the hybrid system due to the ISpS properties. Hence, there exists a $q\geq0$ such that for all (maximal) solutions with $|\xi(0,0)|\leq p$, $|\xi(t,j)|\leq q$ for all $(t,j)\in\dom\xi$. Thus, item vi) of Assumption \ref{ass:lyapcond} holds, and all items of Assumption \ref{ass:lyapcond} are satisfied.
\end{proof}

\subsubsection{Decentralized strategy including time-regularization for undirected graphs \cite{Dolk2019}}\label{sec:Dolk}
Here we consider the setup of \cite{Dolk2019} \emph{without} transmission delays to avoid blurring the exposition with too many technicalities. For this case we consider an undirected, connected graph. For the scheme of \cite{Dolk2019}, we require that each agent has an internal clock, $\tau_i\in\R_{\geq0}$, such that $\dot\tau_i=1$ on flows and $\tau_i^+=0$ at any triggering instant of agent $i$, i.e., the clock  is reset if agent $i$ transmits its state. We denote the hybrid system in which these clocks are integrated in $\mathcal{H}$ with $\mathcal{H}_\mathrm{clock}$. Hence, the state for the hybrid system can be written as $\xi=(\chi,\tau,\eta)$ where $\chi=(x,e,\what)$ is as before in \eqref{eq:chi}, and $\tau:=(\tau_1,\tau_2,\ldots,\tau_N)$.

To prove that Assumption \ref{ass:lyapcond} holds in order to be able to apply Theorem \ref{thm:dynamictrigger} and Corollary \ref{cor:statictrigger}, we analyze the Lyapunov function candidate $W(x)=\frac{1}{2}x^\top Lx$ for any $x\in\R^{N}$. Based on a similar procedure as in Section \ref{sec:Garcia}, we can deduce that
\begin{multline}
    \left\langle\nabla W(x),f(x,e,\what)\right\rangle\\
    \leq\sum_{i\in\mathcal{N}}-(1-2aN_i)u_i^2+\tfrac{1}{a}N_i\left(e_i^2+\what_i^2\right)\label{eq:dWa}
\end{multline}
and
\begin{multline}
    \left\langle\nabla W(x),f(x,e,\what)\right\rangle\\
    \leq\sum_{i\in\mathcal{N}}-(1-2aN_i)u_i^2+\tfrac{1}{a}N_i\left(e_i^2+\what_i^2\right).\label{eq:dWb}
\end{multline}
Combining the two inequalities \eqref{eq:dWa} and \eqref{eq:dWb} results in
\begin{multline}\label{eq:dolkV}
    \left\langle\nabla W(x),f(x,e,\what)\right\rangle\\
    \leq\sum_{i\in\mathcal{N}}-d_iz_i^2-\sigma_iu_i^2+(\gamma_i^2-\mu_i)e_i^2+\frac{1}{a}N_i\what_i^2,
\end{multline}
see also \cite{Dolk2019}, with $d_i:=\varrho(1-2aN_i)$, $\sigma_i:=(1-\varrho)(1-2aN_i)$ and $\gamma_i:=\sqrt{\frac{1}{a}N_i+\mu_i}$ and where $a\in(0,\frac{1}{2N_i})$, $\varrho\in(0,1)$ and $\mu_i\in\R_{>0}$ are tuning parameters. Additionally, we define
\begin{equation}\label{eq:dolkomega}
    \omega_i(\tau_i):=\begin{cases}
        \{1\},&\text{when }\tau_i\in[0,\tau_\text{MIET}^i),\\
        [0,1],&\text{when }\tau_i=\tau_\text{MIET}^i,\\
        \{0\},&\text{when }\tau_i>\tau_\text{MIET}^i,
    \end{cases}
\end{equation}
with constant $\tau_\text{MIET}^i$ as
\begin{equation}\label{eq:dolkmiet}
    \tau_\text{MIET}^i=-\frac{\sqrt{\alpha_i\sigma_i}}{\gamma_i}\arctan\left(\frac{(\lambda_i^2-1)\sqrt{\alpha_i\sigma_i}}{\lambda_i(\alpha_i\sigma_i+1)}\right),
\end{equation}
where $\alpha_i,\lambda_i\in(0,1)$ are tuning parameters.

We show again that Assumption \ref{ass:lyapcond} holds.
\begin{proposition}\label{prop:dolk}
    Assumption \ref{ass:lyapcond} holds for \eqref{eq:gchi} and \eqref{eq:fdynconsensus} with
    \begin{multline}\label{eq:consensusdolk}
        \mathcal{A}^\star=\big\{(\chi,\tau)\in\R^{2N}\times\W\times\R_{\geq0}^N \mid x_i = x_j \text{ for all }\\
        i,j\in\mathcal{N} \land e=\mathbf{0}\land\tau\in\R^N_{\geq0}\big\},
    \end{multline}
    $\beta_i(\widetilde{e}_i,\tau_i)=(1-\omega_i(\tau_i))\times\allowbreak\gamma_i^2\left(\frac{1}{\alpha_i\sigma_i}\lambda_i^2+1\right)\widetilde{e}_i^2$ and $\delta_i(o_i)=(1-\alpha_i)\sigma_iu_i^2$, where $d_i,\varrho,\sigma_i,\gamma_i$ come from \eqref{eq:dolkV} and $\tau_i,\omega_i$ from \eqref{eq:dolkmiet} and \eqref{eq:dolkomega}, respectively.
\end{proposition}
Proposition \ref{prop:dolk} implies that, for any bounded measurement noise as defined by Assumption 1, the ETMs defined by Theorem \ref{thm:dynamictrigger} render the hybrid system $\mathcal{H}_\text{clock}$ ISpS w.r.t. $\mathcal{A}_{\mathcal{H}_\text{clock}}=\{(\chi,\tau,\eta)\in\R^{2N}\times\W\times\R_{\geq0}^N\times\R_{\geq0}^N\mid(\chi,\tau)\in\mathcal{A}^\star\land\eta=\mathbf{0}\}$. Let us note that, due to the inclusion of the timer-dependent function $\omega_i$ in the triggers, the system has a GiMIET (instead of a SGiMIET) in this particular case, and, as it cannot have finite escape times, is therefore persistently flowing. Additionally, there is no requirement (i.e., no lower bound) on the space-regularization constants $c_i$, and, in fact, if $c_i=0$ for all $i\in\mathcal{N}$, we obtain ISS w.r.t. $\mathcal{A}_{\mathcal{H}_\text{clock}}$ (instead of ISpS). The choice specific choice of $\omega_i$ where $\omega_i(\tau_i)$ is set-valued when $\tau_i=\tau_\text{MIET}^i$ makes the function outer semi-continuous, which ensures well-posedness of the hybrid system, see \cite[Theorem 6.30]{hybridsystems}. The fact that $\omega_i$ is set-valued does not matter for solutions, as $\omega_i$ is only set-valued at a measure zero set, hence via Carath\'eodory's existence theorem we still have solutions in the extended sense, i.e., any solution satisfies the differential equation almost everywhere.

\begin{proof}
    We are interested in the stability of the set $\A^\star$ in \eqref{eq:consensusdolk}. To this end, we analyze the Lyapunov function, for any $(\chi,\tau)\in\R^{2N}\times\W\times\R^N_{\geq0}$,
    \begin{equation}
        U(\chi,\tau)=W(x)+\sum_{i\in\mathcal{N}}\gamma_i\phi_i(\tau_i)e_i^2
    \end{equation}
    with
    \begin{equation}
        \frac{d\phi_i}{d\tau_i}=-\omega_i(\tau_i)\gamma_i\left(\frac{1}{\alpha_i\sigma_i}\phi_i^2(\tau_i)+1\right)
    \end{equation}
    and initial condition $\phi_i(0)=\lambda_i^{-1}$ where $\lambda_i\in(0,1)$ is a tuning parameter. Strictly speaking $U$ is Lipschitz and not continuously differentiable and the generalized Clarke derivative should be used here. However, as $\dot{\tau}_i=1$, $\left\langle\nabla U(\chi,\tau),F_\chi(\chi)\right\rangle$ exists almost everywhere, and thus Proposition \ref{prop:lyapunovisps} hold almost everywhere, hence we continue with slight abuse of notation by writing the derivative of $U$ as if it was continuously differentiable. The constant $\tau_\text{MIET}^i$ is chosen such that $\phi_i(\tau_\text{MIET}^i)=\lambda_i$, which ensures that $\phi_i(\tau_i)\geq\lambda_i$ for all $\tau_i\in\R_{\geq0}$. As stated in \cite{Dolk2019}, there exist $\alpha_1,\alpha_2\in\mathcal{K}_\infty$ such that $\alpha_1(|\chi|_\A)\leq U(\xi)\leq\alpha_2(|\chi|_\A)$, hence, item i) of Assumption \ref{ass:lyapcond} holds. For any $(\chi,\tau)\in\R^{2N}\times\W\times\R^N_{\geq0}$,
    \begin{equation}\label{eq:victorubound}
        \begin{aligned}
            &\left\langle\nabla U(\chi,\tau),F_\chi(\chi)\right\rangle\\&\leq\left\langle\nabla W(x),f(x,e,\what)\right\rangle+
            \sum_{i\in\mathcal{N}}\gamma_i\frac{d\phi_i}{d\tau_i}e_i^2+2\gamma_i\phi_ie_iu_i\\
            &\leq\sum_{i\in\mathcal{N}}-d_iz_i^2-\sigma_iu_i^2+(\gamma_i^2-\mu_i)e_i^2+\frac{1}{a}N_i\what_i^2\\
            &~\;\;\;+\gamma_i\frac{d\phi_i}{d\tau_i}e_i^2+\gamma_i^2\frac{1}{\alpha_i\sigma_i}\phi_i^2e_i^2+\alpha_i\sigma_iu_i^2\\
            &\leq\sum_{i\in\mathcal{N}}-d_iz_i^2-\mu_ie_i^2+\frac{1}{a}N_i\what_i^2\\&~\;\;\;-(1-\alpha_i)\sigma_iu_i^2+(1-\omega_i(\tau_i))\gamma_i^2\left(\frac{1}{\alpha_i\sigma_i}\lambda_i^2+1\right)e_i^2.
        \end{aligned}
    \end{equation}
    Due to \eqref{eq:etil}, we can upper-bound $e_i^2$ as
    \begin{multline}
        \begin{aligned}
            e_i^2=&(\widetilde{e}_i-\what_i+w_i)^2\\
            =&\widetilde{e}_i^2+\what_i^2+w_i^2-2\widetilde{e}_i\what_i+2\widetilde{e}_iw_i-2\what_iw_i\\
            =&\widetilde{e}_i^2+\what_i^2+w_i^2-2(e_i+\what_i-w_i)\what_i\\
            &+2(e_i+\what_i-w_i)w_i-2\what_iw_i\\
            =&\widetilde{e}_i^2-\what_i^2-w_i^2+2\what_iw_i-2e_i\what_i+2e_iw_i\\
            \leq&\widetilde{e}_i^2-\what_i^2-w_i^2+\what_i^2+w_i^2+2\kappa_ie_i^2+\frac{1}{\kappa_i}\left(\what_i^2+w_i^2\right)
        \end{aligned}\\
            =\widetilde{e}_i^2+2\kappa_ie_i^2+\frac{1}{\kappa_i}\left(\what_i^2+w_i^2\right)
    \end{multline}
    for any $\kappa_i\in\R_{\geq0}$. %Note that we do not use Young's inequality directly on $2\widetilde{e}_i\what_i$ and $2\widetilde{e}_iw_i$ because that would result in a more conservative trigger.
    Then, we select $\kappa_i$ such that
    \begin{equation}
        \kappa_i:=\frac{\vartheta_i\mu_i}{2}\left(\gamma_i^2\left(\frac{1}{\alpha_i\sigma_i}\lambda_i^2+1\right)\right)^{-1}
    \end{equation}
    for some $\vartheta_i\in(0,1)$. With this, we deduce from \eqref{eq:victorubound} that
    \begin{multline}
        \begin{aligned}
            &\left\langle\nabla U(\chi,\tau),F_\chi(\chi)\right\rangle\\&\leq\sum_{i\in\mathcal{N}}-d_iz_i^2-\mu_ie_i^2+\frac{1}{a}N_i\what_i^2-(1-\alpha_i)\sigma_iu_i^2\\&~\;\;\;+(1-\omega_i(\tau_i))\gamma_i^2\left(\frac{1}{\alpha_i\sigma_i}\lambda_i^2+1\right)e_i^2\\
            &\leq\sum_{i\in\mathcal{N}}-d_iz_i^2-(1-\vartheta_i)\mu_ie_i^2+\frac{1}{a}N_i\what_i^2+\frac{1}{\kappa_i}\left(\what_i^2+w_i^2\right)\\&~\;\;\;-(1-\alpha_i)\sigma_iu_i^2+(1-\omega_i(\tau_i))\gamma_i^2\left(\frac{1}{\alpha_i\sigma_i}\lambda_i^2+1\right)\widetilde{e}_i^2\\
            &\leq\alpha(|\chi|_{\mathcal{A}^\star})+\varpi(|w|)+\sum_{i\in\mathcal{N}}-(1-\alpha_i)\sigma_iu_i^2
        \end{aligned}\\
        \shoveleft~\;\;\;+(1-\omega_i(\tau_i))\gamma_i^2\left(\frac{1}{\alpha_i\sigma_i}\lambda_i^2+1\right)\widetilde{e}_i^2.
    \end{multline}
    for some $\varpi\in\mathcal{K}$, and, indeed, item ii) of Assumption \ref{ass:lyapcond} holds. Additionally, for any $(\chi,\tau)\in\R^{2N}\times\W\times\R^N_{\geq0}$ and $(g,\tau^+)\in G_\chi(\chi,w)$,
    \begin{equation}
        U(g,\tau^+)-U(\chi,\tau)=-\gamma_i\lambda_ie_i^2\leq0,\label{eq:victorreset}
    \end{equation}
    and item iii) of Assumption \ref{ass:lyapcond} also holds. To prove that item iv) of Assumption \ref{ass:lyapcond} holds, we refer to the proof of Proposition \ref{prop:garcia}, as the set $\mathcal{S}:=\{x\in\R^N\mid \bar x=\tfrac{1}{N}\sum_{i\in\mathcal{N}}x_i \}$ is also forward invariant for fixed $\bar{x}$ in this case.

    The terms related to $\what_i$ have been absorbed in the function $\varpi$, as its $\Linf$-norm can be bounded as
    \begin{equation}
        \|\what_i\|_\infty\leq\|w_i\|_\infty,%\leq\overline{w}_i,
    \end{equation}
    hence, we can obtain a similar condition as \eqref{eq:ispsdef} based on $\|w_i\|_\infty$.

    The fact that $\tau_i$ in theory may grow unbounded as $t\to\infty$ is not an issue, as the value of $U(\chi,\tau)$ is not affected by $\tau_i$ when $\tau_i>\tau_\text{MIET}^i$. Hence, we could take the dynamics of $\tau_i$ as $\dot{\tau}_i=\omega_i(\frac{1}{2}\tau_i)$, which would not affect the system behavior, but it would ensure that $\tau_i\in\left[0,2\tau_\text{MIET}^i\right]$. Since, in that case, all states would remain bounded, the satisfaction of item iv) of Assumption \ref{ass:lyapcond} is not affected by our initial choice of the dynamics for $\tau_i$ and the fact that $\A^*$ is unbounded in $\tau$.

    Due to the inclusion of the timer $\omega_i$ in the function $\beta_i$, by definition $\Psi_i(o_i)\geq0$ for all $\tau_i\in[0,\tau_\text{MIET}^i]$, hence, the system has a GiMIET. Thus, the system is persistently flowing and $\mathcal{H}_\text{clock}$ is IS(p)S w.r.t. $\mathcal{A}_{\mathcal{H}_\text{clock}}$. Moreover, due to the GiMIET, we do not require the lower-bound on $c_i$. When $c_i=0$ is selected for all $i\in\mathcal{N}$, the system $\mathcal{H}_\text{clock}$ is ISS.
\end{proof}

\begin{remark}\label{rem:dynreset}
    Equation \eqref{eq:victorreset} in the proof of Proposition \ref{prop:dolk} implies that we can modify the reset of $\eta_i$ as
    \begin{equation}\label{eq:dynamic_reset}
        %\eta_i^0=\gamma_i\lambda_i\widetilde{e}_i^2.
        \eta_i^+(o_i):=\eta_i+\gamma_i\lambda_i\left(\max\left[|\widetilde{e}_i|-2\overline{w},0\right]\right)^2.
    \end{equation}
    As a result we use an estimated lower bound for $e_i$, i.e.,
    \begin{equation}
        \eta_i(o_i)^+\leq\eta_i+\gamma_i\lambda_ie_i^2,
    \end{equation}
    so that item iii) of Assumption \ref{ass:lyapcond} still holds, while only using locally available information (i.e. $\widetilde{e}_i$ and not $e_i$). This may be of interest, as increasing $\eta_i$ after a reset directly increases the inter-event times. Hence, by modifying the reset, the average inter-event time may become significantly larger when the initial condition is sufficiently far from the consensus set.
\end{remark}
\subsubsection{Decentralized strategy for weight-balanced digraphs \cite{Berneburg}}\label{sec:Berneburg}
In the case of \cite{Berneburg}, we consider a network topology described by weight-balanced digraphs. Hence, this scheme requires a less restrictive network topology compared to Sections \ref{sec:Garcia} and \ref{sec:Dolk}.
\begin{proposition}
    Assumption \ref{ass:lyapcond} holds for \eqref{eq:gchi} and \eqref{eq:fdynconsensus} with $\beta_i(s)=\left(\frac{d_i^2}{\vartheta_i}+\gamma_i\right)s^2$ for any $s\geq0$ and $\delta_i(o_i)=\sigma_i(1-\vartheta_i)u_i^2$, where $d_i$ denotes the degree of agent $i$, $\vartheta_i:=\sum_{j\in\mathscr{V}_i^\text{out}}\alpha_{ij}\varrho_{ij}$, $\gamma_i:=\sum_{j\in\mathscr{V}_i^\text{in}}\frac{\alpha_{ji}}{\varrho_{ji}}$, and with $\varrho_{ij}>0$ (chosen such that $\vartheta_i\in(0,1)$) and $\sigma_i\in(0,1)$ tuning parameters.\label{prop:berneburg}
\end{proposition}
\begin{proof}
    We start by analyzing the Lyapunov function $V(\chi)=\frac{1}{2}x^\top L^\top x$ for any $\chi\in\R^{2N}\times\W$. Due to the properties of $L$, item i) of Assumption \ref{ass:lyapcond} holds. Additionally, items iii) holds trivially. Note that from \cite[(13)]{Berneburg}, we know that for any additive disturbance $\varepsilon\in\R^N$ and any $x\in\R^N$, it holds that
    \begin{equation}\label{eq:berneburg_org}
        \left\langle\nabla V(\chi),F_\chi(x,\epsilon,\mathbf{0}_N)\right\rangle\leq\sum_{i\in\mathcal{N}}-(1-\tfrac{1}{2}\vartheta_i)u_i^2-d_i\varepsilon_iu_i+\tfrac{1}{2}\gamma_i\varepsilon_i^2
    \end{equation}
    with $\vartheta_i:=\sum_{j\in\mathscr{V}_i^\text{out}}\alpha_{ij}\varrho_{ij}$, $d_i$ the degree of agent $i$, $\gamma_i:=\sum_{j\in\mathscr{V}_i^\text{in}}\frac{\alpha_{ji}}{\varrho_{ji}}$ and where $\varrho_{ij}>0$ are tuning parameters. Recall that $\alpha_{ij}$ denotes the weights corresponding to the graph. By substitution of $\varepsilon=e+\what=\widetilde{e}+w$ in \eqref{eq:berneburg_org}, we obtain
    \begin{equation}
        \begin{aligned}
            &\left\langle\nabla V(\chi),F_\chi(\chi)\right\rangle\\&\leq\sum_{i\in\mathcal{N}}-(1-\tfrac{1}{2}\vartheta_i)u_i^2-d_i(\widetilde{e}_i+w_i)u_i+\tfrac{1}{2}\gamma_i(\widetilde{e}_i+w_i)^2.
        \end{aligned}
    \end{equation}
    By applying Young's inequality, we derive
    \begin{equation}
        \begin{aligned}
            &\left\langle\nabla V(\chi),F_\chi(\chi)\right\rangle\\&\leq\sum_{i\in\mathcal{N}}-(1-\vartheta_i)u_i^2+\frac{1}{2}\left(\frac{d_i^2}{\vartheta_i}+\gamma_i\right)(\widetilde{e}_i+w_i)^2.
        \end{aligned}
    \end{equation}
    % Recall that, due to Young's inequality,
    % \begin{equation}
    %     \tfrac{1}{2}(p+q)^2\leq p^2+q^2,\label{eq:squared_upperbound}
    % \end{equation}
    As, for any $p,q\in\R$, it holds that $\tfrac{1}{2}(p+q)^2\leq p^2+q^2$, we obtain
    \begin{equation}
        \begin{aligned}
            &\left\langle\nabla V(\chi),F_\chi(\chi)\right\rangle\\&\leq\sum_{i\in\mathcal{N}}-(1-\vartheta_i)z_i^2+\left(\frac{d_i^2}{\vartheta_i}+\gamma_i\right)\widetilde{e}_i^2+\left(\frac{d_i^2}{\vartheta_i}+\gamma_i\right)w_i^2,
        \end{aligned}
    \end{equation}
    where $z_i=L_ix$ with $L_i$ the $i$-th row of matrix $L$. Note that the constants $\varrho_{ij}$ are chosen such that $\vartheta_i\in(0,1)$. Then, for any $\sigma_i\in(0,1)$, it holds that
    \begin{multline}
        \left\langle\nabla V(\chi),F_\chi(\chi)\right\rangle\leq\sum_{i\in\mathcal{N}}-(1-\sigma_i)(1-\vartheta_i)z_i^2+\left(\frac{d_i^2}{\vartheta_i}+\gamma_i\right)w_i^2\\
        -\sigma_i(1-\vartheta_i)u_i^2+\left(\frac{d_i^2}{\vartheta_i}+\gamma_i\right)\widetilde{e}_i^2,
    \end{multline}
    and item ii) of Assumption \ref{ass:lyapcond} also holds with $\delta_i(o_i)$ and $\beta_i(s)$ as specified in Proposition \ref{prop:berneburg}. To prove that item iv) of Assumption \ref{ass:lyapcond} holds we refer to the proof of Proposition \ref{prop:garcia} (due to the graph being weight-balanced, $\bar{x}$ is also invariant under the dynamics in this case). Hence, all items of Assumption \ref{ass:lyapcond} are satisfied.
\end{proof}

% \begin{remark}
%     Observe that in all consensus cases presented above, the quantity $\bar x=\tfrac{1}{N}\sum_{i\in\mathcal{N}}x_i$ is invariant under the dynamics \eqref{eq:consensusdynamics}. Consequently, the level set $\mathcal{S}:=\{x\in\R^N\mid \bar x=\tfrac{1}{N}\sum_{i\in\mathcal{N}}x_i \}$ (which depends on the initial conditions) is forward invariant. Since items i) and ii) of Assumption \ref{ass:lyapcond} are sufficient to prove ISpS-like properties (with the disregard of $t$-completeness), and due to the fact that $\mathcal{S}\cap\mathcal{A}=\{x\in\R^N\mid x_1=\ldots=x_N=\bar x\}$ (which is closed and compact), it is trivial to see that the trajectories of all maximal solutions lie in a compact set. Due to the use of the ZOH, the network-induced error is then necessarily upper-bounded by the maximum of the distance between two points in the compact set of trajectories and the value of $e(0,0)$. Consequently, since $e$ remains bounded and due to the linear dynamics, the system has a (S)GiMIET even when selecting $\mu=0$ in \eqref{eq:trigger_dynamic}.
% \end{remark}
A comparison between the original ETM and robustified one for measurement noise of several examples in this section are summarized in Table \ref{tab:triggers}.
\begin{table}[h!]
    \caption{Comparison of the original triggering rules vs. the modified ones to be robust to measurement noise using Theorem \ref{thm:dynamictrigger}.}
    \label{tab:triggers}
    \def\arraystretch{1.3}
    \begin{tabular}[t]{ll}
        % \bf{Trigger}
        \hline\hline
        \cite{Tabuada2007,Girard2015} & $\Psi(o)=\sigma\alpha(|x|)-\varrho(|e|)$ \emph{(static)}\\
        Modified & $\Psi(o)=\sigma\alpha(\tfrac{1}{2}|\widetilde{x}|)-\varrho(2|\widetilde{e}|)+c$ with $c>\varrho(4\overline{w})$\\\hline
        \cite{Garcia2013} & $\Psi_i(o_i)=\sigma_i(1-aN_i)u_i^2-\tfrac{1}{a}N_ie_i^2$ \emph{(static)}\\
        Modified & $\Psi_i(o_i)=\sigma_i(1-2aN_i)u_i^2-\tfrac{1}{a}N_i\widetilde{e}_i^2+c_i$\\
        &with $c_i>\frac{4}{a}N_i\overline{w}_i^2$\\\hline
        \makecell[l]{\cite{Dolk2019}\vspace{2.07em}} & $\begin{aligned}\Psi_i(o_i)=&(1-\alpha_i)\sigma_iu_i^2\\&-(1-\omega_i(\tau_i))\gamma_i^2\left(\tfrac{1}{\alpha_i\sigma_i}\lambda_i^2+1\right)e_i^2,\end{aligned}$\\
            & $\qquad\sigma_i=(1-\varrho)(1-aN_i)$\\
            \makecell[l]{Modified\vspace{2.07em}} & $\begin{aligned}\Psi_i(o_i)=&(1-\alpha_i)\sigma_iu_i^2+c_i\\&-(1-\omega_i(\tau_i))\gamma_i^2\left(\tfrac{1}{\alpha_i\sigma_i}\lambda_i^2+1\right)\widetilde{e}_i^2,\end{aligned}$\\
            & $\qquad\sigma_i=(1-\varrho)(1-2aN_i)$\\
            &with $c_i\geq0$\\\hline
        \cite{Berneburg} & $\Psi_i(o_i)=\sigma_i(1-\tfrac{1}{2}\vartheta_i)u_i^2+d_ie_iu_i-\tfrac{1}{2}\gamma_ie_i^2$\\
            Modified & $\Psi_i(o_i)=\sigma_i(1-\vartheta_i)u_i^2-\big(\tfrac{d_i^2}{\vartheta_i}+\gamma_i\big)\widetilde{e}_i^2+c_i$\\
            &with $c_i>\bigg(\frac{d_i^2}{\vartheta_i}+\gamma_i\bigg)4\overline{w}_i^2$\\\hline
    \end{tabular}
\end{table}

\section{Numerical examples}\label{ch:7}
\begin{figure}[tb]
    \centering
    \begin{tikzpicture}[rotate=-35.0502,transform shape,scale=0.75]
        \def\x{{1.344165949439182,1.147802907952021,-0.344341120813885,-1.941501203559574,-1.145764402277264,-1.348887815518533,1.945464534718671,0.343061150059382}};
        \def\y{{1.580677994789488,-0.051210719786944,-1.096047240976449,-0.722333594743439,0.051185125218539,-1.582586107107431,0.72469923109782,1.095615311508416}};
        \def\rot{35.0502}
        \def\factor{1.5}
        \tikzstyle{agent} = [circle, fill=blue!20, inner sep=0pt, minimum width=0.75cm, minimum height=0.75cm, rotate=\rot]
        \tikzstyle{link} = [-, black, ultra thick]
        \node[agent] (1) at ({\factor*\x[0]},{\factor*\y[0]}) {4};
        \node[agent] (2) at ({\factor*\x[1]},{\factor*\y[1]}) {3};
        \node[agent] (3) at ({\factor*\x[2]},{\factor*\y[2]}) {2};
        \node[agent] (4) at ({\factor*\x[3]},{\factor*\y[3]}) {1};
        \node[agent] (5) at ({\factor*\x[4]},{\factor*\y[4]}) {8};
        \node[agent] (6) at ({\factor*\x[5]},{\factor*\y[5]}) {7};
        \node[agent] (7) at ({\factor*\x[6]},{\factor*\y[6]}) {6};
        \node[agent] (8) at ({\factor*\x[7]},{\factor*\y[7]}) {5};
        \draw[link] (1) -- (2);
        \draw[link] (1) -- (8);
        \draw[link] (2) -- (3);
        \draw[link] (2) -- (7);
        \draw[link] (3) -- (4);
        \draw[link] (3) -- (6);
        \draw[link] (4) -- (5);
        \draw[link] (5) -- (6);
        \draw[link] (5) -- (8);
        \draw[link] (7) -- (8);
    \end{tikzpicture}
    \caption{The undirected communication topology used in the numerical examples.}
    \label{fig:topology}
\end{figure}
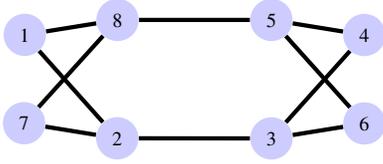

In this section, we illustrate the results of Sections \ref{sec:Garcia}-2 with $N=8$ agents that are connected as described in Fig. \ref{fig:topology}. %by a graph $\mathcal{G}$ with undirected edges $(1,2)$, $(1,8)$, $(2,3)$, $(2,7)$, $(3,4)$, $(3,6)$, $(4,5)$, $(5,6)$, $(5,8)$ and $(7,8)$. %See Figure \ref{fig:topology} for a graphical representation of the network topology.
In both cases we include uniformly distributed piecewise constant noise in the interval $10^{-4}\cdot[-1,1]$ as measurement noise, hence, $\overline{w}_i=1\cdot10^{-4}$ for all $i\in\mathcal{N}$. The noise is sampled at a rate of $1\cdot10^{4}$Hz and a ZOH is applied between samples.

\subsection{Decentralized static ETM of Proposition \ref{prop:garcia}}
To illustrate the results of Proposition \ref{prop:garcia}, we select $\sigma_i=0.5$ for all $i\in\mathcal{N}$ and $a=0.1$. Note that, for these parameters, $\max_i(\beta_i(2\overline{w}_i))=1.2\cdot10^{-6}$, hence we select $c_i>1.2\cdot10^{-6}$ to guarantee Zeno-freeness. We demonstrate the results of Corollary \ref{cor:statictrigger}, i.e., we apply static triggering. Two cases are simulated, first with no space-regularization for all $i\in\mathcal{N}$ (i.e. $c_i=0$), to demonstrate that we indeed obtain Zeno-like behavior, and second with $c_i=2\cdot10^{-6}>\max_i(\beta_i(2\overline{w}_i))$. In Fig. \ref{fig:garcia}, the evolution of the states $x_i$, $i\in\mathcal{N}$ and the corresponding inter-event times for $c_i=0$ are shown for the initial condition $x(0,0)=(8,6,4,2,-2,-4,-6,-8)$, $e(0,0)=\mathbf{0}_N$, $\what(0,0)=w(0)$ and $\eta(0,0)=\mathbf{0}_N$. Fig. \ref{fig:garcia_c} depicts the same simulations for $c_i=2\cdot10^{-6}$.

\begin{figure}[tb]
    \centering
    \begin{subfigure}[b]{\linewidth}
        \input{Figures/Garcia2013_StateEvolution.tex}
    \end{subfigure}

    \begin{subfigure}[b]{\linewidth}
        \input{Figures/Garcia2013_InterEventTimes.tex}
    \end{subfigure}
    \caption{\footnotesize Evolution of the states (top) and inter-event times (bottom) of the MAS using the dynamic trigger obtained by applying Corollary \ref{cor:statictrigger} to Proposition \ref{prop:garcia} with $c_i=0$ and initial condition $x(0,0)=(8,6,4,2,-2,-4,-8)$.}\label{fig:garcia}
\end{figure}
\begin{figure}[tb]
    \centering
    \begin{subfigure}[b]{\linewidth}
        \input{Figures/Garcia2013_StateEvolution_c.tex}
    \end{subfigure}

    \begin{subfigure}[b]{\linewidth}
        \input{Figures/Garcia2013_InterEventTimes_c.tex}
    \end{subfigure}
    \caption{\footnotesize Evolution of the states (top) and inter-event times (bottom) of the MAS using the dynamic trigger obtained by applying Corollary \ref{cor:statictrigger} to Proposition \ref{prop:garcia} with $c_i=2\cdot10^{-6}$ and initial condition $x(0,0)=(8,6,4,2,-2,-4,-8)$.}\label{fig:garcia_c}
\end{figure}

We note that when $c_i=0$ for all $i\in\mathcal{N}$, we indeed obtain ``Zeno-like'' behavior, i.e., the inter-event times converge to zero. If the space-regularization constant $c_i$ is designed properly (e.g. as in Fig. \ref{fig:garcia_c}), we can see that indeed the inter-event times close to the consensus set remain relatively large, and desirable behavior for the overall system is obtained.
\subsection{Decentralized dynamic ETM of Proposition \ref{prop:dolk}}
For the simulation of the dynamic triggering condition of Proposition \ref{prop:dolk}, the tuning parameters of \cite{Dolk2019} are used, i.e., $\varrho=\mu_i=\epsilon_{\eta,i}=0.05$, $a=0.1$ and $\alpha_i=0.5$ for all $i\in\mathcal{N}$. We thus obtain $\gamma_i=4.478$ and $\sigma_i=0.76$ for agents $i\in\mathcal{N}$ with two neighbors (i.e., $N_i=2$, thus agents $P_1$, $P_4$, $P_6$ and $P_7$) and $\gamma_i=5.482$ and $\sigma_i=0.665$ for agents $i\in\mathcal{N}$ with three neighbors (i.e., $N_i=3$, thus agents $P_2$, $P_3$, $P_5$ and $P_8$). We choose $\lambda_i=0.2$ for all agents. For these values, we obtain $\tau^i_\text{MIET}=0.1562$ for agents $i\in\mathcal{N}$ for which $N_i=2$ and $\tau^i_\text{MIET}=0.1180$ for agents $i\in\mathcal{N}$ for which $N_i=3$.

We demonstrate the results of Theorem \ref{thm:dynamictrigger}, i.e., we apply dynamic triggering. Two cases are simulated, first with no space-regularization for all $i\in\mathcal{N}$, for which we obtain ISS w.r.t. the consensus set, second with space-regularization constant $c_i=1\cdot10^{-5}$ for all $i\in\mathcal{N}$, for which we have ISpS w.r.t. the consensus set. To compare with the results to \cite{Dolk2019} (not considering measurement noise), in all cases we select $\theta_i=0$. In Fig. \ref{fig:state}, the evolution of the states $x_i$, $i\in\mathcal{N}$, with $c_i=0$ and the corresponding inter-event times are shown for the initial condition $x(0,0)=(8,6,4,2,-2,-4,-6,-8)$, $e(0,0)=\mathbf{0}_N$, $\what(0,0)=w(0)$, $\tau(0,0)=\mathbf{0}_N$ and $\eta(0,0)=\mathbf{0}_N$. Fig. \ref{fig:state_sigma} depicts the same simulations for $c_i=1\cdot10^{-7}$.

\begin{figure}[tb]
    \centering
    \begin{subfigure}[b]{\linewidth}
        \input{Figures/Dolk2019_StateEvolution.tex}
    \end{subfigure}

    \begin{subfigure}[b]{\linewidth}
        \input{Figures/Dolk2019_InterEventTimes.tex}
    \end{subfigure}
    \caption{\footnotesize Evolution of the states (top) and inter-event times (bottom) of the MAS using the dynamic trigger obtained by applying Theorem \ref{thm:dynamictrigger} to Proposition \ref{prop:dolk} with $c_i=0$ and initial condition $x(0,0)=(8,6,4,2,-2,-4,-8)$.}\label{fig:state}
\end{figure}
\begin{figure}[tb]
    \centering
    \begin{subfigure}[b]{\linewidth}
        \input{Figures/Dolk2019_StateEvolution_c.tex}
    \end{subfigure}

    \begin{subfigure}[b]{\linewidth}
        \input{Figures/Dolk2019_InterEventTimes_c.tex}
    \end{subfigure}
    \caption{\footnotesize Evolution of the states (top) and inter-event times (bottom) of the MAS using the dynamic trigger obtained by applying Theorem \ref{thm:dynamictrigger} to Proposition \ref{prop:dolk} with $c_i=1\cdot10^{-7}$ and initial condition $x(0,0)=(8,6,4,2,-2,-4,-8)$.}\label{fig:state_sigma}
\end{figure}

From Fig. \ref{fig:state} and \ref{fig:state_sigma} we can make a few observations. For $c_i=0$, close to the consensus set the inter-event times are generally close to $\tau^i_\text{MIET}$. This can be explained from the observation that, in these cases, $\eta_i^+=0$ and $u_i$ is generally small, and consequently, the increase in $\eta_i$ for $\tau\in[0,\tau^i_\text{MIET})$ is limited. Additionally, we observe that by selecting a $c_i>0$, the inter-event times are generally significantly larger than the enforced minimum inter-event time. Moreover, because there is no lower-bound on $c_i$, a relatively small $c_i$ is often sufficient to obtain desirable average inter-event times. We want to stress that this is a beneficial aspect of this particular scheme, since in general there are constraints on the minimum size of the space-regularization constants $c_i$ to ensure Zeno-freeness.

Even though the inclusion of $c_i$ leads to ISpS instead of ISS properties, applying space-regularization leads to triggering conditions that are not only robust to measurement noise, but also have, on average, larger inter-event times for the considered simulations. Since ISS only leads to asymptotic behavior of the consensus set for vanishing noise, and since most measurement noise is non-vanishing, practical stability or ISpS with larger inter-event times may be more desirable when having communication limitations in mind.

\begin{figure}[tb]
    \input{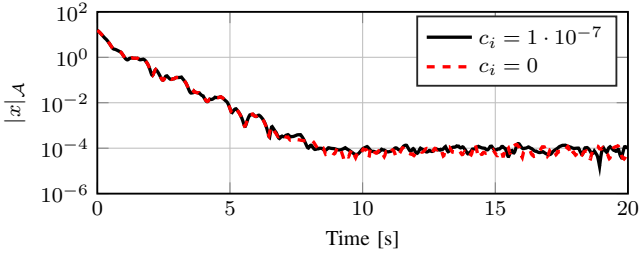}
    \caption{Distance of agents to the consensus set $\mathcal{A}$.}
    \label{fig:Dolk_e_norm}
\end{figure}

In Fig. \ref{fig:Dolk_e_norm}, the distance of the solution to the consensus set is depicted. We note that even though the inter-event times are more favorable if we apply space-regularization, the remaining distance to the consensus set has the same order of magnitude, which underlines the effectiveness of applying both space- and time-regularization at the same time.

\section{Conclusions}\label{ch:8}
In this paper, we presented a general ``prescriptive'' framework for set stabilization of event-triggered control systems affected by measurement noise. It is shown how, by careful design, to obtain both \emph{dynamic} and \emph{static} triggering conditions that render a closed set input-to-state (practically) stable with a guaranteed positive (semi-)global individual minimum inter-event time. Key to obtaining this framework is a novel hybrid model that describes the behavior of event-triggered control systems and the careful application of space-regularization. Due to this model and the space-regularization, differentiability conditions are not required on the measurement noises, as opposed to the existing works in the literature. The strengths and generality of the framework were demonstrated on several interesting event-triggered control problems, such as the stabilization of the origin for single-plant systems and consensus problems for multi-agent systems, robustifying them for measurement noise.

\appendix

\section{Proofs}
The first step in proving Theorem \ref{thm:dynamictrigger} and Lemma \ref{lem:xibound} is to ensure the satisfaction of the Lyapunov conditions in Proposition \ref{prop:lyapunovisps}. To this end, we introduce the Lyapunov function candidate $U$, defined for all $\xi\in\mathbb X$, where we recall that $\X=\R^{n_x}\times\R^{n_y}\times\W\times\R^{N}_{\geq0}$, as
\begin{equation}
    U(\xi):=V(x)+\sum_{i\in\mathcal{N}}\eta_i.
\end{equation}

The following lemma will be useful in the sequel.
\begin{lemma}\label{lem:isps}
    Consider system $\mathcal{H}$ as given by \eqref{eq:flowmap}, \eqref{eq:flowset}, \eqref{eq:jumpset}, \eqref{eq:jumpmap}, and \eqref{eq:trigger_dynamic}. When Assumption \ref{ass:noise} and items i)-iii) of Assumption \ref{ass:lyapcond} are satisfied, items i)-iv) of Proposition \ref{prop:lyapunovisps} are also satisfied.
\end{lemma}
\begin{proof}
    Recall that $V$ is continuously differentiable by Assumption \ref{ass:lyapcond}, hence the function $U$ is also continuously differentiable. Since $\Pi_\xi(\mathcal{C}\cup\mathcal{D})\subseteq\X$ and $U$ is continuously differentiable, item i) of Proposition \ref{prop:lyapunovisps} holds. Recall that $\A_\mathcal{H}:=\{\xi\in\X\mid\chi\in\A\land\eta=0\}$. Due to item i) of Assumption \ref{ass:lyapcond}, there exist functions $\alpha_1,\alpha_2\in\mathcal{K}_\infty$ such that for all $\xi\in\X$ %and $\nu\in\mathcal{PC}_\V$,
    \begin{equation}\label{eq:ubound4}
        \alpha_1(|\xi|_{\mathcal{A}_\mathcal{H}})\leq U(\xi)\leq\alpha_2(|\xi|_{\mathcal{A}_\mathcal{H}})
    \end{equation}
    and thus item ii) of Proposition \ref{prop:lyapunovisps} holds. Next, we have for all $\xi\in\X$ and $\nu\in\V$,
    \begin{multline}\label{eq:ubound2}
        \left\langle \nabla U(\xi),F(\xi,\nu)\right\rangle\leq\left\langle\nabla V(\chi),F_\chi(\chi,\nu)\right\rangle+\sum_{i\in\mathcal{N}}\bar\Psi_i(o_i)\\
        \shoveleft{\stackrel{\eqref{eq:barpsi},\eqref{eq:vbound},\eqref{eq:trigger_dynamic}}{\leq} -\alpha(|\chi|_\A) + \gamma(|\nu|) + \sum_{i\in\mathcal{N}}\Big(c_i-\varphi_i(\eta_i)\Big)}\\
        \leq -\alpha^d(|\xi|_{\A_\mathcal{H}})+\gamma(|\nu|)+c
    \end{multline}
    with $c:=\sum_{i\in\mathcal{N}}c_i$ and for some $\alpha^d\in\mathcal{K}_\infty$. Hence, item iii) of Proposition \ref{prop:lyapunovisps} holds. In view of \eqref{eq:jumpmap} and \eqref{eq:vbound2}, we note that for all $(\xi,\nu)\in\mathcal{D}$ and $g\in G(\xi,\nu)$,
    \begin{equation}
        U(g)-U(\xi)\leq0,\label{eq:ubound3}
    \end{equation}
    thus, item iv) of Proposition \ref{prop:lyapunovisps} also holds.
\end{proof}

Since we have not established that the system is persistently flowing, we cannot claim that $\A_\mathcal{H}$ is IS(p)S, however, the bound \eqref{eq:ispsdef} of Definition \ref{def:isps} holds w.r.t. $\A_\mathcal{H}$ as long as the solution is defined (i.e., for all $(t,j)\in\dom\xi$). A similar argument can be constructed for the hybrid system $\mathcal{H}^s$ given by \eqref{eq:fchi}, \eqref{eq:gchi}, \eqref{eq:sflowjump} and the static triggering condition \eqref{eq:trigger_static}.

\subsection{Proof of Theorem \ref{thm:dynamictrigger}}\label{app:thm1}
Due to Lemma \ref{lem:isps}, we are left with proving that all maximal solutions to $\mathcal{H}$ are complete and that $\mathcal{H}$ has a SGiMIET, which implies that $\mathcal{H}$ is persistently flowing. We continue with proving completeness of maximal solutions first.

\subsubsection{Completeness of maximal solutions}
To prove that all maximal solutions are complete, we use the following proposition, which is taken from \cite[Proposition 9]{Heemels2021}, using closedness of $\mathcal{C}$.

\begin{proposition}\label{prop:210_v2}%\cite{Heemels2021}
    Consider the hybrid system $\mathcal{H}$ in Section \ref{ch:4} where $\Psi_i$ is given by \eqref{eq:trigger_dynamic}. Given an input $\nu\in\PC_\V$, there exists a non-trivial solution $\phi$ to $\mathcal{H}$ with $\phi(0,0)=\xi\in\X$ if and only if $(\xi,\nu(0))\in\mathcal{D}$ or
    \renewcommand{\labelitemi}{(VC)}
    \begin{itemize}
        \item there exist $\epsilon>0$ and an absolutely continuous function $z:[0,\epsilon]\rightarrow\R^{n_x}$ such that $z(0)=\xi$, $\dot{z}(t) \in F(z(t),w(t))$ for almost all $t \in [0,\epsilon]$ and $(z(t),w(t))\in \mathcal{C} $ for all $t\in[0,\epsilon]$.
    \end{itemize}
    \renewcommand{\labelitemi}{$\textbullet$}
    If (VC) holds for all  $\xi\in\R^{n_x}$ and all $\nu\in\PC_\V$ with $(\xi,\nu(0))\in\mathcal{C}\setminus\mathcal{D}$, then for all $\nu\in\PC_\V$ every maximal solution $\phi\in\mathcal{S}_\mathcal{H}(\nu)$ satisfies exactly one of the following properties:
    \begin{enumerate}[label={(\alph*)}]
        \item $\phi$ is complete;
        \item $\phi$ is not complete and ``ends with flow'': $\dom \phi$ is bounded and the interval $I^J:=\{t:(t,J)\in\dom\phi\}$ with $J=\sup_j\dom\phi$ is open to the right, and there does not exist an absolutely continuous function $z:\overline{I^J}\rightarrow\X$ satisfying $\dot z(t)\in F(z(t),\nu(t))$ for  almost all $t\in I^J$ and $(z(t),\nu(t))\in\mathcal{C}$ for all $t\in \interior I^J$, and such that  $z(t)=\phi(t,J)$ for all $t\in I^J$;
        \item $\phi$ is not complete and ``ends with a jump'' or a ``discontinuity'' of $w$: $\dom\phi$ is bounded with $(T,J):=\sup\dom\phi\in \dom \phi$,  $(\phi(T,J),\nu(T))\not\in{\mathcal{C}}\cup {\mathcal{D}} $.
    \end{enumerate}
\end{proposition}
%We would like to note that the sets $\mathcal{C}$ and $\mathcal{D}$ are independent of the process disturbances $v$, hence, we do not require that $v\in\mathcal{PC}$.
Let $\nu\in\PC_\V$ and $(\xi,\nu(0))\in\mathcal{C}\setminus\mathcal{D}$ be given.
%$\phi\in\mathcal{S}_\mathcal{H}(\nu)$ with $\xi=\phi(0,0)\in\X$.
To ensure completeness of $\phi\in\mathcal{S}_\mathcal{H}(\nu)$, we first prove that (VC) holds. Let $t_1>t_0=0$ denote the time at which the first discontinuity in $\nu$ occurs. There exists an $\epsilon_1\in(0,t_1)$ such that $v$ and $w$ are continuous on $[0,\epsilon_1]$. Since $F$ in \eqref{eq:flowmap} is continuous in both the state and time, we can use Peano’s existence theorem to conclude that there exist (possibly multiple) solutions $z$ to $\dot{z}(t)\in F(z(t),\nu(t))$ with $z(0)=\xi$ defined on $[0,\epsilon_2]$ where $\epsilon_2\leq\epsilon_1$. Next, we have to show that the solutions $z(\cdot)$ remain in $\mathcal{C}$ on $[0,\epsilon]$ for some $\epsilon\in(0,\epsilon_2]$. We write $z=(x,e,\what,\eta)$. Observe that if we have for all $i\in\mathcal{N}$ that $\eta_i(0)>0$, $|\what_i(0)|<\overline{w}_i$ and $\eta_i(0)+\theta_i\Psi_i(o_i(0))>0$ then certainly the solution will remain in $\mathcal{C}_i$ for some nontrivial time window due to the ``gap'' to the boundary of $\mathcal{C}$ in \eqref{eq:flowset} and continuity of solutions. Consider all $\mathcal{M}\subseteq\mathcal{N}$ for which one of these inequalities does not hold and let $i\in\mathcal{M}$. For $i\in\mathcal{M}$ and since $\xi\in\mathcal{C}\setminus\mathcal{D}$ we can distinguish three cases, which may or may not hold simultaneously:
\begin{enumerate}[label={\arabic*)}]
    \item $\eta_i(0)+\theta_i\Psi_i(o_i(0))=0$ and $\Psi_i(o_i(0))>0$ (implying that $\theta_i=0$),
    \item $\eta_i(0)=0$ and $\eta_i(0)+\theta_i\Psi_i>0$ (implying that $\Psi_i>0$),
    \item $|\what_i(0)|=\overline{w}_i$.
%    \item $\nu\in\partial\V$, where $\partial\V$ denotes the boundary of $\V$.
\end{enumerate}
For cases 1) and 2), we note that $\dot\eta_i(0)=\bar\Psi_i(\eta_i(0),o_i(0))=\Psi_i(o_i(0))>0$. Recall that $x$, $e$, $\what$, $\eta$ and $w$ are continuous on the interval $[0,\epsilon_2]$. Since $\eta_i(0)=0$, $\bar\Psi_i$ is continuous, and $\Psi_i(o_i)>0$ for some nontrivial time window, we find that $\eta_i\geq0$ on this time window by means of the comparison lemma, and consequently $\eta_i+\theta_i\Psi_i(o_i)\geq0$ on some time window. %For case 2), we can follow a similar argument, since, as $\eta_i+\theta_i\Psi_i(o_i)>0$ due to $(\xi,\nu(0))\in\mathcal{C}\setminus\mathcal{D}$, and due to $\Psi_i(o_i)$ and $\eta_i$ being absolutely continuous on the interval $[0,\epsilon_2]$, hence, $\eta_i+\theta_i\Psi_i(o_i)\geq0$ for some nontrivial time window and solutions remain in $\mathcal{C}$ for some nontrivial amount of time.
For case 3), we note that, in view of \eqref{eq:modelsplit}, $\dot{\what}=0$. Consequently, in all cases, there exists $\epsilon_i'\in[0,\epsilon_2]$ such that $z(t)\in\mathcal{C}_i$ for any $t\in[0,\epsilon_i']$.
%a non-trivial interval of time during which $\phi$ remains in $\mathcal{C}_i$.
Since $\mathcal{C}$ is the intersection of $\mathcal{C}_1,\ldots,\mathcal{C}_N$ with $N\in\N_{>0}$, there exists $\epsilon>0$ such that $z(t)\in\mathcal{C}$ for any $t\in[0,\epsilon]$.

%a non-trivial interval of time during which $\phi$ remains in $\mathcal{C}$ and (VC) holds for all $(\xi,\nu(0))\in\mathcal{C}\setminus\mathcal{D}$. %For item 4), (VC) holds trivially due to $\nu\in\V$.

Since (VC) holds for $(\xi,\nu(0))\in\mathcal{C}\setminus\mathcal{D}$ and $\nu\in\PC_\V$, there exists a non-trivial solution to $\mathcal{H}$ for any $\xi$ and $\nu\in\PC_\V$ with $(\xi,\nu(0))\in\mathcal{C}\cup\mathcal{D}$. Hence, any maximal $\phi$ satisfies exactly one of the three cases (a)-(c) in Proposition \ref{prop:210_v2}.

Item (b) cannot occur due to item iv) of Assumption \ref{ass:lyapcond}, as the states of the hybrid system remain bounded for all $(t,j)\in\dom\phi$, hence, there are no finite escape times.
%the $x$-system does not have finite escape times. Moreover, due to Lemma \ref{lem:ebound}, $|e(t,j)|$ remains bounded bounded for all $(t,j)\in\dom\phi$, which means that finite escape times cannot occur due to $e$. Since $\dot\eta_i=\Psi_i(o_i)-\varphi_i(\eta_i)$, $\varphi_i\in\mathcal{K}_\infty$, $\eta_i\geq0$ and $\Psi_i(o_i)$ being continuous and independent of $\eta_i$ in view of \eqref{eq:trigger_dynamic}, this, in turn, implies that $\eta$ does not have finite escape times.

Item (c) only occurs if either $G(\mathcal{D})\times\V\not\subset\mathcal{C}\cup\mathcal{D}$ or if $(\xi(T,J),\nu(T))\not\in\mathcal{C}\cup\mathcal{D}$ due to a discontinuity in $w$. For the former, we note that $\mathcal{C}\cup\mathcal{D}=\mathbb X\times\V$. In view of \eqref{eq:jumpmap}, we note that $\what_i^+=w_i$ if $i$ broadcasts its state and $\what_i^+=\what_i$ otherwise. Additionally, $\eta_i^+=\eta_i$. Consequently, $G(\mathcal{D})\times\V\subset\mathcal{C}\cup\mathcal{D}$. Furthermore, since $\mathcal{C}\cup\mathcal{D}=\mathbb X\times\V$, item (c) cannot occur due to a discontinuity in signal $w$, since by Assumption \ref{ass:noise}, $w(t)\in\W$ for all $t\geq0$. Thus we deduce that $\phi$ is complete. Since $\nu$ and $\phi\in\mathcal{S}_\mathcal{H}(\nu)$ have been arbitrarily selected, we have proved that all maximal solutions are complete. Next we show that maximal solutions are also $t$-complete.

\subsubsection{$t$-completeness and SGiMIET}
We prove $t$-completeness by showing that system $\mathcal{H}$ has the SGiMIET property. We proceed by examining the time between two successive jumps generated by triggering condition $i\in\mathcal{N}$. To do so, note that the ``static triggering condition'' $\Psi_i(o_i)\leq0$ always holds when the (mixed) dynamic triggering condition $\eta_i+\theta_i\Psi_i(o_i)\leq0\land\Psi_i(o_i)\leq0$ is satisfied. Let $\nu\in\PC_\V$, $\Delta\geq0$ and $\phi\in\mathcal{S}_\mathcal{H}(\nu,\mathcal{B})$ with $\mathcal{B}=\{\Xi\in\X\mid|\Xi|\leq\Delta\}$. Since $\eta_i(t,j)\geq0$ for all $(t,j)\in\dom\phi$, we analyze when $\Psi_i(o_i)\leq0$, i.e., when
\begin{equation}
    \delta_i(o_i)+c_i-\beta_i(|\widetilde{e}_i|)\leq0\label{eq:conservativetrigger}
\end{equation}
holds to under-estimate the inter-event times generated by triggering condition $i$. Since $\delta_i$ takes non-negative values only, we can under-estimate the inter-event times for triggering condition $i$ by analyzing when
\begin{equation}\label{eq:tile_bound}
    c_i\leq\beta_i(|\widetilde{e}_i|),
\end{equation}
i.e., when $|\widetilde{e}_i|\geq\beta_i^{-1}(c_i)$.
% By defining $z_i(s)=\beta_i(s)+\mu\zeta(s)$ and by observing that $z$ is class-$\mathcal{K}_\infty$, we obtain by rewriting
% \begin{equation}
    % z_i^{-1}(c_i)\leq|\widetilde{e}_i|.\label{eq:tile_bound}
% \end{equation}
Note that we can upper-bound the right-hand side of the latter inequality, in view of Assumption \ref{ass:noise}, as
\begin{equation}
    \beta_i^{-1}(c_i)\leq|\widetilde{e}_i|\leq|e_i|+|w_i|+|\what_i|\leq|e_i|+2\overline{w}_i.
\end{equation}
Hence, we can under-estimate the inter-event times by analyzing when
\begin{equation}
    \beta_i^{-1}(c_i)-2\overline{w}_i=|e_i|.\label{eq:semiglobaltrigger}
\end{equation}
Recall that, by the condition on $c_i$ in Theorem \ref{thm:dynamictrigger}, we have $c_i>\beta_i(2\overline{w}_i)$, thus, the left-hand side of \eqref{eq:semiglobaltrigger} is strictly positive. In view of \eqref{eq:semiglobaltrigger}, we define
\begin{equation}\label{eq:cbar}
    \overline{c}_i:=\beta_i^{-1}(c_i)-2\overline{w}_i>0.
\end{equation}
Since $|e_i|$ is set to 0 after a transmission due to triggering rule $i$, the inter-event time for triggering rule $i$ is lower bounded by the time it takes for $|e_i|$ to grow from $0$ to $\overline{c}_i$ in view of \eqref{eq:semiglobaltrigger}. Note that the bound in \eqref{eq:cbar} is \emph{not} dependent on actual values of $w_i$, only on the upper-bounds presented in Assumption \ref{ass:noise}. In the following, we provide a lower-bound on this inter-event time. In view of Lemma \ref{lem:isps}, we have that there exists a $q>0$ such that $|\phi(t,j)|\leq q$ for all $(t,j)\in\dom\phi$. Since $F$ in \eqref{eq:flowmap} is continuous and $\|\nu\|_\infty$ is finite by Assumption \ref{ass:noise}, there exists $\mu>0$ such that $|F(\phi(t,j),\nu(t))|\leq\mu$ for all $(t,j)\in\dom\phi$. Thus, for almost all $j\in\N_{\geq 0}$ and almost all $t\in I^j$ where $I^j=\{t \,:\, (t,j)\in\dom \xi\}$, $\frac{d|e_i(t)|}{dt}\leq\mu$. Consequently, the time between any two transmissions generated by triggering rule $i$ is larger than or equal to $\overline{c}_i / \mu$. Hence, $\mathcal{H}$ has the SGiMIET property and all maximal solutions are $t$-complete. In other words, $\mathcal{H}$ is persistently flowing.

%In view of Lemma \ref{lem:isps}, item iv) of Assumption \ref{ass:lyapcond} and Assumption \ref{ass:noise}, we have that $|g(x,e,\what,v,w)|\leq m+\ell(|\nu|)$ for all $(t,j)\in\dom\phi$. Thus, for almost all $j\in\N_{\geq 0}$ and almost all $t\in I^j$ where $I^j=\{t \,:\, (t,j)\in\dom \xi\}$, $\frac{d|e_i(t)|}{dt}\leq m+\ell(\|\nu\|_\infty)$. Consequently, the time between any two transmissions generated by triggering rule $i$ is larger than or equal to $\overline{c}_i / (m+\ell(\|\nu\|_\infty))$. Hence, $\mathcal{H}$ has the SGiMIET property and all maximal solutions are $t$-complete. In other words, $\mathcal{H}$ is persistently flowing.

Since the system is persistently flowing, combined with the result of Lemma \ref{lem:isps}, we also have that $\mathcal{H}$ is ISpS w.r.t. the set $\mathcal{A}_\mathcal{H}$ according to Proposition \ref{prop:lyapunovisps}.
\hfill$\blacksquare$

\subsection*{Proof of Lemma \ref{lem:xibound}}\label{app:lem1}
Let $\nu\in\PC_\V$ and $p>0$ be given. Using Lemma \ref{lem:isps}, we establish that for all $\xi\in\mathcal{S}_\mathcal{H}(\nu,\mathcal{B})$ with $\mathcal{B}:=\{\Xi\in\X\mid|\Xi|\leq p\}$, $|\xi(t,j)|_{\A_\mathcal{H}}\leq\lambda$ for all $(t,j)\in\dom\xi$, where $\lambda>0$ depends on $p$, $\|\nu\|_\infty$ and $c$ in \eqref{eq:ubound2}. For the first case, i.e., when $\A$ is compact (and thus $\A_\mathcal{H}$ is compact), the set $\mathcal{S}:=\{\Xi\in\mathbb{X}\mid|\Xi|_{\A_\mathcal{H}}\leq\lambda\}$ is compact. Since $\xi(t,j)\in\mathcal{S}$ for all $(t,j)\in\dom\xi$, there exists $q>0$ such that $|\xi(t,j)|\leq q$ for all $(t,j)\in\dom\xi$, hence, item iv) of Assumption \ref{ass:lyapcond} is satisfied. For the second case, i.e., when $\proj_x(\A)$ is compact and a ZOH is employed, we have that $|x(t,j)|_{\proj_x(\A)}\leq\lambda$ for all $(t,j)\in\dom\xi$ due to \eqref{eq:ispsdef}. Hence, since $\proj_x(\A)$ is compact, there exists a $r>0$ such that $|x(t,j)|\leq r$ for all $(t,j)\in\dom\xi$. Since all maps $g_{p,i}$ in \eqref{eq:sysdyn} are continuous, there exists $s>0$ such that $|y(t,j)|\leq s$ for all $(t,j)\in\dom\xi$, and, consequently, $|\widetilde{y}(t,j)|\leq \bar{s}$ for all $(t,j)\in\dom\xi$, where $\bar{s}=s+\overline{w}_i$. Moreover, due to the ZOH, $e_i(t,j)$ is the difference between two points on the trajectories of $\widetilde{y}_i(t,j)$ after the first jump. Consequently, there exists a $\varsigma>0$ (possibly dependent on $e(0,0)$) such that $|e(t,j)|\leq \varsigma$ for all $e(0,0)\in\proj_e(\mathcal{B})$ and all $(t,j)\in\dom\xi$. Lastly, $\what\in\W$, which is compact, and $|\eta(t,j)|\leq\lambda$ due to \eqref{eq:ispsdef} (as $\proj_\eta(\A_\mathcal{H})$ is compact). Thus, since $\xi=(x,e,\what,\eta)$, there exists $q>0$ such that $|\xi(t,j)|\leq q$ for all $(t,j)\in\dom\xi$. \hfill$\blacksquare$

\bibliographystyle{plain}        % Include this if you use bibtex
\bibliography{references.bib}           % and a bib file to produce the
                                 % bibliography (preferred). The
                                 % correct style is generated by
                                 % Elsevier at the time of printing.

%\begin{thebibliography}{99}     % Otherwise use the
                                 % thebibliography environment.
                                 % Insert the full references here.
                                 % See a recent issue of Automatica
                                 % for the style.
%  \bibitem[Heritage, 1992]{Heritage:92}
%     (1992) {\it The American Heritage.
%     Dictionary of the American Language.}
%     Houghton Mifflin Company.
%  \bibitem[Able, 1956]{Abl:56}
%     B.~C.~Able (1956). Nucleic acid content of macroscope.
%     {\it Nature 2}, 7--9.
%  \bibitem[Able {\em et al.}, 1954]{AbTaRu:54}
%     B.~C. Able, R.~A. Tagg, and M.~Rush (1954).
%     Enzyme-catalyzed cellular transanimations.
%     In A.~F.~Round, editor,
%     {\it Advances in Enzymology Vol. 2} (125--247).
%     New York, Academic Press.
%  \bibitem[R.~Keohane, 1958]{Keo:58}
%     R.~Keohane (1958).
%     {\it Power and Interdependence:
%     World Politics in Transition.}
%     Boston, Little, Brown \& Co.
%  \bibitem[Powers, 1985]{Pow:85}
%     T.~Powers (1985).
%     Is there a way out?
%     {\it Harpers, June 1985}, 35--47.

%\end{thebibliography}

\end{document}

%% file: Figures/Garcia2013_StateEvolution_c.tex
% !TEX root = ../ieeetac.tex
\begin{tikzpicture}[baseline=(current bounding box.center)]
    \footnotesize
\begin{axis}[%
width=\columnwidth,
height=4cm,
at={(0,0)},
%scale only axis,
xmin=0,
xmax=8,
xlabel style={font=\color{white!15!black}},
xlabel={Time [s]},
ymin=-8,
ymax=8,
ylabel style={font=\color{white!15!black}},
ylabel={$x_i$},
%axis background/.style={fill=white},
xmajorgrids,
ymajorgrids,
legend columns=4,
legend style={legend cell align=left, align=left, draw=white!15!black},
ytick = {-8, -4, 0, 4, 8},
]
\addplot [color=black]
  table[row sep=crcr]{%
0	8\\
0.0601519961646206	6.91726609542521\\
0.117621217771277	5.90445843861971\\
0.174000221065236	5.1461423949187\\
0.229151401495203	4.5011647680592\\
0.28786680199119	3.87356548421196\\
0.343853052847511	3.42859777047797\\
0.404061505582155	2.94464662031485\\
0.457313348769833	2.63499223721436\\
0.517299303171965	2.28970306441674\\
0.575515143700868	2.01972465138044\\
0.632678126050925	1.77901777575274\\
0.689298545478555	1.56075169452438\\
0.747512149273467	1.38566216365498\\
0.808557611947797	1.20336103372618\\
0.864929098110121	1.0746176495161\\
0.924967090987112	0.941252981913162\\
0.982603929724419	0.82669720875616\\
1.04138129251223	0.734783588738941\\
1.10131646566076	0.640617289227402\\
1.15963029968841	0.574083662368738\\
1.22006508955089	0.504557875872978\\
1.27901774522118	0.449447536760696\\
1.3389	0.398341631421994\\
1.40032960290062	0.352328110794849\\
1.462	0.313731350041464\\
1.52214078004887	0.277406701881005\\
1.58332532630938	0.249255450331027\\
1.64287270637993	0.222084678434384\\
1.70300000000001	0.199795922387107\\
1.7644	0.178792691724941\\
1.82480000000001	0.159527723513791\\
1.8864	0.144450500644428\\
1.94606989993623	0.129118560183113\\
2.0066080111338	0.117895667630701\\
2.06656338967949	0.1063614156273\\
2.12790000000004	0.0967755504659357\\
2.18830000000005	0.0880167798486741\\
2.24900000000003	0.0800706541995691\\
2.31035525726748	0.0733227185061356\\
2.37180000000001	0.0667302046973263\\
2.43120000000003	0.0614498146210584\\
2.49250000000004	0.0565461614347628\\
2.55311101827205	0.0517317242024289\\
2.61520000000002	0.048065662384573\\
2.67770000000001	0.043974492317504\\
2.73920000000001	0.0410371625811499\\
2.80090000000004	0.0379709811286132\\
2.86110000000001	0.0353062100811079\\
2.9231	0.0328702119806277\\
2.98520000000004	0.0303863024766308\\
3.04560000000002	0.0285412001300396\\
3.10810000000001	0.0266146220831996\\
3.16880000000002	0.024874713310855\\
3.23130000000001	0.0234011840267236\\
3.29280061851412	0.0218707798859905\\
3.355	0.0206326645953491\\
3.41680000000002	0.0194961303858536\\
3.47920000000002	0.0181268307478556\\
3.5411	0.0171752055407549\\
3.60260000000002	0.0162026549869151\\
3.66350000000002	0.0151927794001237\\
3.72600000000001	0.0144378908121255\\
3.78750000000002	0.0136570647883611\\
3.84910000000002	0.0127699500718454\\
3.91080000000002	0.0121350526563273\\
3.97240000000001	0.0114877786609196\\
4.03490000000002	0.0108015738879097\\
4.09640000000003	0.0102838526668603\\
4.15880000000006	0.00975137098535543\\
4.22110000000003	0.00917957442531112\\
4.28350000000006	0.00856779195105187\\
4.34600000000002	0.00818156149986777\\
4.40850000000004	0.00779585520303001\\
4.47100000000007	0.00740771953499733\\
4.53350000000009	0.00701958386696466\\
4.59400000000002	0.00657942705519005\\
4.65620000000003	0.00623262453458036\\
4.71870000000005	0.00587913413687197\\
4.78120000000007	0.00553061057821958\\
4.84370000000009	0.0051820870195672\\
4.90620000000005	0.00499129070177317\\
4.96870000000002	0.00482929154281656\\
5.03060000000004	0.00459496125929079\\
5.09300000000002	0.00431858806083436\\
5.15550000000004	0.00399524685041817\\
5.21780000000007	0.00367220927858331\\
5.28030000000004	0.00348918747632768\\
5.34280000000006	0.00336733131788952\\
5.40530000000008	0.00323888291720638\\
5.46770000000005	0.00310910282029493\\
5.53020000000007	0.00296696003962647\\
5.59270000000003	0.00280371586067305\\
5.65520000000005	0.00257736448246085\\
5.71770000000007	0.00235101310424863\\
5.78020000000004	0.00221105400646741\\
5.84270000000006	0.00217415938358585\\
5.90520000000002	0.00204839333484077\\
5.96770000000004	0.00190885608218973\\
6.03020000000001	0.00176815783163212\\
6.09260000000004	0.00162609302806376\\
6.15510000000006	0.00149454322991052\\
6.21760000000009	0.00136159043249413\\
6.28010000000011	0.00122863763507774\\
6.34260000000013	0.00109568483766134\\
6.40510000000015	0.000962732040244943\\
6.46760000000006	0.000857782302239671\\
6.53010000000008	0.000899848626141246\\
6.5926000000001	0.000941914950042821\\
6.65510000000012	0.000983981273944397\\
6.71760000000014	0.00102604759784598\\
6.78010000000005	0.00106718202182978\\
6.84250000000001	0.00107475637049668\\
6.90500000000003	0.00101357574463318\\
6.96750000000006	0.000952395118769668\\
7.03000000000008	0.00089121449290616\\
7.0925000000001	0.000830033867042653\\
7.15500000000012	0.000768853241179146\\
7.21670000000005	0.000716846081335957\\
7.27900000000005	0.000671047327657035\\
7.34150000000001	0.000622267968007753\\
7.40400000000003	0.000490264458925305\\
7.46650000000005	0.000358260949842862\\
7.52900000000007	0.000226257440760419\\
7.5915000000001	9.4253931677972e-05\\
7.65400000000012	-3.77495774044748e-05\\
7.71650000000014	-0.00016975308648692\\
7.77900000000016	-0.000301756595569365\\
7.84150000000018	-0.000433760104651814\\
7.90400000000003	-0.000443593379642848\\
7.96650000000005	-0.000319588948068861\\
};
\addlegendentry{$P_1$}

\addplot [color=red, dashed]
  table[row sep=crcr]{%
0	6\\
0.0601519961646206	5.2781956411331\\
0.117621217771277	4.59006825328309\\
0.174000221065236	4.04013263808564\\
0.229151401495203	3.51245663335258\\
0.28786680199119	3.08586051497431\\
0.343853052847511	2.66575891409787\\
0.404061505582155	2.35065710658206\\
0.457313348769833	2.03203900081741\\
0.517299303171965	1.79761871647275\\
0.575515143700868	1.53766975460448\\
0.632678126050925	1.36686112075359\\
0.689298545478555	1.19926192134709\\
0.747512149273467	1.0466569837301\\
0.808557611947797	0.920808511734549\\
0.864929098110121	0.802278871220311\\
0.924967090987112	0.712702351250902\\
0.982603929724419	0.629381764246581\\
1.04138129251223	0.556080677585584\\
1.10131646566076	0.493744761120716\\
1.15963029968841	0.432861408188121\\
1.22006508955089	0.385465985765226\\
1.27901774522118	0.336917413528799\\
1.3389	0.302619715035902\\
1.40032960290062	0.26547312895163\\
1.462	0.235761084650962\\
1.52214078004887	0.210064443800691\\
1.58332532630938	0.185794931772566\\
1.64287270637993	0.166706711986068\\
1.70300000000001	0.147817859302417\\
1.7644	0.132836465440542\\
1.82480000000001	0.11937813246527\\
1.8864	0.106685127776028\\
1.94606989993623	0.0968546553146769\\
2.0066080111338	0.0863803170478966\\
2.06656338967949	0.07874329708062\\
2.12790000000004	0.0704272161450957\\
2.18830000000005	0.0643436020463286\\
2.24900000000003	0.0579529689871559\\
2.31035525726748	0.0529584944281408\\
2.37180000000001	0.0480311712235909\\
2.43120000000003	0.0443238151604575\\
2.49250000000004	0.0401853295540145\\
2.55311101827205	0.0373076387305737\\
2.61520000000002	0.0336982959074248\\
2.67770000000001	0.0313386986754164\\
2.73920000000001	0.028448730345955\\
2.80090000000004	0.0264112733668473\\
2.86110000000001	0.024299618103983\\
2.9231	0.0225122595322725\\
2.98520000000004	0.0207959101618194\\
3.04560000000002	0.0192381627105867\\
3.10810000000001	0.0178708690396039\\
3.16880000000002	0.01671700023933\\
3.23130000000001	0.0155488408550674\\
3.29280061851412	0.0146091597058942\\
3.355	0.0136827722924014\\
3.41680000000002	0.012613277155185\\
3.47920000000002	0.0119490173168441\\
3.5411	0.0111457822135612\\
3.60260000000002	0.0104214150945773\\
3.66350000000002	0.0099240459811605\\
3.72600000000001	0.00926300820450614\\
3.78750000000002	0.00864032674588157\\
3.84910000000002	0.00827314409646412\\
3.91080000000002	0.00776982213840735\\
3.97240000000001	0.00728048395291008\\
4.03490000000002	0.00691717135609931\\
4.09640000000003	0.00652994970777017\\
4.15880000000006	0.00607994310627966\\
4.22110000000003	0.00565681880376178\\
4.28350000000006	0.00550487872283469\\
4.34600000000002	0.00522202011566588\\
4.40850000000004	0.0048730675350252\\
4.47100000000007	0.00452849885555949\\
4.53350000000009	0.0041839301760938\\
4.59400000000002	0.00405223159303277\\
4.65620000000003	0.00385925903306313\\
4.71870000000005	0.00367491068510874\\
4.78120000000007	0.00348011586526624\\
4.84370000000009	0.00328532104542374\\
4.90620000000005	0.0030064575506034\\
4.96870000000002	0.00276397050084079\\
5.03060000000004	0.00264980359683248\\
5.09300000000002	0.0025055055358653\\
5.15550000000004	0.00235197125227029\\
5.21780000000007	0.00231960422738648\\
5.28030000000004	0.00220722351567755\\
5.34280000000006	0.00205813988340423\\
5.40530000000008	0.0019129773413021\\
5.46770000000005	0.0016975168742167\\
5.53020000000007	0.00147457504961037\\
5.59270000000003	0.00132262428223208\\
5.65520000000005	0.0013360167362678\\
5.71770000000007	0.00134940919030353\\
5.78020000000004	0.00132423167052809\\
5.84270000000006	0.00125304084866218\\
5.90520000000002	0.00119516590407931\\
5.96770000000004	0.00113797732502245\\
6.03020000000001	0.00109323945884493\\
6.09260000000004	0.0011121055898391\\
6.15510000000006	0.00105604942413354\\
6.21760000000009	0.00100156829851342\\
6.28010000000011	0.000947087172893299\\
6.34260000000013	0.000892606047273178\\
6.40510000000015	0.000838124921653057\\
6.46760000000006	0.000769958731060964\\
6.53010000000008	0.000629945949365366\\
6.5926000000001	0.000489933167669766\\
6.65510000000012	0.000349920385974167\\
6.71760000000014	0.000209907604278569\\
6.78010000000005	6.9325487621384e-05\\
6.84250000000001	4.67434025316768e-06\\
6.90500000000003	0.00013897342572824\\
6.96750000000006	0.000273272511203314\\
7.03000000000008	0.000407571596678389\\
7.0925000000001	0.000541870682153461\\
7.15500000000012	0.000676169767628531\\
7.21670000000005	0.0007870454564531\\
7.27900000000005	0.000857648684189039\\
7.34150000000001	0.000875952886684692\\
7.40400000000003	0.000895381529751447\\
7.46650000000005	0.000914810172818201\\
7.52900000000007	0.000934238815884956\\
7.5915000000001	0.000953667458951713\\
7.65400000000012	0.000973096102018475\\
7.71650000000014	0.000992524745085236\\
7.77900000000016	0.001011953388152\\
7.84150000000018	0.00103138203121876\\
7.90400000000003	0.00101141947995691\\
7.96650000000005	0.000947580203100412\\
};
\addlegendentry{$P_2$}

\addplot [color=red, dashdotted]
  table[row sep=crcr]{%
0	4\\
0.0601519961646206	3.51877817442359\\
0.117621217771277	3.08065817906172\\
0.174000221065236	2.75732687235971\\
0.229151401495203	2.44658604460711\\
0.28786680199119	2.16782003628608\\
0.343853052847511	1.94970495995646\\
0.404061505582155	1.69849682214589\\
0.457313348769833	1.53290455091757\\
0.517299303171965	1.33523938944966\\
0.575515143700868	1.18164668615564\\
0.632678126050925	1.03936503252135\\
0.689298545478555	0.910730245558434\\
0.747512149273467	0.803735624045081\\
0.808557611947797	0.689615213130234\\
0.864929098110121	0.611436315584671\\
0.924967090987112	0.522384690144091\\
0.982603929724419	0.458022804787732\\
1.04138129251223	0.393139552328547\\
1.10131646566076	0.33998347816675\\
1.15963029968841	0.292540332455341\\
1.22006508955089	0.249110528275517\\
1.27901774522118	0.215165597959768\\
1.3389	0.182231702323227\\
1.40032960290062	0.154951274348322\\
1.462	0.13035522238987\\
1.52214078004887	0.109697230094\\
1.58332532630938	0.0919395064987537\\
1.64287270637993	0.0760945105704902\\
1.70300000000001	0.0634938300178835\\
1.7644	0.0511478377585116\\
1.82480000000001	0.0414471850267491\\
1.8864	0.0329595993781591\\
1.94606989993623	0.0252710700932088\\
2.0066080111338	0.0199580617495684\\
2.06656338967949	0.0139164405995751\\
2.12790000000004	0.0101798656839504\\
2.18830000000005	0.00603290578631622\\
2.24900000000003	0.0032703502740009\\
2.31035525726748	0.000453547384898456\\
2.37180000000001	-0.00154041986807546\\
2.43120000000003	-0.00339355082393726\\
2.49250000000004	-0.00481958017976326\\
2.55311101827205	-0.00616298394386505\\
2.61520000000002	-0.00694525283394258\\
2.67770000000001	-0.00801690920985266\\
2.73920000000001	-0.00832279713391726\\
2.80090000000004	-0.00887758283095006\\
2.86110000000001	-0.00920733089077134\\
2.9231	-0.00956881127233089\\
2.98520000000004	-0.00957348348469733\\
3.04560000000002	-0.00954659426114771\\
3.10810000000001	-0.00956253562814002\\
3.16880000000002	-0.00945440282417834\\
3.23130000000001	-0.0092903265049375\\
3.29280061851412	-0.00916581052253615\\
3.355	-0.00881117424529522\\
3.41680000000002	-0.00843549800907273\\
3.47920000000002	-0.00817333354745268\\
3.5411	-0.00780466997989259\\
3.60260000000002	-0.00759257727896029\\
3.66350000000002	-0.00746031154792463\\
3.72600000000001	-0.00707557707990352\\
3.78750000000002	-0.00668342363223316\\
3.84910000000002	-0.00650599545708102\\
3.91080000000002	-0.00632883469575055\\
3.97240000000001	-0.006037651108549\\
4.03490000000002	-0.0057968038989256\\
4.09640000000003	-0.00553521264393623\\
4.15880000000006	-0.00515896043633709\\
4.22110000000003	-0.00484654387293083\\
4.28350000000006	-0.00479564791249598\\
4.34600000000002	-0.00466768982544119\\
4.40850000000004	-0.00443538390082284\\
4.47100000000007	-0.00416530381214282\\
4.53350000000009	-0.00389522372346281\\
4.59400000000002	-0.00367532070666181\\
4.65620000000003	-0.0034793293824154\\
4.71870000000005	-0.00338917663069155\\
4.78120000000007	-0.00324795928133825\\
4.84370000000009	-0.00310674193198495\\
4.90620000000005	-0.00294080736061399\\
4.96870000000002	-0.00271978356451054\\
5.03060000000004	-0.00254816058858671\\
5.09300000000002	-0.00230307694089534\\
5.15550000000004	-0.00205605193901041\\
5.21780000000007	-0.00210045103446157\\
5.28030000000004	-0.00213489565081076\\
5.34280000000006	-0.00209083116068855\\
5.40530000000008	-0.00200308956321159\\
5.46770000000005	-0.00188785660246965\\
5.53020000000007	-0.00172293647141593\\
5.59270000000003	-0.00157771627723067\\
5.65520000000005	-0.00147109168145542\\
5.71770000000007	-0.00136446708568017\\
5.78020000000004	-0.00126738746279359\\
5.84270000000006	-0.0011816948251073\\
5.90520000000002	-0.00105540672684541\\
5.96770000000004	-0.00087497083318874\\
6.03020000000001	-0.000727731396111856\\
6.09260000000004	-0.000823536187480972\\
6.15510000000006	-0.000828443378309119\\
6.21760000000009	-0.000821963460116678\\
6.28010000000011	-0.000815483541924237\\
6.34260000000013	-0.000809003623731796\\
6.40510000000015	-0.000802523705539355\\
6.46760000000006	-0.000796601937664366\\
6.53010000000008	-0.000793610458955928\\
6.5926000000001	-0.000790618980247491\\
6.65510000000012	-0.000787627501539053\\
6.71760000000014	-0.000784636022830615\\
6.78010000000005	-0.000767353715064388\\
6.84250000000001	-0.000702118601960089\\
6.90500000000003	-0.000711702775515296\\
6.96750000000006	-0.000721286949070504\\
7.03000000000008	-0.000730871122625711\\
7.0925000000001	-0.000740455296180917\\
7.15500000000012	-0.000750039469736124\\
7.21670000000005	-0.000692723194866064\\
7.27900000000005	-0.000589263437710224\\
7.34150000000001	-0.000491278615531465\\
7.40400000000003	-0.000405037718760439\\
7.46650000000005	-0.000318796821989414\\
7.52900000000007	-0.00023255592521839\\
7.5915000000001	-0.000146315028447365\\
7.65400000000012	-6.0074131676341e-05\\
7.71650000000014	2.61667650946816e-05\\
7.77900000000016	0.000112407661865704\\
7.84150000000018	0.000198648558636727\\
7.90400000000003	0.000290975585827205\\
7.96650000000005	0.0003895512734643\\
};
\addlegendentry{$P_3$}

\addplot [color=black, dotted]
  table[row sep=crcr]{%
0	2\\
0.0601519961646206	1.8796858022843\\
0.117621217771277	1.75228827157192\\
0.174000221065236	1.56645781134657\\
0.229151401495203	1.42334439933407\\
0.28786680199119	1.25043496810853\\
0.343853052847511	1.07597240463707\\
0.404061505582155	0.941850978848987\\
0.457313348769833	0.8202825071067\\
0.517299303171965	0.706271650595041\\
0.575515143700868	0.607847533062153\\
0.632678126050925	0.517638767091077\\
0.689298545478555	0.448237894038904\\
0.747512149273467	0.373726357497147\\
0.808557611947797	0.319242640613673\\
0.864929098110121	0.263974886200846\\
0.924967090987112	0.224543324985976\\
0.982603929724419	0.182263794854423\\
1.04138129251223	0.153013053364168\\
1.10131646566076	0.120607443291898\\
1.15963029968841	0.0990631207180014\\
1.22006508955089	0.0762026662089746\\
1.27901774522118	0.0589612145866892\\
1.3389	0.0428591628053167\\
1.40032960290062	0.0298967989036218\\
1.462	0.0181219209391321\\
1.52214078004887	0.00919613675781238\\
1.58332532630938	0.00103280925788859\\
1.64287270637993	-0.00501118473205362\\
1.70300000000001	-0.0104258977314116\\
1.7644	-0.0146704316071493\\
1.82480000000001	-0.0180647360805584\\
1.8864	-0.0210103329070468\\
1.94606989993623	-0.0227984543008629\\
2.0066080111338	-0.0248916328435802\\
2.06656338967949	-0.0256744803934589\\
2.12790000000004	-0.0269540327254958\\
2.18830000000005	-0.0271942907081623\\
2.24900000000003	-0.0277687242749501\\
2.31035525726748	-0.0277032577066236\\
2.37180000000001	-0.027884052763027\\
2.43120000000003	-0.0275355298847236\\
2.49250000000004	-0.0272673135561721\\
2.55311101827205	-0.0266600976658238\\
2.61520000000002	-0.0261114907274469\\
2.67770000000001	-0.0252556872808431\\
2.73920000000001	-0.0246347344448474\\
2.80090000000004	-0.0238324351025032\\
2.86110000000001	-0.0230368510466489\\
2.9231	-0.022060454423567\\
2.98520000000004	-0.0214033663830574\\
3.04560000000002	-0.0205629463422337\\
3.10810000000001	-0.0197554909978448\\
3.16880000000002	-0.019030312902716\\
3.23130000000001	-0.0182448041954655\\
3.29280061851412	-0.0174163895615073\\
3.355	-0.0167931400388852\\
3.41680000000002	-0.0161070886482811\\
3.47920000000002	-0.0153342614032596\\
3.5411	-0.0147650420454366\\
3.60260000000002	-0.0140918048344698\\
3.66350000000002	-0.0133866142689818\\
3.72600000000001	-0.0128552793494366\\
3.78750000000002	-0.0123146705425501\\
3.84910000000002	-0.0116930197259649\\
3.91080000000002	-0.0110574713433375\\
3.97240000000001	-0.010649620327969\\
4.03490000000002	-0.0101106762733102\\
4.09640000000003	-0.00957778826764149\\
4.15880000000006	-0.00916203015378345\\
4.22110000000003	-0.00881212477958269\\
4.28350000000006	-0.00831697858924217\\
4.34600000000002	-0.00780072438578608\\
4.40850000000004	-0.00742664002486732\\
4.47100000000007	-0.00712903457566096\\
4.53350000000009	-0.00683142912645461\\
4.59400000000002	-0.00645907547334063\\
4.65620000000003	-0.00603899476941228\\
4.71870000000005	-0.00570397536436419\\
4.78120000000007	-0.00545510922194649\\
4.84370000000009	-0.00520624307952879\\
4.90620000000005	-0.00494829188547188\\
4.96870000000002	-0.00468275736280302\\
5.03060000000004	-0.0043593326101363\\
5.09300000000002	-0.0042006558251173\\
5.15550000000004	-0.00403534400171069\\
5.21780000000007	-0.00377513490320293\\
5.28030000000004	-0.00351496076026533\\
5.34280000000006	-0.00326233905205836\\
5.40530000000008	-0.003013105475278\\
5.46770000000005	-0.00282710400774249\\
5.53020000000007	-0.002767579074495\\
5.59270000000003	-0.00270954910108153\\
5.65520000000005	-0.00257876868330947\\
5.71770000000007	-0.00244798826553741\\
5.78020000000004	-0.0023046574865632\\
5.84270000000006	-0.00214635434685635\\
5.90520000000002	-0.00199797236053652\\
5.96770000000004	-0.00185368183683109\\
6.03020000000001	-0.00169923406207665\\
6.09260000000004	-0.00146834763062538\\
6.15510000000006	-0.00140346321026752\\
6.21760000000009	-0.00136048816493193\\
6.28010000000011	-0.00131751311959634\\
6.34260000000013	-0.00127453807426075\\
6.40510000000015	-0.00123156302892515\\
6.46760000000006	-0.00118663114832809\\
6.53010000000008	-0.00113142588260785\\
6.5926000000001	-0.00107622061688761\\
6.65510000000012	-0.00102101535116737\\
6.71760000000014	-0.000965810085447131\\
6.78010000000005	-0.000909675334147705\\
6.84250000000001	-0.000812563424609749\\
6.90500000000003	-0.00067115408246695\\
6.96750000000006	-0.000529744740324148\\
7.03000000000008	-0.000388335398181346\\
7.0925000000001	-0.00024692605603855\\
7.15500000000012	-0.000105516713895752\\
7.21670000000005	-7.5171898520841e-05\\
7.27900000000005	-0.000119303330854623\\
7.34150000000001	-0.000155480177600239\\
7.40400000000003	-0.000192661049760839\\
7.46650000000005	-0.000229841921921439\\
7.52900000000007	-0.00026702279408204\\
7.5915000000001	-0.000304203666242643\\
7.65400000000012	-0.000341384538403244\\
7.71650000000014	-0.000378565410563843\\
7.77900000000016	-0.000415746282724441\\
7.84150000000018	-0.00045292715488504\\
7.90400000000003	-0.000497695878153207\\
7.96650000000005	-0.000550762426034537\\
};
\addlegendentry{$P_4$}

\addplot [color=red]
  table[row sep=crcr]{%
0	-2\\
0.0601519961646206	-2.24058923937088\\
0.117621217771277	-2.43310048098074\\
0.174000221065236	-2.37950477138595\\
0.229151401495203	-2.32628181603125\\
0.28786680199119	-2.17984173526144\\
0.343853052847511	-1.99971260681693\\
0.404061505582155	-1.84818861752615\\
0.457313348769833	-1.67243652291381\\
0.517299303171965	-1.51696205990709\\
0.575515143700868	-1.36292754055394\\
0.632678126050925	-1.22050204706049\\
0.689298545478555	-1.10069283439681\\
0.747512149273467	-0.97270680996994\\
0.808557611947797	-0.870701329793151\\
0.864929098110121	-0.768548972568292\\
0.924967090987112	-0.688297487341112\\
0.982603929724419	-0.605751420367253\\
1.04138129251223	-0.542996644877198\\
1.10131646566076	-0.476272099756559\\
1.15963029968841	-0.42557031350848\\
1.22006508955089	-0.378677233400141\\
1.27901774522118	-0.33697313652988\\
1.3389	-0.299386994839163\\
1.40032960290062	-0.265610653706248\\
1.462	-0.235537941403786\\
1.52214078004887	-0.20972201183648\\
1.58332532630938	-0.186554665768005\\
1.64287270637993	-0.166776862390547\\
1.70300000000001	-0.148991458327526\\
1.7644	-0.133021460634512\\
1.82480000000001	-0.119072710000725\\
1.8864	-0.106407262854387\\
1.94606989993623	-0.0968927413304042\\
2.0066080111338	-0.0863927989566377\\
2.06656338967949	-0.0791454114520308\\
2.12790000000004	-0.0705272362022657\\
2.18830000000005	-0.064591123987952\\
2.24900000000003	-0.0579571859959482\\
2.31035525726748	-0.0530547389093918\\
2.37180000000001	-0.0478220829137436\\
2.43120000000003	-0.0439147857844955\\
2.49250000000004	-0.0400163581886286\\
2.55311101827205	-0.0369650499160291\\
2.61520000000002	-0.0336662026248977\\
2.67770000000001	-0.0312362581409333\\
2.73920000000001	-0.0284162540138482\\
2.80090000000004	-0.0264274792919413\\
2.86110000000001	-0.0242975131300588\\
2.9231	-0.0227227703126955\\
2.98520000000004	-0.0209501803191776\\
3.04560000000002	-0.0195562076015142\\
3.10810000000001	-0.0181193943195828\\
3.16880000000002	-0.0169244085182781\\
3.23130000000001	-0.0156834644691148\\
3.29280061851412	-0.0147557158401905\\
3.355	-0.0136462335264446\\
3.41680000000002	-0.0126885641519015\\
3.47920000000002	-0.0120679900842442\\
3.5411	-0.0111674315549627\\
3.60260000000002	-0.0105154342114439\\
3.66350000000002	-0.00997014597294372\\
3.72600000000001	-0.00917287753476822\\
3.78750000000002	-0.00865054606922239\\
3.84910000000002	-0.00828248920199979\\
3.91080000000002	-0.00777678451647961\\
3.97240000000001	-0.00714669914753158\\
4.03490000000002	-0.00684625990682368\\
4.09640000000003	-0.0064819473869878\\
4.15880000000006	-0.0060473374520716\\
4.22110000000003	-0.00561419011876974\\
4.28350000000006	-0.00540095217651654\\
4.34600000000002	-0.00517536675764713\\
4.40850000000004	-0.00484172137004766\\
4.47100000000007	-0.00447691628718306\\
4.53350000000009	-0.00411211120431845\\
4.59400000000002	-0.00405431848710399\\
4.65620000000003	-0.00393600735918881\\
4.71870000000005	-0.00373818928386117\\
4.78120000000007	-0.00350012606702138\\
4.84370000000009	-0.0032620628501816\\
4.90620000000005	-0.00300861287117171\\
4.96870000000002	-0.00268564571125791\\
5.03060000000004	-0.00260970146072527\\
5.09300000000002	-0.00250383300025174\\
5.15550000000004	-0.00245717952892277\\
5.21780000000007	-0.00239445657068762\\
5.28030000000004	-0.00232019919834044\\
5.34280000000006	-0.0021909153748422\\
5.40530000000008	-0.00202964791882906\\
5.46770000000005	-0.00183607165512849\\
5.53020000000007	-0.00158810983017086\\
5.59270000000003	-0.0013371521355804\\
5.65520000000005	-0.00133391739687372\\
5.71770000000007	-0.00133068265816703\\
5.78020000000004	-0.00133315940650776\\
5.84270000000006	-0.00134244985869452\\
5.90520000000002	-0.0013587432913815\\
5.96770000000004	-0.00131765058190883\\
6.03020000000001	-0.00127959821336944\\
6.09260000000004	-0.00126481030224722\\
6.15510000000006	-0.0011641285419563\\
6.21760000000009	-0.00105223970414317\\
6.28010000000011	-0.000940350866330038\\
6.34260000000013	-0.000828462028516906\\
6.40510000000015	-0.000716573190703774\\
6.46760000000006	-0.000606477146554131\\
6.53010000000008	-0.000505793269136523\\
6.5926000000001	-0.000405109391718913\\
6.65510000000012	-0.000304425514301304\\
6.71760000000014	-0.000203741636883695\\
6.78010000000005	-8.74085340621437e-05\\
6.84250000000001	-2.14032000938046e-05\\
6.90500000000003	-9.24416811369189e-05\\
6.96750000000006	-0.000163480162180034\\
7.03000000000008	-0.00023451864322315\\
7.0925000000001	-0.000305557124266262\\
7.15500000000012	-0.000376595605309375\\
7.21670000000005	-0.000387101238884614\\
7.27900000000005	-0.000356675472389908\\
7.34150000000001	-0.000334966916347432\\
7.40400000000003	-0.000409179383238792\\
7.46650000000005	-0.000483391850130152\\
7.52900000000007	-0.000557604317021508\\
7.5915000000001	-0.000631816783912865\\
7.65400000000012	-0.000706029250804223\\
7.71650000000014	-0.000780241717695579\\
7.77900000000016	-0.000854454184586935\\
7.84150000000018	-0.000928666651478292\\
7.90400000000003	-0.000949885851378813\\
7.96650000000005	-0.000914609480993508\\
};
\addlegendentry{$P_5$}

\addplot [color=black, dashed]
  table[row sep=crcr]{%
0	-4\\
0.0601519961646206	-3.39848149425897\\
0.117621217771277	-2.83624093145324\\
0.174000221065236	-2.46530436201222\\
0.229151401495203	-2.12456226489575\\
0.28786680199119	-1.84205373749052\\
0.343853052847511	-1.62436988036794\\
0.404061505582155	-1.39287831789354\\
0.457313348769833	-1.25282883142156\\
0.517299303171965	-1.09150330840703\\
0.575515143700868	-0.96231328049247\\
0.632678126050925	-0.858130925859984\\
0.689298545478555	-0.757602735322124\\
0.747512149273467	-0.681287035344472\\
0.808557611947797	-0.595369268022089\\
0.864929098110121	-0.54061138686842\\
0.924967090987112	-0.476660709706545\\
0.982603929724419	-0.431147603694026\\
1.04138129251223	-0.383929860525478\\
1.10131646566076	-0.34521746212066\\
1.15963029968841	-0.309968641300375\\
1.22006508955089	-0.275306506273289\\
1.27901774522118	-0.249412700676058\\
1.3389	-0.224644084112027\\
1.40032960290062	-0.202671434086874\\
1.462	-0.183371196861795\\
1.52214078004887	-0.166130033300346\\
1.58332532630938	-0.151359397304773\\
1.64287270637993	-0.137384965750062\\
1.70300000000001	-0.126164350757694\\
1.7644	-0.114644990329877\\
1.82480000000001	-0.105552610629845\\
1.8864	-0.096633785070786\\
1.94606989993623	-0.088295484112344\\
2.0066080111338	-0.0820363993284473\\
2.06656338967949	-0.0748869646036894\\
2.12790000000004	-0.0698938979021817\\
2.18830000000005	-0.0644497876586941\\
2.24900000000003	-0.0601167036686218\\
2.31035525726748	-0.0557480835621287\\
2.37180000000001	-0.0518848491298948\\
2.43120000000003	-0.0485066014670831\\
2.49250000000004	-0.045117664840643\\
2.55311101827205	-0.0420928271646581\\
2.61520000000002	-0.0394688622339692\\
2.67770000000001	-0.0366563753056129\\
2.73920000000001	-0.0346654761215135\\
2.80090000000004	-0.0323595175559215\\
2.86110000000001	-0.0305519064339177\\
2.9231	-0.0286647691530714\\
2.98520000000004	-0.026980234372351\\
3.04560000000002	-0.0255317021288356\\
3.10810000000001	-0.0240066898366721\\
3.16880000000002	-0.0226313481021912\\
3.23130000000001	-0.0214246917433206\\
3.29280061851412	-0.0201398179282431\\
3.355	-0.0191519679864056\\
3.41680000000002	-0.0181723373097821\\
3.47920000000002	-0.0171218668814785\\
3.5411	-0.0163255852843581\\
3.60260000000002	-0.0154023518467648\\
3.66350000000002	-0.0144938421617819\\
3.72600000000001	-0.0138512316631724\\
3.78750000000002	-0.0131248344247642\\
3.84910000000002	-0.0123378637249091\\
3.91080000000002	-0.0117590998124756\\
3.97240000000001	-0.011187519808935\\
4.03490000000002	-0.0105020808470421\\
4.09640000000003	-0.00997797354458312\\
4.15880000000006	-0.00954824272098787\\
4.22110000000003	-0.00907085306717409\\
4.28350000000006	-0.00846565560232426\\
4.34600000000002	-0.00800569519625098\\
4.40850000000004	-0.00762091995577273\\
4.47100000000007	-0.00723360737553074\\
4.53350000000009	-0.00684629479528875\\
4.59400000000002	-0.00646464145821193\\
4.65620000000003	-0.00618274412633986\\
4.71870000000005	-0.00584271824429108\\
4.78120000000007	-0.00550797343168111\\
4.84370000000009	-0.00517322861907114\\
4.90620000000005	-0.00488459634120437\\
4.96870000000002	-0.00474526903233914\\
5.03060000000004	-0.00452916905744929\\
5.09300000000002	-0.00429272916182893\\
5.15550000000004	-0.00405616933981393\\
5.21780000000007	-0.00372993173906238\\
5.28030000000004	-0.00341621905547273\\
5.34280000000006	-0.0032322233300579\\
5.40530000000008	-0.00312629829574548\\
5.46770000000005	-0.00301627053903554\\
5.53020000000007	-0.00289320151094369\\
5.59270000000003	-0.00277230844895104\\
5.65520000000005	-0.00256910833917732\\
5.71770000000007	-0.0023659082294036\\
5.78020000000004	-0.00216242761140936\\
5.84270000000006	-0.00195861235202882\\
5.90520000000002	-0.00184634680509331\\
5.96770000000004	-0.00184585965563409\\
6.03020000000001	-0.00183166712468762\\
6.09260000000004	-0.00171717616594764\\
6.15510000000006	-0.00160750557531277\\
6.21760000000009	-0.00149845644585413\\
6.28010000000011	-0.0013894073163955\\
6.34260000000013	-0.00128035818693687\\
6.40510000000015	-0.00117130905747823\\
6.46760000000006	-0.00106240142158518\\
6.53010000000008	-0.000954236626910181\\
6.5926000000001	-0.000846071832235184\\
6.65510000000012	-0.000737907037560187\\
6.71760000000014	-0.000629742242885191\\
6.78010000000005	-0.000551947796378933\\
6.84250000000001	-0.000577420098156448\\
6.90500000000003	-0.000544393241431911\\
6.96750000000006	-0.000511366384707372\\
7.03000000000008	-0.000478339527982834\\
7.0925000000001	-0.000445312671258296\\
7.15500000000012	-0.000412285814533759\\
7.21670000000005	-0.000386650187393441\\
7.27900000000005	-0.000366524863688652\\
7.34150000000001	-0.000346917416845656\\
7.40400000000003	-0.000308982620001324\\
7.46650000000005	-0.000271047823156992\\
7.52900000000007	-0.00023311302631266\\
7.5915000000001	-0.000195178229468327\\
7.65400000000012	-0.000157243432623995\\
7.71650000000014	-0.000119308635779662\\
7.77900000000016	-8.13738389353296e-05\\
7.84150000000018	-4.34390420909962e-05\\
7.90400000000003	-1.24535022470271e-05\\
7.96650000000005	1.08130695119755e-05\\
};
\addlegendentry{$P_6$}

\addplot [color=black, dashdotted]
  table[row sep=crcr]{%
0	-6\\
0.0601519961646206	-5.39848759045865\\
0.117621217771277	-4.80216292869128\\
0.174000221065236	-4.23987900013168\\
0.229151401495203	-3.77738861069033\\
0.28786680199119	-3.27437985042223\\
0.343853052847511	-2.89319771327457\\
0.404061505582155	-2.51010520639213\\
0.457313348769833	-2.16922117508997\\
0.517299303171965	-1.90575571077195\\
0.575515143700868	-1.62718373187676\\
0.632678126050925	-1.43293693057992\\
0.689298545478555	-1.23928583580181\\
0.747512149273467	-1.07088118539468\\
0.808557611947797	-0.921193550574077\\
0.864929098110121	-0.790005041839349\\
0.924967090987112	-0.684401191382116\\
0.982603929724419	-0.581809145897995\\
1.04138129251223	-0.502446047364582\\
1.10131646566076	-0.42553115662171\\
1.15963029968841	-0.358216164416875\\
1.22006508955089	-0.305626296747967\\
1.27901774522118	-0.251656288718626\\
1.3389	-0.21541685207155\\
1.40032960290062	-0.17643977833006\\
1.462	-0.147114400919524\\
1.52214078004887	-0.12025618769794\\
1.58332532630938	-0.0981440095185013\\
1.64287270637993	-0.0792589651809259\\
1.70300000000001	-0.0626557273161188\\
1.7644	-0.0489468543950203\\
1.82480000000001	-0.0360370166350836\\
1.8864	-0.0264398701892806\\
1.94606989993623	-0.0178666495127095\\
2.0066080111338	-0.0107211772259872\\
2.06656338967949	-0.00504152449144073\\
2.12790000000004	0.000442725237381502\\
2.18830000000005	0.00409155451789244\\
2.24900000000003	0.0080092321545609\\
2.31035525726748	0.0103105077417728\\
2.37180000000001	0.0127414339363464\\
2.43120000000003	0.0141415491331319\\
2.49250000000004	0.0156291308821055\\
2.55311101827205	0.0163995601390044\\
2.61520000000002	0.0173498636884853\\
2.67770000000001	0.0177176226686888\\
2.73920000000001	0.0181333261384683\\
2.80090000000004	0.0182593958351536\\
2.86110000000001	0.0183580012525215\\
2.9231	0.018238544135252\\
2.98520000000004	0.0180708238497031\\
3.04560000000002	0.0178067728611432\\
3.10810000000001	0.0174736268876437\\
3.16880000000002	0.017052893875064\\
3.23130000000001	0.0165501772342283\\
3.29280061851412	0.0159877356944651\\
3.355	0.0153889923345377\\
3.41680000000002	0.01500638979154\\
3.47920000000002	0.0143818282613365\\
3.5411	0.0137668073319704\\
3.60260000000002	0.0132422020812667\\
3.66350000000002	0.0127418929357747\\
3.72600000000001	0.0122300177293949\\
3.78750000000002	0.0116863979010263\\
3.84910000000002	0.0111208096630847\\
3.91080000000002	0.0106703862119713\\
3.97240000000001	0.0102131078787161\\
4.03490000000002	0.00971636600394607\\
4.09640000000003	0.00932066068528369\\
4.15880000000006	0.00892881502987234\\
4.22110000000003	0.00850407777330374\\
4.28350000000006	0.00801417075364415\\
4.34600000000002	0.00759829204905747\\
4.40850000000004	0.00731519684962094\\
4.47100000000007	0.00702429955846527\\
4.53350000000009	0.0067334022673096\\
4.59400000000002	0.00628311342884434\\
4.65620000000003	0.00591625834798291\\
4.71870000000005	0.00564494234673016\\
4.78120000000007	0.00537056909857286\\
4.84370000000009	0.00509619585041555\\
4.90620000000005	0.0048309074048835\\
4.96870000000002	0.00454963486841218\\
5.03060000000004	0.00434644797602993\\
5.09300000000002	0.00415708160428818\\
5.15550000000004	0.00392330649569272\\
5.21780000000007	0.00367179649654209\\
5.28030000000004	0.00342459085162867\\
5.34280000000006	0.00318594184833338\\
5.40530000000008	0.0029509100036196\\
5.46770000000005	0.00286010442508263\\
5.53020000000007	0.00280345840276085\\
5.59270000000003	0.00272152083728983\\
5.65520000000005	0.00259735366067318\\
5.71770000000007	0.00247318648405653\\
5.78020000000004	0.00234726463086456\\
5.84270000000006	0.00221924947930179\\
5.90520000000002	0.00202178384012593\\
5.96770000000004	0.00181649144078764\\
6.03020000000001	0.00160912802882407\\
6.09260000000004	0.00143361092466178\\
6.15510000000006	0.00140051242623796\\
6.21760000000009	0.00136677851967795\\
6.28010000000011	0.00133304461311793\\
6.34260000000013	0.00129931070655792\\
6.40510000000015	0.00126557679999791\\
6.46760000000006	0.00123022236931601\\
6.53010000000008	0.00118636018699382\\
6.5926000000001	0.00114249800467163\\
6.65510000000012	0.00109863582234944\\
6.71760000000014	0.00105477364002725\\
6.78010000000005	0.0010129418856562\\
6.84250000000001	0.000949262711535769\\
6.90500000000003	0.000817370559466431\\
6.96750000000006	0.000685478407397092\\
7.03000000000008	0.000553586255327752\\
7.0925000000001	0.000421694103258416\\
7.15500000000012	0.000289801951189081\\
7.21670000000005	0.000168006125363256\\
7.27900000000005	0.000108763329620531\\
7.34150000000001	0.000162215133904602\\
7.40400000000003	0.000114581176001413\\
7.46650000000005	6.69472180982261e-05\\
7.52900000000007	1.93132601950386e-05\\
7.5915000000001	-2.83206977081495e-05\\
7.65400000000012	-7.59546556113378e-05\\
7.71650000000014	-0.000123588613514526\\
7.77900000000016	-0.000171222571417714\\
7.84150000000018	-0.000218856529320901\\
7.90400000000003	-0.000217004306382831\\
7.96650000000005	-0.000163168772678231\\
};
\addlegendentry{$P_7$}

\addplot [color=red, dotted]
  table[row sep=crcr]{%
0	-8\\
0.0601519961646206	-6.55636738917769\\
0.117621217771277	-5.25596880141119\\
0.174000221065236	-4.42537158318081\\
0.229151401495203	-3.65531915373569\\
0.28786680199119	-3.08140568040675\\
0.343853052847511	-2.60275384870999\\
0.404061505582155	-2.18447938608004\\
0.457313348769833	-1.92573176663078\\
0.517299303171965	-1.61461174184821\\
0.575515143700868	-1.39446407227964\\
0.632678126050925	-1.19131279261844\\
0.689298545478555	-1.02140034994815\\
0.747512149273467	-0.884906098218309\\
0.808557611947797	-0.745763250815417\\
0.864929098110121	-0.653142321245968\\
0.924967090987112	-0.551523959864457\\
0.982603929724419	-0.477657402685722\\
1.04138129251223	-0.407644319250086\\
1.10131646566076	-0.347932253307944\\
1.15963029968841	-0.304793404504575\\
1.22006508955089	-0.255727019701402\\
1.27901774522118	-0.222449636911495\\
1.3389	-0.186604280563806\\
1.40032960290062	-0.157927446875347\\
1.462	-0.131946038836429\\
1.52214078004887	-0.110256279698847\\
1.58332532630938	-0.0919646252690612\\
1.64287270637993	-0.0764539229374586\\
1.70300000000001	-0.0628701775747634\\
1.7644	-0.0514932579575424\\
1.82480000000001	-0.0416259676597058\\
1.8864	-0.0336039767772225\\
1.94606989993623	-0.0253909563347857\\
2.0066080111338	-0.0201920380736206\\
2.06656338967949	-0.0142727723669821\\
2.12790000000004	-0.0104501907025266\\
2.18830000000005	-0.00624963984450908\\
2.24900000000003	-0.00346059167587308\\
2.31035525726748	-0.000539187882910225\\
2.37180000000001	0.0016285948173703\\
2.43120000000003	0.00343528904548458\\
2.49250000000004	0.00486029489421729\\
2.55311101827205	0.00644203561826239\\
2.61520000000002	0.00707798643966656\\
2.67770000000001	0.0081344162755257\\
2.73920000000001	0.00842004264844617\\
2.80090000000004	0.00885536445059499\\
2.86110000000001	0.00912977206367746\\
2.9231	0.00939578951340587\\
2.98520000000004	0.0096542280710234\\
3.04560000000002	0.00961131463185497\\
3.10810000000001	0.00948499277168565\\
3.16880000000002	0.00939586492200787\\
3.23130000000001	0.0091430847967126\\
3.29280061851412	0.00901005856602095\\
3.355	0.00869808657463598\\
3.41680000000002	0.00828769078635243\\
3.47920000000002	0.00823977559029248\\
3.5411	0.00797493377825724\\
3.60260000000002	0.00773589600877339\\
3.66350000000002	0.00745219563446672\\
3.72600000000001	0.00702404888114777\\
3.78750000000002	0.00678968523339462\\
3.84910000000002	0.00665546427845441\\
3.91080000000002	0.00634692936123098\\
3.97240000000001	0.00604011990033246\\
4.03490000000002	0.00582070967804025\\
4.09640000000003	0.00543845878312817\\
4.15880000000006	0.00515644164156626\\
4.22110000000003	0.00500324083597439\\
4.28350000000006	0.00489239285294184\\
4.34600000000002	0.00464760250042787\\
4.40850000000004	0.00434054566372803\\
4.47100000000007	0.00404434410138914\\
4.53350000000009	0.00374814253905025\\
4.59400000000002	0.00373858404814486\\
4.65620000000003	0.00362893372162359\\
4.71870000000005	0.00347507235439078\\
4.78120000000007	0.00332987245982227\\
4.84370000000009	0.00318467256525375\\
4.90620000000005	0.00295365280109562\\
4.96870000000002	0.00269055875873482\\
5.03060000000004	0.00245515088463812\\
5.09300000000002	0.00231911972699922\\
5.15550000000004	0.00233422021097042\\
5.21780000000007	0.00233636424479643\\
5.28030000000004	0.00226527282114914\\
5.34280000000006	0.00216489586791364\\
5.40530000000008	0.00206937099082983\\
5.46770000000005	0.00190057868467567\\
5.53020000000007	0.00172683339492156\\
5.59270000000003	0.00154886498254244\\
5.65520000000005	0.00144215122130789\\
5.71770000000007	0.00133543746007333\\
5.78020000000004	0.00118508165930766\\
5.84270000000006	0.000982661671030962\\
5.90520000000002	0.000993126104704505\\
5.96770000000004	0.00102883805945673\\
6.03020000000001	0.00106770547683826\\
6.09260000000004	0.00110206074363038\\
6.15510000000006	0.0010524356254575\\
6.21760000000009	0.00100321052425421\\
6.28010000000011	0.000953985423050928\\
6.34260000000013	0.000904760321847647\\
6.40510000000015	0.000855535220644366\\
6.46760000000006	0.000794148251408889\\
6.53010000000008	0.000668911475003825\\
6.5926000000001	0.000543674698598759\\
6.65510000000012	0.000418437922193694\\
6.71760000000014	0.000293201145788634\\
6.78010000000005	0.000166935984439622\\
6.84250000000001	8.48119024282866e-05\\
6.90500000000003	4.9772050617044e-05\\
6.96750000000006	1.47321988058003e-05\\
7.03000000000008	-2.03076530054437e-05\\
7.0925000000001	-5.53475048166875e-05\\
7.15500000000012	-9.03873566279316e-05\\
7.21670000000005	-0.000130251143593535\\
7.27900000000005	-0.000205692236929378\\
7.34150000000001	-0.000331792862378431\\
7.40400000000003	-0.000184366393022953\\
7.46650000000005	-3.69399236674759e-05\\
7.52900000000007	0.000110486545688001\\
7.5915000000001	0.00025791301504348\\
7.65400000000012	0.00040533948439896\\
7.71650000000014	0.000552765953754439\\
7.77900000000016	0.000700192423109918\\
7.84150000000018	0.000847618892465399\\
7.90400000000003	0.000818237851914455\\
7.96650000000005	0.00060018508159229\\
};
\addlegendentry{$P_8$}

\end{axis}
\end{tikzpicture}%

%% file: Figures/Garcia2013_InterEventTimes_c.tex
% !TEX root = ../ieeetac.tex
\begin{tikzpicture}[baseline=(current bounding box.center)]
    \footnotesize
\begin{axis}[%
width=\columnwidth,
height=4cm,
at={(0,0)},
%scale only axis,
xmin=0,
xmax=8,
xlabel style={font=\color{white!15!black}},
xlabel={Time [s]},
ymin=0,
ymax=3,
ylabel style={font=\color{white!15!black}},
ylabel={Inter-event times},
%axis background/.style={fill=white},
axis x line*=bottom,
axis y line*=left,
xmajorgrids,
ymajorgrids,
legend columns=4,
legend style={legend cell align=left, align=left, draw=white!15!black},
legend pos=north west
]
\addplot [only marks, mark=x, mark options={solid, black}]
  table[row sep=crcr]{%
0.273000000000003	0.138717754618419\\
0.41463942359177	0.141639423591768\\
0.546700000000007	0.132060576408237\\
0.681200000000004	0.134499999999997\\
0.816575680002774	0.13537568000277\\
0.963620628662987	0.147044948660213\\
1.10891369648462	0.145293067821637\\
1.24636929998591	0.137455603501283\\
1.38244339183048	0.136074091844576\\
1.52014078004887	0.13769738821839\\
1.6585	0.138359219951131\\
1.80840000000001	0.149900000000005\\
1.95063328978731	0.1422332897873\\
2.09230000000003	0.141666710212718\\
2.23400000000003	0.141700000000005\\
2.38610000000001	0.152099999999977\\
2.53661101827205	0.150511018272041\\
2.68690000000001	0.15028898172796\\
2.83700000000002	0.150100000000005\\
2.98830000000002	0.151300000000002\\
3.14630000000002	0.158\\
3.3041	0.157799999999984\\
3.47620000000002	0.172100000000021\\
3.65170000000003	0.175500000000003\\
3.83810000000002	0.186399999999989\\
4.04540000000003	0.20730000000001\\
4.28000000000006	0.234600000000032\\
4.56400000000001	0.283999999999955\\
4.85870000000004	0.294700000000025\\
5.23880000000002	0.380099999999985\\
5.75170000000003	0.512900000000003\\
6.45760000000005	0.705900000000026\\
7.87450000000002	1.41689999999997\\
};
\addlegendentry{$P_1$}

\addplot [only marks, mark=o, mark options={solid, red}]
  table[row sep=crcr]{%
0.228709476487134	0.113436536199808\\
0.341274326210167	0.112564849723034\\
0.451538388001139	0.110264061790972\\
0.580351447633337	0.128813059632197\\
0.709920759964214	0.129569312330877\\
0.839530223876305	0.129609463912092\\
0.952155304748258	0.112625080871953\\
1.06330000000002	0.111144695251762\\
1.17570533090086	0.112405330900838\\
1.28650000000001	0.110794669099155\\
1.41688438651476	0.130384386514744\\
1.54782532630939	0.13094093979463\\
1.67868127336756	0.130855947058175\\
1.7923	0.113618726632438\\
1.90451962941933	0.112219629419329\\
2.01900000000001	0.114480370580677\\
2.13670000000002	0.117700000000015\\
2.2579	0.121199999999981\\
2.37190000000004	0.114000000000041\\
2.48850000000004	0.116599999999998\\
2.61320000000002	0.124699999999976\\
2.74240000000005	0.129200000000029\\
2.87200000000003	0.129599999999986\\
3.00700000000001	0.13499999999998\\
3.1342	0.127199999999989\\
3.26430000000001	0.130100000000004\\
3.4173	0.152999999999996\\
3.59160000000002	0.174300000000015\\
3.78200000000002	0.190400000000003\\
4.00540000000001	0.223399999999993\\
4.23600000000004	0.230600000000031\\
4.53700000000003	0.300999999999987\\
4.95470000000001	0.417699999999982\\
5.57570000000002	0.621000000000011\\
6.82450000000001	1.24879999999998\\
};
\addlegendentry{$P_2$}

\addplot [only marks, mark=triangle, mark options={solid, red}]
  table[row sep=crcr]{%
0.220045430039777	0.112029708916387\\
0.323524939437172	0.103479509397395\\
0.427648650661638	0.104123711224467\\
0.544903325222141	0.117254674560502\\
0.665718670943806	0.120815345721665\\
0.78620561170006	0.120486940756254\\
0.906515622769758	0.120310011069698\\
1.02286298555237	0.116347362782614\\
1.13710000000001	0.114237014447638\\
1.2458	0.108699999999994\\
1.36566938386834	0.119869383868332\\
1.48830000000001	0.122630616131671\\
1.6097	0.121399999999997\\
1.7304	0.1207\\
1.84540000000001	0.115000000000004\\
1.95260000000001	0.1072\\
2.07500000000001	0.122400000000001\\
2.20050000000002	0.125500000000007\\
2.3217407858909	0.121240785890884\\
2.44630000000002	0.124559214109125\\
2.56240000000002	0.116099999999992\\
2.68190000000001	0.119499999999995\\
2.9086	0.226699999999992\\
3.50660000000001	0.598000000000003\\
3.82210000000005	0.31550000000004\\
4.19810000000003	0.37599999999998\\
4.63670000000002	0.438599999999994\\
5.15780000000005	0.521100000000033\\
6.0227	0.86489999999995\\
};
\addlegendentry{$P_3$}

\addplot [only marks, mark=square, mark options={solid, black}]
  table[row sep=crcr]{%
0.351308538343733	0.167378598566955\\
0.484500000000003	0.133191461656271\\
0.614721260203055	0.130221260203051\\
0.74238798603813	0.127666725835076\\
0.868000000000009	0.125612013961879\\
0.99180000000001	0.123800000000001\\
1.11190000000002	0.120100000000008\\
1.23534038024148	0.123440380241461\\
1.35290916338567	0.117568783144194\\
1.47020000000001	0.117290836614335\\
1.58510000000001	0.114899999999998\\
1.69750000000001	0.1124\\
1.8038	0.106299999999997\\
1.91050000000002	0.106700000000014\\
2.01069015522482	0.100190155224801\\
2.13020000000002	0.119509844775205\\
2.70490000000001	0.574699999999985\\
2.907	0.202099999999991\\
3.09010000000002	0.183100000000016\\
3.28230061851412	0.192200618514105\\
3.48480000000001	0.202499381485893\\
3.68200000000002	0.197200000000004\\
3.90030000000003	0.218300000000007\\
4.12430000000005	0.224000000000022\\
4.36700000000003	0.242699999999983\\
4.67970000000004	0.312700000000004\\
5.02310000000004	0.343400000000005\\
5.44870000000004	0.425599999999998\\
6.09960000000005	0.650900000000007\\
7.18070000000003	1.08109999999999\\
};
\addlegendentry{$P_4$}

\addplot [only marks, mark=x, mark options={solid, red}]
  table[row sep=crcr]{%
0.194600000000002	0.0865842788766121\\
0.351308538343733	0.156708538343731\\
0.479318293497217	0.128009755153485\\
0.606115506227587	0.12679721273037\\
0.728583332755935	0.122467826528348\\
0.849868438790376	0.121285106034441\\
0.970744335693927	0.120875896903551\\
1.09121299953766	0.120468663843729\\
1.19970757598796	0.108494576450306\\
1.3134	0.113692424012042\\
1.4295	0.116099999999997\\
1.54736552046584	0.117865520465843\\
1.6666	0.119234479534162\\
1.78680000000001	0.120200000000002\\
1.89650000000001	0.109700000000001\\
2.01069015522482	0.114190155224812\\
2.13020000000002	0.119509844775205\\
2.2509	0.120699999999979\\
2.375	0.1241\\
2.48690000000003	0.111900000000021\\
2.60440000000001	0.117499999999987\\
2.73320000000001	0.128799999999998\\
2.85810000000001	0.124900000000002\\
2.99300000000004	0.134900000000031\\
3.12110000000001	0.12809999999997\\
3.25030000000001	0.129199999999995\\
3.39170000000002	0.141400000000012\\
3.5606	0.168899999999982\\
3.74320000000001	0.182600000000008\\
3.97240000000001	0.229200000000005\\
4.23600000000004	0.26360000000003\\
4.53740000000003	0.301399999999981\\
4.97110000000002	0.433699999999997\\
5.59170000000003	0.620600000000007\\
6.8126	1.22089999999997\\
};
\addlegendentry{$P_5$}

\addplot [only marks, mark=o, mark options={solid, black}]
  table[row sep=crcr]{%
0.256724988494914	0.131374932000171\\
0.399500000000003	0.142775011505089\\
0.544903325222141	0.145403325222137\\
0.677488521275318	0.132585196053177\\
0.809946414298527	0.132457893023209\\
0.942168241908029	0.132221827609502\\
1.07511858138377	0.132950339475743\\
1.2286	0.153481418616231\\
1.36566938386834	0.137069383868333\\
1.50099815126631	0.13532876739797\\
1.63587270637993	0.13487455511362\\
1.77020000000002	0.134327293620091\\
1.9235	0.153299999999986\\
2.07500000000001	0.151500000000006\\
2.21560000000001	0.140600000000003\\
2.35780000000001	0.142199999999998\\
2.51560000000002	0.15780000000001\\
2.66820000000001	0.152599999999992\\
2.81970000000002	0.151500000000011\\
2.96820000000004	0.148500000000019\\
3.13831251867721	0.170112518677174\\
3.30310000000003	0.164787481322817\\
3.47220000000003	0.169099999999993\\
3.65600000000002	0.183799999999996\\
3.85310000000002	0.197099999999994\\
4.06140000000002	0.208300000000005\\
4.30500000000007	0.243600000000047\\
4.57500000000002	0.26999999999995\\
4.89370000000005	0.318700000000033\\
5.30480000000005	0.411099999999996\\
5.87770000000001	0.572899999999966\\
6.76960000000004	0.891900000000031\\
};
\addlegendentry{$P_6$}

\addplot [only marks, mark=triangle, mark options={solid, black}]
  table[row sep=crcr]{%
0.296084247151208	0.143960997155278\\
0.447554731947804	0.151470484796595\\
0.583656386991931	0.136101655044127\\
0.719029351178433	0.135372964186502\\
0.853124242679105	0.134094891500671\\
0.989900000000006	0.136775757320902\\
1.13474764603807	0.144847646038068\\
1.28122350466672	0.146475858628647\\
1.41688438651476	0.135660881848037\\
1.54782532630939	0.13094093979463\\
1.67868127336756	0.130855947058175\\
1.82030000000001	0.141618726632446\\
1.96160801113381	0.141308011133797\\
2.10490000000005	0.14329198886624\\
2.25080000000002	0.145899999999977\\
2.37190000000004	0.121100000000022\\
2.48850000000004	0.116599999999998\\
2.63120000000002	0.142699999999974\\
3.34400000000001	0.712799999999989\\
3.57570000000003	0.231700000000025\\
3.81680509481674	0.241105094816707\\
4.05220000000005	0.235394905183311\\
4.32100000000002	0.268799999999967\\
4.62470000000007	0.303700000000058\\
4.98910000000003	0.364399999999955\\
5.42070000000003	0.4316\\
6.07760000000004	0.656900000000009\\
7.25850000000004	1.1809\\
};
\addlegendentry{$P_7$}

\addplot [only marks, mark=square, mark options={solid, red}]
  table[row sep=crcr]{%
0.216936956460442	0.108921235337052\\
0.325413151444241	0.1084761949838\\
0.417854211099601	0.0924410596553596\\
0.53571265007452	0.117858438974919\\
0.655493540475638	0.119780890401117\\
0.775987555448026	0.120494014972389\\
0.896370933337506	0.12038337788948\\
1.01372619338223	0.11735526004472\\
1.10891369648462	0.0951875031023985\\
1.23220000000002	0.123286303515399\\
1.3567142860831	0.124514286083073\\
1.47895730094168	0.122243014858583\\
1.59854275028582	0.119585449344137\\
1.7156	0.117057249714188\\
1.80840000000001	0.0928000000000044\\
1.93256989993624	0.124169899936227\\
2.05406338967949	0.121493489743251\\
2.17480000000005	0.12073661032056\\
2.29485525726748	0.120055257267434\\
2.39770000000003	0.102844742732552\\
2.53661101827205	0.138911018272018\\
2.68690000000001	0.15028898172796\\
3.00710000000003	0.320200000000014\\
3.41630000000003	0.409200000000005\\
3.76500000000002	0.348699999999992\\
4.12630000000005	0.361300000000026\\
4.53730000000007	0.411000000000024\\
5.06500000000001	0.527699999999934\\
5.8537	0.788699999999998\\
7.33850000000001	1.48480000000001\\
7.87350000000008	0.535000000000065\\
};
\addlegendentry{$P_8$}

\end{axis}
\end{tikzpicture}%

%% file: Figures/Dolk2019_InterEventTimes.tex
% !TEX root = ../ieeetac.tex
\begin{tikzpicture}[baseline=(current bounding box.center)]
    \footnotesize
\begin{axis}[%
width=\columnwidth,
height=4cm,
at={(0,0)},
%scale only axis,
xmin=0,
xmax=8,
xlabel style={font=\color{white!15!black}},
xlabel={Time [s]},
ymin=0,
ymax=3,
ylabel style={font=\color{white!15!black}},
ylabel={Inter-event times},
%axis background/.style={fill=white},
axis x line*=bottom,
axis y line*=left,
xmajorgrids,
ymajorgrids,
legend columns=5,
legend style={legend cell align=left, align=left, draw=white!15!black}
]
\addplot [only marks, mark=x, mark options={solid, black}]
  table[row sep=crcr]{%
1.52125701853967	1.14379143297839\\
2.1332870472232	0.612030028683536\\
3.17028067806785	1.03699363084464\\
3.63415461990593	0.463873941838087\\
4.41668372196025	0.78252910205432\\
5.0997562192388	0.683072497278546\\
5.96517380567667	0.865417586437871\\
6.52392895453157	0.558755148854899\\
6.71690362944708	0.192974674915511\\
6.8859689087962	0.169065279349121\\
7.0839689760719	0.198000067275698\\
7.26523054646929	0.181261570397386\\
7.43200872126241	0.166778174793127\\
7.60304299128533	0.171034270022922\\
7.79743658790258	0.194393596617245\\
7.97281251162751	0.175375923724935\\
};
\addlegendentry{$P_1$}

\addplot [only marks, mark=o, mark options={solid, red}]
  table[row sep=crcr]{%
0.287762416448596	0.145462723867453\\
0.430270429614359	0.142508013165763\\
0.703891011050342	0.273620581435982\\
1.19736590478616	0.49347489373582\\
1.35005303013643	0.152687125350271\\
1.47616789495605	0.126114864819616\\
1.70866592294195	0.232498027985898\\
2.25704149324324	0.548375570301288\\
2.65285211295746	0.395810619714219\\
3.00835786766566	0.355505754708209\\
3.14743141246059	0.139073544794928\\
3.45441564518148	0.306984232720883\\
4.92970169019984	1.47528604501837\\
5.21810972708626	0.288408036886416\\
5.74905177990619	0.530942052819926\\
5.94622972506885	0.197177945162666\\
6.0996737549321	0.153444029863248\\
6.21783598047832	0.11816222554622\\
6.36971825834621	0.151882277867886\\
6.49871650648711	0.1289982481409\\
6.61928812489568	0.120571618408571\\
6.76535349130077	0.14606536640509\\
6.89964368039393	0.13429018909316\\
7.03503233708493	0.135388656691003\\
7.19656481613415	0.161532479049217\\
7.32034828271067	0.123783466576526\\
7.44020898450196	0.11986070179129\\
7.59663372445392	0.156424739951952\\
7.71614027838128	0.119506553927359\\
7.83638610636271	0.120245827981433\\
7.96347793600878	0.127091829646069\\
};
\addlegendentry{$P_2$}

\addplot [only marks, mark=triangle, mark options={solid, red}]
  table[row sep=crcr]{%
0.491575573276444	0.349237551290011\\
1.09092021890952	0.599344645633072\\
1.68318612268064	0.592265903771121\\
2.04789146067591	0.364705337995277\\
2.82830592540774	0.780414464731825\\
2.95061523549961	0.12230931009187\\
3.22694380706101	0.276328571561396\\
3.46324742303154	0.236303615970535\\
3.66062015226263	0.197372729231094\\
4.08817893180506	0.427558779542426\\
4.2150774158136	0.126898484008539\\
4.42426501836056	0.209187602546958\\
4.59011816852429	0.16585315016373\\
4.77987686022119	0.189758691696901\\
4.90986077697535	0.129983916754159\\
5.12545119540723	0.21559041843188\\
5.2648027534353	0.139351558028076\\
5.3909112905605	0.126108537125194\\
5.53774455876619	0.146833268205693\\
5.65882682894673	0.121082270180541\\
5.78980630080386	0.130979471857135\\
5.91663286567158	0.12682656486772\\
6.04270715279601	0.126074287124427\\
6.16909725715319	0.126390104357182\\
6.28856549599472	0.119468238841525\\
6.43429784045544	0.145732344460721\\
6.55999931410769	0.125701473652255\\
6.67948148206384	0.119482167956148\\
6.82706535200856	0.147583869944718\\
6.94719018143391	0.120124829425351\\
7.06877197185971	0.121581790425797\\
7.217225433574	0.148453461714294\\
7.34468697947838	0.127461545904375\\
7.47186962621992	0.127182646741547\\
7.59930058690785	0.127430960687923\\
7.72240246966414	0.123101882756294\\
7.88888269565439	0.166480225990249\\
};
\addlegendentry{$P_3$}

\addplot [only marks, mark=square, mark options={solid, black}]
  table[row sep=crcr]{%
0.577724782146903	0.413658357056502\\
1.77282163557216	1.19509685342526\\
2.14727661007275	0.374454974500587\\
2.59523335158275	0.447956741510005\\
3.09894282886569	0.503709477282943\\
3.48465831655011	0.385715487684411\\
3.76007427862478	0.275415962074677\\
4.16624320463584	0.406168926011056\\
4.58732062759287	0.421077422957034\\
5.15890695889747	0.571586331304594\\
5.41498413026951	0.25607717137204\\
5.58254245269878	0.167558322429271\\
5.78728492010112	0.204742467402347\\
5.95039947555829	0.163114555457161\\
6.13214547233833	0.181745996780044\\
6.29078984391793	0.158644371579604\\
6.46069908810392	0.169909244185988\\
6.6410373131435	0.180338225039578\\
6.80855728794805	0.167519974804547\\
6.9869182624363	0.178360974488253\\
7.15942600795014	0.17250774551384\\
7.31965991603333	0.160233908083188\\
7.47749962122535	0.157839705192025\\
7.65009566678951	0.172596045564159\\
7.86208608321747	0.21199041642796\\
};
\addlegendentry{$P_4$}

\addplot [color=black]
  table[row sep=crcr]{%
0	0.156171044770359\\
8	0.156171044770359\\
};
\addlegendentry{$\tau_\text{MIET}^1$}

\addplot [only marks, mark=x, mark options={solid, red}]
  table[row sep=crcr]{%
0.31924478810566	0.158958865406885\\
0.65529847117451	0.33605368306885\\
1.30653985757907	0.65124138640456\\
1.64668597457069	0.340146116991625\\
2.2032651565443	0.556579181973604\\
2.5709828914307	0.367717734886404\\
2.93888802455831	0.367905133127608\\
3.31870987284143	0.379821848283114\\
3.83123197651929	0.512522103677867\\
4.23940372988106	0.408171753361771\\
4.72946564957697	0.490061919695906\\
5.23604639818551	0.506580748608539\\
5.72190271668198	0.48585631849647\\
6.06244547538389	0.340542758701908\\
6.22462798058183	0.162182505197947\\
6.45148168062917	0.226853700047334\\
6.64058095921026	0.189099278581097\\
6.84784018129663	0.207259222086362\\
7.04005354095618	0.192213359659554\\
7.20714531749297	0.167091776536787\\
7.3682427665445	0.161097449051534\\
7.5280376789694	0.159794912424896\\
7.72844335319582	0.20040567422642\\
7.91890302890552	0.190459675709705\\
};
\addlegendentry{$P_5$}

\addplot [only marks, mark=o, mark options={solid, black}]
  table[row sep=crcr]{%
1.42933567655214	1.03917702378769\\
1.90962311791842	0.480287441366282\\
2.81994692621099	0.91032380829257\\
3.26595757078876	0.446010644577767\\
3.76883444340481	0.502876872616053\\
4.23302543258793	0.464190989183113\\
4.64140714154774	0.408381708959817\\
5.10955968407908	0.468152542531336\\
5.59418424262613	0.484624558547048\\
5.77590014532235	0.181715902696223\\
5.97559937630899	0.199699230986639\\
6.15521187560303	0.179612499294043\\
6.31324162956541	0.158029753962376\\
6.50163680725373	0.188395177688319\\
6.6690914742158	0.167454666962071\\
6.84404354844691	0.17495207423111\\
7.00226032047399	0.158216772027084\\
7.16397941252579	0.161719092051795\\
7.33601039474468	0.172030982218894\\
7.50776104785818	0.171750653113497\\
7.69034342180976	0.182582373951576\\
7.90403279709314	0.213689375283387\\
};
\addlegendentry{$P_6$}

\addplot [only marks, mark=triangle, mark options={solid, black}]
  table[row sep=crcr]{%
0.624990193452181	0.418638933181623\\
1.72934903669551	1.10435884324333\\
2.18638013912249	0.457031102426973\\
2.8540794138308	0.667699274708308\\
3.62360999496067	0.769530581129874\\
4.65662167659033	1.03301168162966\\
5.14603187794233	0.489410201351995\\
5.876441579669	0.730409701726667\\
6.09082531037623	0.214383730707236\\
6.26593358295218	0.175108272575947\\
6.51501624103263	0.249082658080452\\
6.70254442012145	0.187528179088821\\
6.8900107260638	0.187466305942347\\
7.08175294961443	0.191742223550635\\
7.24860227113933	0.166849321524897\\
7.41552925164427	0.166926980504939\\
7.62460231076543	0.209073059121166\\
7.80684622923696	0.182243918471523\\
7.99499435875309	0.188148129516136\\
};
\addlegendentry{$P_7$}

\addplot [only marks, mark=square, mark options={solid, red}]
  table[row sep=crcr]{%
0.891234307856392	0.564220739187054\\
1.21712817904797	0.325893871191577\\
1.79426731609255	0.577139137044577\\
2.18989663056539	0.395629314472848\\
2.6188859083203	0.428989277754906\\
2.97249482859992	0.353608920279623\\
3.55027407633695	0.57777924773703\\
3.98919627495547	0.438922198618519\\
4.4892337471213	0.500037472165833\\
5.14728990209461	0.658056154973304\\
5.5511262124535	0.403836310358894\\
5.86940450828468	0.318278295831177\\
6.06736878591419	0.197964277629514\\
6.2817021304345	0.214333344520303\\
6.47215438016533	0.190452249730839\\
6.67510043984949	0.202946059684158\\
6.873716375562	0.198615935712508\\
7.09353203903619	0.219815663474188\\
7.25655194853715	0.163019909500966\\
7.45332427725431	0.196772328717158\\
7.62391810378955	0.17059382653524\\
7.82755299663322	0.203634892843668\\
};
\addlegendentry{$P_8$}

\addplot [color=red]
  table[row sep=crcr]{%
0	0.118035300737759\\
8	0.118035300737759\\
};
\addlegendentry{$\tau_\text{MIET}^2$}

\end{axis}
\end{tikzpicture}%

%% file: Figures/Dolk2019_InterEventTimes_c.tex
% !TEX root = ../ieeetac.tex
\begin{tikzpicture}[baseline=(current bounding box.center)]
    \footnotesize
\begin{axis}[%
width=\columnwidth,
height=4cm,
at={(0,0)},
%scale only axis,
xmin=0,
xmax=8,
xlabel style={font=\color{white!15!black}},
xlabel={Time [s]},
ymin=0,
ymax=3,
ylabel style={font=\color{white!15!black}},
ylabel={Inter-event times},
%axis background/.style={fill=white},
axis x line*=bottom,
axis y line*=left,
xmajorgrids,
ymajorgrids,
legend columns=5,
legend style={legend cell align=left, align=left, draw=white!15!black},
]
\addplot [only marks, mark=x, mark options={solid, black}]
  table[row sep=crcr]{%
1.52176544498105	1.14430694053557\\
2.13420787774576	0.612442432764711\\
3.16984339450141	1.03563551675565\\
3.63744748969332	0.467604095191902\\
4.43704040806315	0.79959291836983\\
5.10280968743133	0.665769279368181\\
6.04378576399927	0.940976076567947\\
6.5515690337467	0.50778326974743\\
7.10918755067377	0.557618516927063\\
7.70120422853986	0.592016677866091\\
7.99583176513335	0.294627536593492\\
};
\addlegendentry{$P_1$}

\addplot [only marks, mark=o, mark options={solid, red}]
  table[row sep=crcr]{%
0.287827745666839	0.145526620912954\\
0.430389022997375	0.142561277330536\\
0.704098952803524	0.273709929806148\\
1.19826484201493	0.494165889211411\\
1.35185108826342	0.153586246248482\\
1.47840952656711	0.126558438303694\\
1.70921748793937	0.230807961372258\\
2.24712287061706	0.537905382677692\\
2.64316045898805	0.396037588370986\\
3.01292620828453	0.369765749296483\\
3.15391464972765	0.14098844144312\\
3.44895689685971	0.295042247132063\\
4.91432018341364	1.46536328655393\\
5.22962900856284	0.315308825149199\\
5.78141802088029	0.551789012317449\\
6.05157635642002	0.270158335539737\\
6.5343430597613	0.482766703341274\\
6.87271942200464	0.338376362243348\\
7.33541076656134	0.462691344556697\\
7.46817496372257	0.132764197161224\\
7.59285754061024	0.124682576887671\\
7.89853003297296	0.305672492362729\\
};
\addlegendentry{$P_2$}

\addplot [only marks, mark=triangle, mark options={solid, red}]
  table[row sep=crcr]{%
0.491595832487648	0.349290496051458\\
1.09154485696997	0.599949024482324\\
1.6830647150377	0.591519858067728\\
2.04794231290407	0.36487759786637\\
3.29155500378845	1.24361269088438\\
3.70041785526224	0.408862851473789\\
4.06682951476355	0.366411659501317\\
4.51440605356048	0.447576538796927\\
4.86664907309505	0.352243019534574\\
5.2407233279532	0.374074254858142\\
5.89028902241316	0.649565694459968\\
6.74674323313293	0.856454210719766\\
6.88936990448562	0.142626671352687\\
7.07898586478745	0.189615960301833\\
7.34807768562964	0.269091820842185\\
7.55398345927103	0.205905773641391\\
7.9098526282147	0.355869168943672\\
};
\addlegendentry{$P_3$}

\addplot [only marks, mark=square, mark options={solid, black}]
  table[row sep=crcr]{%
0.577778940716862	0.413736361363744\\
1.77242691634511	1.19464797562825\\
2.14769059618733	0.375263679842216\\
2.59651526526225	0.448824669074923\\
3.10300167727805	0.506486412015804\\
3.47814619227445	0.375144514996392\\
3.74293786763381	0.264791675359365\\
4.13650513863891	0.393567271005096\\
4.549112426988	0.41260728834909\\
5.16905652211292	0.61994409512492\\
5.67117103623277	0.502114514119858\\
6.14935901313227	0.478187976899499\\
6.65168528482933	0.502326271697054\\
7.7947893085096	1.14310402368028\\
};
\addlegendentry{$P_4$}

\addplot [color=black]
  table[row sep=crcr]{%
0	0.156171044770359\\
8	0.156171044770359\\
};
\addlegendentry{$\tau_\mathrm{MIET}^1$}

\addplot [only marks, mark=x, mark options={solid, red}]
  table[row sep=crcr]{%
0.319247298595033	0.158971147447055\\
0.655284655520461	0.336037356925428\\
1.3060304620257	0.650745806505236\\
1.64667644712801	0.340645985102312\\
2.2045883820782	0.557911934950189\\
2.57349908242744	0.368910700349245\\
2.94049034821464	0.366991265787196\\
3.32362277049576	0.383132422281116\\
3.83232957968733	0.508706809191573\\
4.23673298450479	0.404403404817462\\
4.72362289605419	0.486889911549403\\
5.21300980583768	0.489386909783486\\
5.77974666778908	0.566736861951398\\
6.26838176736584	0.488635099576765\\
6.73166100567257	0.46327923830673\\
7.13645676248098	0.404795756808412\\
7.60920087805282	0.472744115571837\\
};
\addlegendentry{$P_5$}

\addplot [only marks, mark=o, mark options={solid, black}]
  table[row sep=crcr]{%
1.42933888567865	1.03915093963319\\
1.90924658028979	0.479907694611138\\
2.8261578166128	0.916911236323011\\
3.2682192593594	0.442061442746598\\
3.74397454570625	0.475755286346849\\
4.2134050304967	0.469430484790453\\
4.62403040535711	0.410625374860413\\
5.1659501200777	0.541919714720585\\
5.73610748848062	0.570157368402922\\
6.23864281340674	0.502535324926124\\
6.61409051861762	0.375447705210881\\
7.1055165113075	0.491425992689872\\
7.43266404355545	0.327147532247949\\
};
\addlegendentry{$P_6$}

\addplot [only marks, mark=triangle, mark options={solid, black}]
  table[row sep=crcr]{%
0.624769211179286	0.418503391314157\\
1.72682244673599	1.10205323555671\\
2.18535752816807	0.458535081432073\\
2.85694639289728	0.671588864729214\\
3.63152127380729	0.774574880910007\\
4.64635143232274	1.01483015851545\\
5.14962289638236	0.503271464059627\\
5.87399705657459	0.724374160192221\\
6.44892234085924	0.574925284284658\\
7.66271039711953	1.21378805626028\\
};
\addlegendentry{$P_7$}

\addplot [only marks, mark=square, mark options={solid, red}]
  table[row sep=crcr]{%
0.891367151890074	0.564365917557736\\
1.21741059723649	0.326043445346412\\
1.79492629665916	0.577515699422679\\
2.19072014220199	0.395793845542825\\
2.62173361341646	0.43101347121447\\
2.97498612526771	0.353252511851245\\
3.54699118141197	0.572005056144266\\
3.98851897689954	0.441527795487571\\
4.50663718404901	0.518118207149471\\
5.21462973168742	0.707992547638408\\
5.75994550227142	0.545315770584003\\
6.12835978798107	0.368414285709649\\
6.54822070403622	0.419860916055145\\
6.89925954777231	0.351038843736092\\
7.3222828261538	0.423023278381494\\
7.67316839879008	0.350885572636277\\
7.93457598102466	0.261407582234582\\
};
\addlegendentry{$P_8$}

\addplot [color=red]
  table[row sep=crcr]{%
0	0.118035300737759\\
8	0.118035300737759\\
};
\addlegendentry{$\tau_\mathrm{MIET}^2$}

\end{axis}
\end{tikzpicture}%